\documentclass[journal,onecolumn]{IEEEtran}

\usepackage{graphicx}
\usepackage{amsmath,bm}
\usepackage{amssymb,mathrsfs}
\newtheorem{remark}{Remark}[section]
\newtheorem{lemma}{Lemma}[section]
\newtheorem{theorem}{Theorem}[section]

\newtheorem{example}{Example}[section]
\usepackage[colorlinks=true, allcolors=blue]{hyperref}
\usepackage[caption=false]{subfig}
\usepackage{balance}
\usepackage[switch,pagewise]{lineno}
\newcommand{\m}[1]{\mathbf{#1}}
\newcommand{\mc}[1]{\mathcal{#1}}
\newcommand{\mb}[1]{\mathbb{#1}}

\begin{document}
\title{Matrix-Scaled Consensus over Undirected Networks}

\author{ Minh Hoang Trinh\IEEEauthorrefmark{1},~\IEEEmembership{Member,~IEEE},~Hoang Huy Vu\IEEEauthorrefmark{2},~Nhat-Minh Le-Phan\IEEEauthorrefmark{2}\IEEEauthorrefmark{3}, Quyen Ngoc Nguyen\IEEEauthorrefmark{2}
\thanks{$^{*}$AI Department, FPT University, An Thinh Phu New Urban Area, Nhon Binh Ward, Quy Nhon City, Binh Dinh 55117, Vietnam. Corresponding author. Email: \texttt{minhtrinh@ieee.org}}
\thanks{$^{\ddagger}$Department of Automation Engineering, School of Electrical and Electronic Engineering, Hanoi University of Science and Technology (HUST), 1 Dai Co Viet Str., Hai Ba Trung Dist., Hanoi 11615, Vietnam. Emails: \texttt{\{Hoang.VH200250,quyen.NN200518\}@sis.hust.edu.vn}}
\thanks{$^{\sharp}$Research and Development in Navigation, Guidance and Control
Technologies Center, Viettel Aerospace Institute, Hanoi 13151, Vietnam. Email: \texttt{minh.lpn221013m@sis.hust.edu.vn}}
\thanks{Manuscript received ...}}

\maketitle

\begin{abstract}
In this paper, we propose matrix-scaled consensus algorithms for linear dynamical agents interacting over an undirected network. Under the proposed algorithms, the state vectors of all agents to asymptotically agree up to some matrix scaling weights. First, the {algebraic properties of the} matrix-scaled Laplacian { and the geometry of the matrix-scaled consensus space are studied.} Second, we examine matrix-scaled consensus algorithms for networks of single-integrators with or without constant parametric uncertainties. {Nonlinear and finite-time matrix-scaled consensus algorithms} are also proposed. Third, observer-based matrix-scaled consensus algorithms for homogeneous or heterogeneous {linear-time invariant} agents are designed. The convergence of the proposed algorithms is asserted by rigorous mathematical analysis and supported by numerical simulations.
\end{abstract}

\begin{IEEEkeywords}
multi-agent systems, consensus, multi-dimensional opinion dynamics, adaptive control
\end{IEEEkeywords}


\section{Introduction}
\label{sec:introduction}

Collective behaviors displayed in nature such as bird flocking, fish schooling, synchronous fireflies, have inspired a lot of research works. Though simple, the consensus algorithm has been widely used for modeling and studying such striking phenomena \cite{Olfati2007consensuspieee}. It is interesting that a lot of multiagent systems such as autonomous vehicle formations, electrical system, sensor networks, or social networks can be coordinated by appropriately applying consensus algorithms and its modifications \cite{Ren2007magazine,Proskurnikov2017tutorial}.

Consider a network in which the interactions between subsystems, or agents, is modeled by a graph. In the consensus algorithm, each agent updates its state based on the sum of the relative states with its nearby agents. If the interaction graph is connected, the agents' states asymptotically converge to a common point in the space, and {the system is said to asymptotically achieve} a consensus \cite{Olfati2007consensuspieee}. 

A (scalar) scaled consensus model was proposed in \cite{Roy2015scaled}, in which each agent has a {nonzero} scalar scaling gain and updates its state variable based on the sum of differences in the scaled states. Under the scaled consensus model, the scaled states of all agents agree to the same virtual consensus value, and the state variable of each agent converges to a value differing from the virtual consensus value by an inverse of the scaling gain. {Thus, if each agent is characterized by a $d$-dimensional state vector, the state vectors of the agents asymptotically converges to a straight line through the origin under the scalar scaled consensus algorithm \cite{Roy2015scaled}.} The scaled consensus system can describe a cooperative network, where agents have different levels of consensus on a single topic. Further studies of the scalar scaled consensus algorithm with consideration to switching or signed graphs \cite{Meng2015scaled,Hanada2019new}, time delays \cite{Aghbolagh2016scaled,Shang2017delayed}, disturbance rejection \cite{Meng2015TIE}, or different agents' models can be found in the literature \cite{Wu2021adaptive,Chen2022scaled}. 

{Matrix-scaled consensus (MSC), a multi-dimensional generalization of the scaled consensus model in \cite{Roy2015scaled}, has been recently proposed in \cite{Trinh2022matrix}. Under the MSC algorithm, each agent is associated with a positive/negative definite scaling matrix, and the matrix-scaled states of all agents eventually agree to a virtual consensus point. The state vector of each agent asymptotically converges to a point, which can be obtained from the virtual consensus point by a linear transformation determined by the inverse of its scaling matrix. Thus, clustering behaviors usually happen due to the homogeneity of the scaling matrices. 
Several studies, e.g., MSC algorithms with state constraints \cite{Shang2023MSCconstraint}, resilient protocols \cite{Shang2023matrix}, some uncertain dynamics \cite{Chen2023matrix,Cheng2023matrix}, or for double-integrator without velocity measurements \cite{Vu2024VCCA} have been recently proposed.}

{In this paper, we firstly characterize the geometry of the matrix-scaled consensus space and the algebraic properties of the matrix scaled Laplacian. Second, we study matrix-scaled consensus algorithm for a network of single integrators interacting over an undirected network. Besides revisiting the basic algorithm in \cite{Trinh2022matrix}, we also consider nonlinear MSC algorithms and providing a corresponding Lyapunov-based analysis. %
{Third, we study the MSC algorithm for a system of single integrators with parametric uncertainties. An adaptive scheme is developed, which guarantees the system to eventually achieve matrix-scaled consensus.} If in addition, a persistently exciting condition is satisfied, the estimate variables eventually {converge to the precise} parameters. Fourth, we consider a network of homogeneous linear agents and propose an observer-based matrix-scaled {consensus} algorithm. It is shown that the agents asymptotically achieve matrix-scaled consensus with regard to a {trajectory of a virtual agent}. If we {specify} the agents to be linear oscillators in 2D, and the scaling matrices to {be a composition of rotation and expansion/compression matrices, the asymptotic biases in phase and magnitude of each agent's trajectory with a common oscillator can be determined by the scaling matrices}. Finally, as practical networks often consists of agents with different sizes and capacities, we propose a matrix-scaled consensus algorithm for a network of heterogeneous linear agents. The proposed algorithm combines the corresponding MSC algorithm for homogeneous linear agents with a disturbance observer which compensates the differences between each agent's model and a pre-specified one. The main challenges and differences in the analysis of our proposed algorithms with regard to the existing {consensus algorithms for general linear agents} in the literature, e.g., \cite{Scardovi2009synchronization,Kim2010output,Li2009consensus,Li2013distributed,Panteley2017synchronization,Tuna2016synchronization,Burbano2019distributed}, are originated from the asymmetry of the matrix scaled Laplacian and the multi-dimension of the problem.}

{As a generalization of the scaled consensus algorithm, the MSC algorithm can replace scaled consensus in any application that scaled consensus are being used. First, the algorithm can be used as a biased resource allocation algorithm. Second, in \cite{Trinh2022matrix}, the MSC was interpreted as a multi-dimensional model opinion dynamics \cite{Ye2020Aut,Parsegov2016novel}. The positive/negative definite scaling matrix weight{s} capture the private belief system of an individual on $d$ logically dependent topics and clustering happens as the private belief system of each individual is usually different from each other and not perfectly aligned with a social norm. In this paper, we further demonstrate how to design the MSC space for generating a desired shape, thus, extent the applicability of the MSC algorithm to formation control \cite{Oh2015survey}.}

The remainder of this paper is organized as follows. Section~\ref{sec:2} provides theoretical background, main assumptions and definitions. The matrix-scaled consensus space and properties of the matrix-scaled Laplacian are discussed in detail in Section~\ref{sec:3}. Matrix-scaled consensus algorithms for single  integrator agents with and without parametric uncertainties are proposed and examined in Section~\ref{sec:single_integrator}. Section \ref{sec:linear_agents} studies matrix-scaled consensus for networks of homogeneous and heterogeneous linear agents. Simulations  are given in Section~\ref{sec:5} to support the theoretical results. Finally, Section~\ref{sec:6} concludes the paper. 

\section{Preliminaries}
\label{sec:2}
\subsection{Notations}
In this paper, the sets of real, complex, and natural numbers are denoted by $\mb{R}, \mb{C}$ and $\mb{N}$, respectively. Scalars are denoted by lowercase letters, while bold font normal and capital letters are used for vectors and matrices, respectively. {We denote $\m{1}_d \in \mb{R}^d$ the $d$-dimensional vector of all 1 an $\m{0}_{m\times n} \in \mb{R}^{m\times n}$ the matrix of all zeros.} The transpose of a matrix $\m{A} \in \mb{R}^{m\times n}$ is denoted by $\m{A}^\top$. The kernel, image, rank, and determinant of $\m{A}$ are respectively denoted as ker$(\m{A})$, im$(\m{A})$, rank($\m{A}$), and $\text{det}(\m{A})$. A diagonal matrix with elements $a_1, a_2, \ldots, a_n$ is denoted by $\text{diag}(a_1, a_2, \ldots, a_n)$. The 2-norm ($\infty$-norm) of a vector $\m{x} = [x_1, \ldots, x_d]^\top$ is denoted by $\|\m{x}\| = \sqrt{\sum_{i=1}^{d} x_i^2}$ (correspondingly, $\|\m{x}\|_{\infty} = \max_{i\in \{1,\ldots,d\}} |x_i|$). The 2-norm of a real matrix $\m{A}$, denoted by $\|\m{A}\|$, is defined as $\max_{\|\m{x}\|=1}\|\m{A}\m{x}\|$. A matrix $\m{A} \in \mb{R}^{d\times d}$ is positive definite (negative definite) if and only if $\forall \m{x} \in \mb{R}^d$, $\m{x}\neq \m{0}_d$, then $\m{x}^\top \m{A} \m{x}>0$ (resp., $\m{x}^\top \m{A} \m{x} < 0$). For a real, symmetric positive definite matrix $\m{W}\in \mb{R}^{d\times d}$, which can be diagonalized as $\m{T}^{-1}\m{A}\m{T} = \text{diag}(\omega_1,\ldots,\omega_d)$, we use $\sqrt{\m{A}}$ to denote its square root $\m{T}\text{diag}(\sqrt{\omega_1},\ldots,\sqrt{\omega_d})\m{T}^{-1}$. Let $\m{x}_1, \ldots, \m{x}_n \in \mb{R}^d$, the vectorization operator is defined as vec$(\m{x}_1,\ldots,\m{x}_n) = [\m{x}_1^\top,\ldots,\m{x}_n^\top]^\top \in \mb{R}^{dn}$. Given matrices $\m{A}_1,\ldots,\m{A}_n$, we use blkdiag$(\m{A}_1,\ldots,\m{A}_n)$ as the block diagonal matrix with $\m{A}_k$, $k=1,\ldots,n$ in the main diagonal. {For $x\in \mb{R}$, the signum function is defined as $\text{sgn}(x)=1$ if $x>0$, $\text{sgn}(x)=-1$ if $x<0$, and $\text{sgn}(x)=0$ if $x=0$.}

\subsection{Graph theory}
Consider a undirected graph $\mc{G} = (\mc{V},\mc{E},\mc{W})$, where $\mc{V}=\{1,\ldots,n\}$ is the set of $n$ vertices, {$\mc{E}=\{e_k=(i,j),~i\neq j \}_{k=1,\ldots,m} \subset \mc{V}\times \mc{V}$} is the set of $m$ edges, and {$\mc{W}=\{w_{ij}>0\}_{(i,j) \in \mc{E}}\equiv \{w_{k}\}_{k = 1, \ldots, m}$} is the set of positive scalar weights corresponding to each edge of $\mc{G}$. By undirectedness, if $(i,j) \in \mc{E}$, then $(j,i)\in \mc{E}$. {Further, it is assumed that there is no self-loop (an edge with the same end vertices $(i,i)$, $i \in \mc{V}$) in the graph.} The neighbor set of a vertex $i$ is defined as $\mc{N}_i = \{j|~(j,i) \in \mc{E}\}$. A path is a sequences of vertices connected by edges in $\mc{E}$, where each vertex appears one time, except for possibly the starting and the ending vertices. For examples, the path $\mc{P}=i_1i_2\ldots i_p$ from $i_1$ to $i_p$ has $(i_k,i_{k+1})\in \mc{E}$, $k=1,\ldots,p-1$. A cycle is a path with the same starting and ending vertices. A graph is connected if there exists a path between any two vertices in $\mc{V}$. 

A spanning subgraph $\mc{G}'=(\mc{V}',\mc{E}',\mc{W}')$ of $\mc{G}$ has $\mc{V}'= \mc{V}$ and $\mc{E}\subseteq \mc{E}'$. A spanning tree of $\mc{G}$ is a connected spanning sub-graph of $\mc{G}$ with $n-1$ edges. Let $\mc{T}$ be a spanning tree of $\mc{G}$ and consider an arbitrary labeling and orientation of the edges in $\mc{E}$ so that $n-1$ edges of $\mc{E}(\mc{T})$ are $e_1,\ldots,e_{n-1}$. 

Corresponding to this labeling and edge orientation, the incidence matrix of $\mc{G}$ is defined as $\m{H}=[h_{ki}]\in \mb{R}^{m\times n}$, where $h_{ki} = -1$ if $e_k=(i,j)$, {$h_{ki} = 1$ if $e_k=(j,i)$,  and $h_{ki} = 0$, otherwise.} It is well-known that if $\mc{G}$ is a connected graph, we have rank$(\m{H})=n-1$ and $\m{H}\m{1}_n = \m{0}_{m}$ \cite{MesbahiEgerstedt}.

Let $\m{H}_{\mc{T}} \in \mb{R}^{(n-1)\times n}$, and $\m{H}_{\mc{C}} \in \mb{R}^{(m-n+1) \times n}$ denote the corresponding incidence matrices of the subgraphs $\mc{T}$, and the subgraph $(\mc{V},\mc{E}\setminus\mc{E}(\mc{T}),\mc{W}\setminus\mc{W}(\mc{T}))$. Then, we can express the incidence matrix as
\begin{align*}
\m{H} = \begin{bmatrix}
\m{H}_{\mc{T}} \\ 
\m{H}_{\mc{C}}
\end{bmatrix} 
= \begin{bmatrix}
\m{I}_{n-1} \\ 
\m{T}
\end{bmatrix} \m{H}_{\mc{T}} = \m{R} \m{H}_{\mc{T}},
\end{align*}
where $\m{H}_{\mc{T}} \in \mb{R}^{(n-1)\times n}$, $\m{H}_{\mc{C}} \in \mb{R}^{(m-n+1)\times n}$, and $\m{T} = \m{H}_{\mc{C}} \m{H}_{\mc{T}}^\top (\m{H}_{\mc{T}} \m{H}_{\mc{T}}^\top)^{-1}$ \cite{Zelazo2010tac}. The graph Laplacian of $\mc{G}$ can be defined as $\m{L} =[l_{ij}]_{{n\times n}} = \m{H}^\top \m{W} \m{H} \in \mb{R}^{n\times n},$ 
where $\m{W} = \text{diag}(\ldots,w_{ij},\ldots) = \text{diag}(w_1,\ldots,w_m)$. The matrix $\m{L}$ of a connected graph is symmetric positive semidefinite, with spectrum $0=\lambda_1 <\lambda_2 \leq \ldots \leq \lambda_n$, and ker$(\m{L})=\text{im}(\m{1}_n)$. 

\subsection{{The multi-agent system and the objective}}
\begin{figure*}[th]
    \centering
    \subfloat[{Axes scaling}]{\includegraphics[width=.21\linewidth]{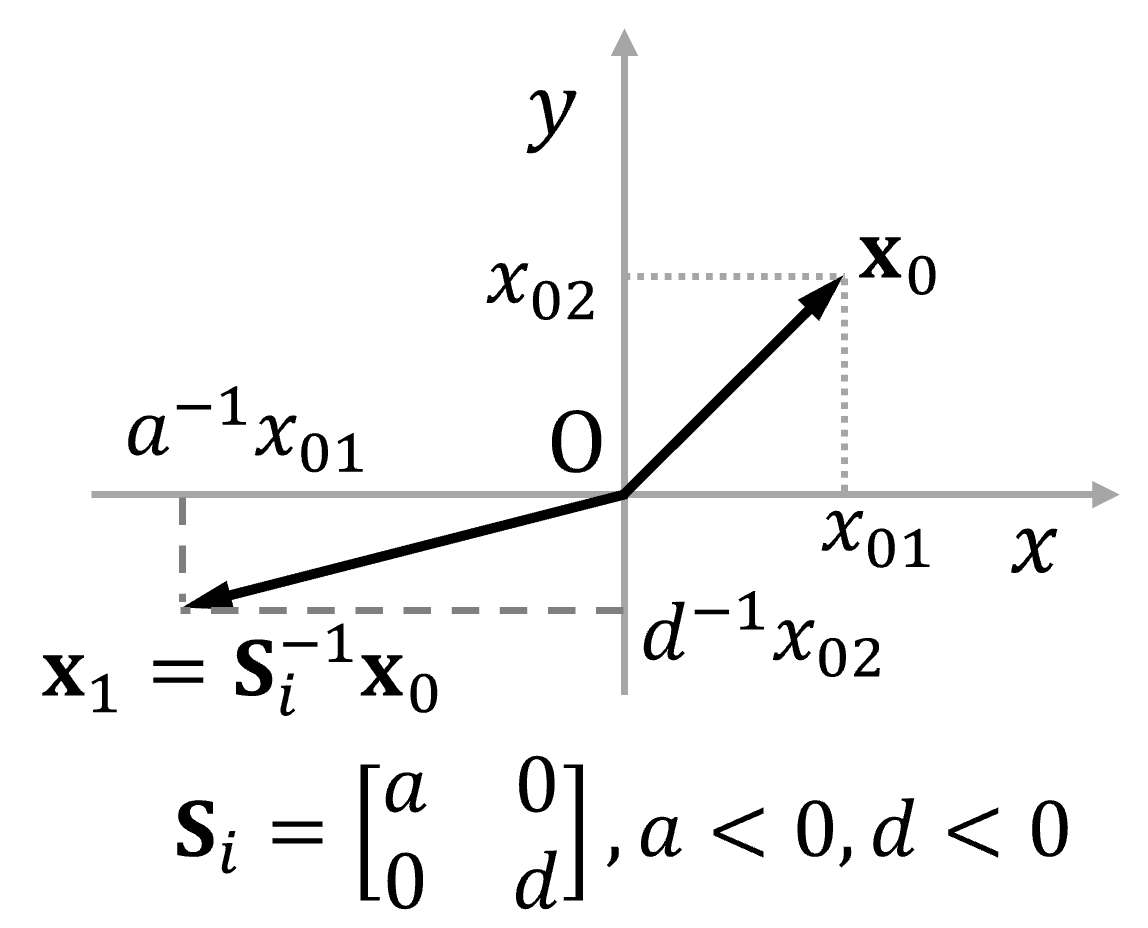}}
    \hfill
    \subfloat[{Rotation}]{\includegraphics[width=.18\linewidth]{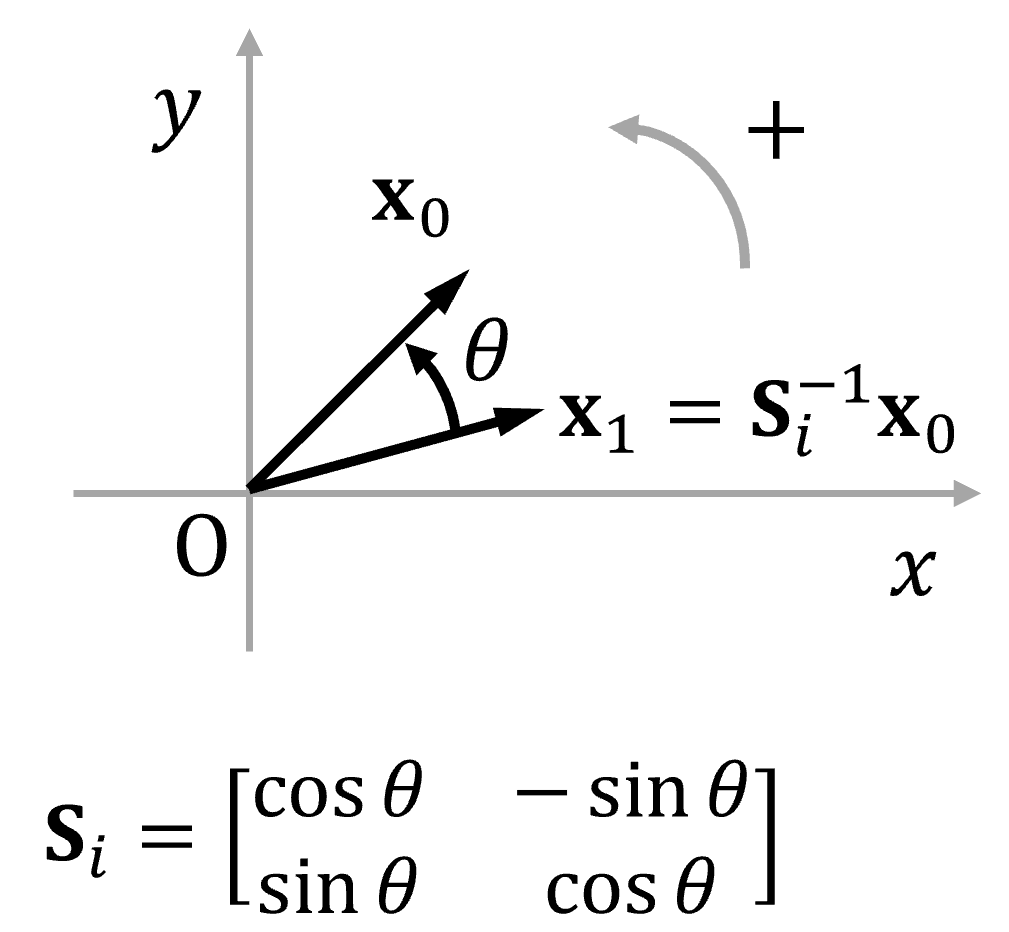}}
    \hfill
    \subfloat[{$x$-shear transform}]{\includegraphics[width=.2\linewidth]{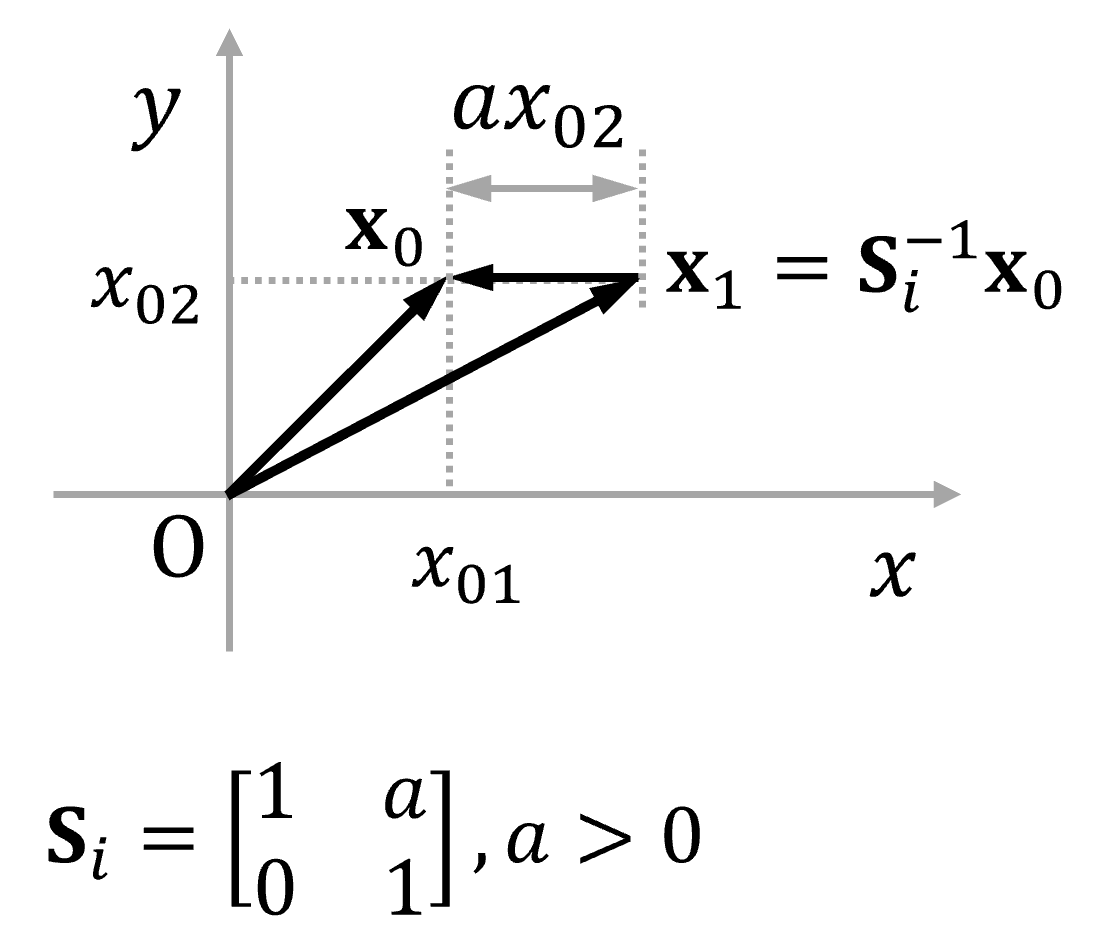}}
    \hfill
    \subfloat[{$y$-shear transform}]{\includegraphics[width=.18\linewidth]{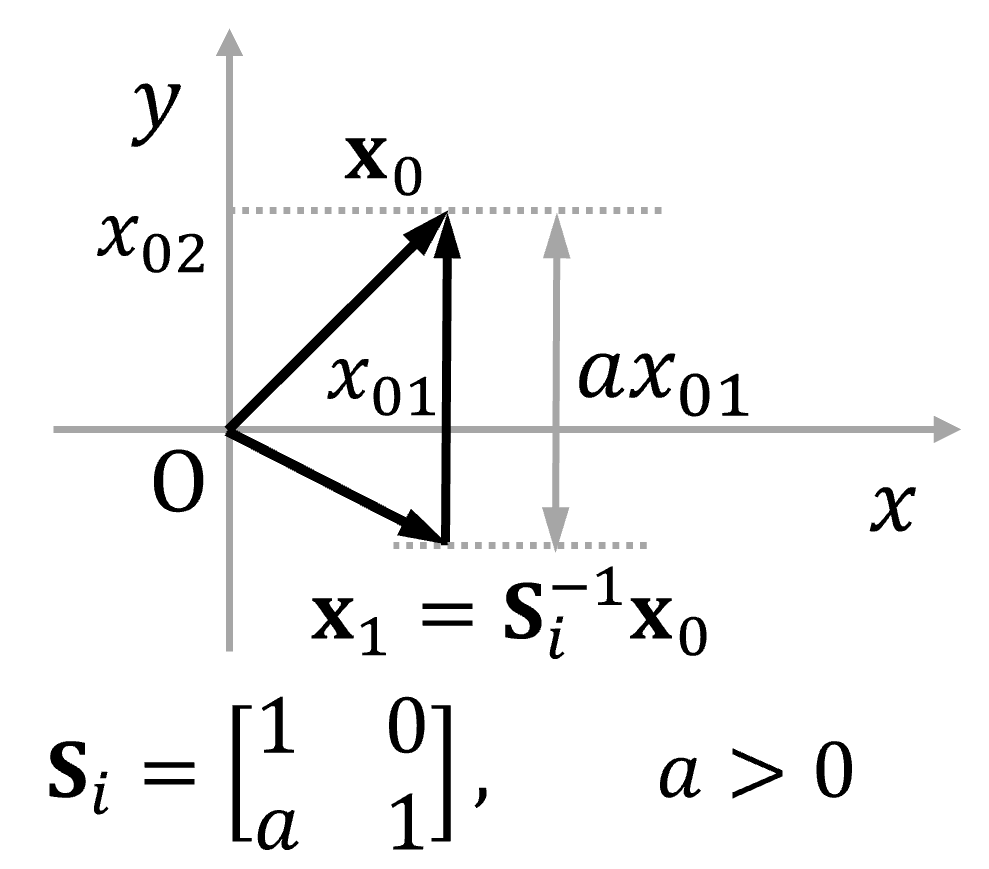}}
    \hfill
    \subfloat[{Virtual translation}]{\includegraphics[width=.2\linewidth]{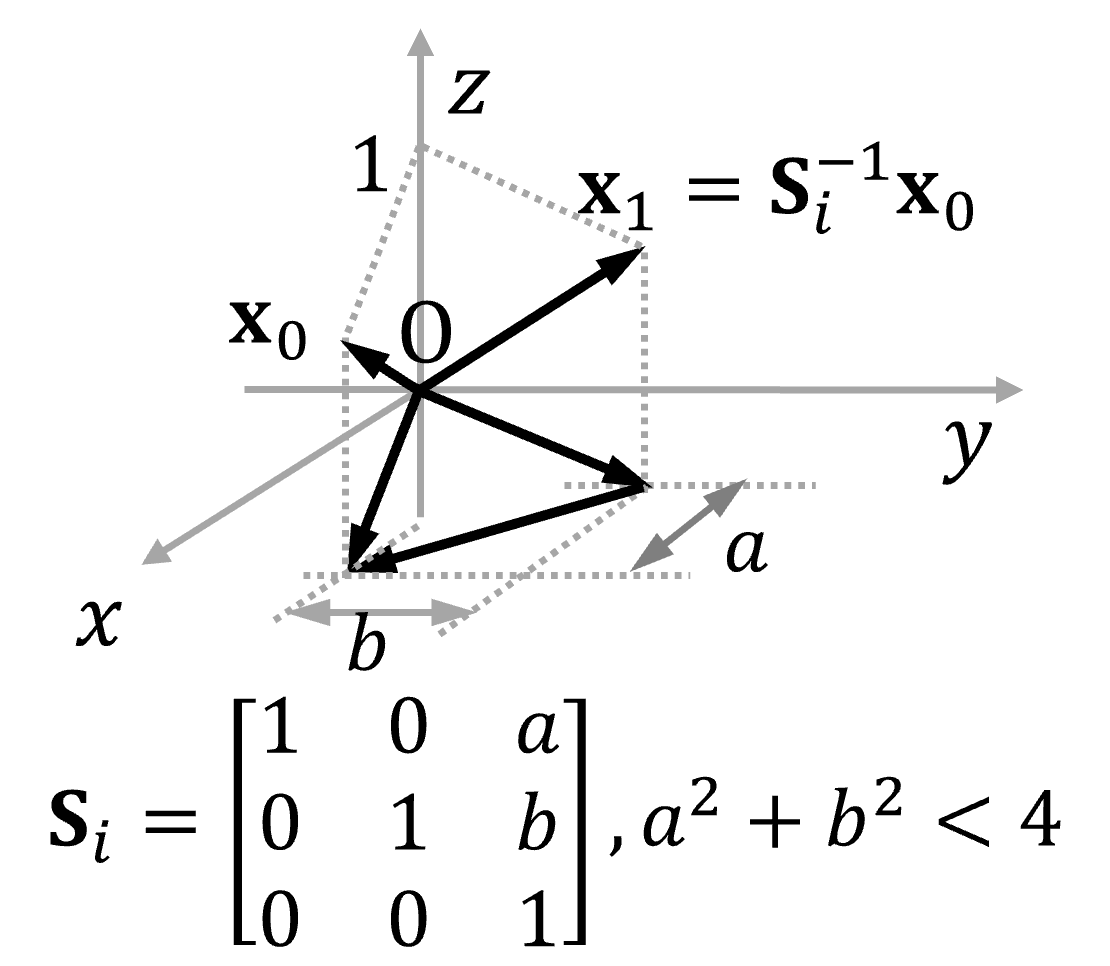}}
    \caption{{Geometry of the MSC space in $\mb{R}^2$ for several types of the scaling matrices $\m{S}_i$. \label{fig:geometryMSCS}}}
\end{figure*}
In this paper, we consider an $n$-agent system interacted via a {connected} undirected weighted graph $\mc{G}$. The dynamics of each agent is modeled by either
\begin{enumerate}
\item[(i)]  single-integrator with parametric uncertainty:
\begin{align} \label{eq:model}
\dot{\m{x}}_i = \m{u}_i + \bm{\phi}_i(t,\m{x}_i) \bm{\theta}_i,~i=1,\ldots, n,
\end{align}
where $\m{x}_i,\m{u}_i \in \mb{R}^d$ are the state variable vector and the control input, $\bm{\phi}_i(t,\m{x}_i),~\dot{\bm{\phi}}_i(t,\m{x}_i)\in \mb{R}^{d\times r}$ are the matrices of known bounded continuous functions, and $\bm{\theta}_i\in \mb{R}^r$ is a vector of constant unknown parameters; or
\item[(ii)] homogeneous {time-invariant} linear agents
\begin{subequations} \label{eq:homogeneous_linear}
\begin{align}
\dot{\m{x}}_i &= \m{A} \m{x}_i + \m{B} \m{u}_i,\\
\m{y}_i &= \m{C} \m{x}_i,~i=1,\ldots,n,
\end{align}
\end{subequations}
where $(\m{A},\m{B})$ is controllable and $(\m{A},\m{C})$ is observable. 
\item[(iii)]  heterogeneous {time-invariant} linear agents
\begin{subequations} \label{eq:heterogeneous_linear}
\begin{align}
\dot{\m{x}}_i &= \m{A}_i \m{x}_i + \m{B}_i \m{u}_i,\\
\m{y}_i &= \m{C}_i \m{x}_i,~i=1,\ldots,n,
\end{align}
\end{subequations}
where {$\m{A}_i \in \mb{R}^{d\times d},~\m{B}_i \in \mb{R}^{d\times r}$, $\m{C}_i \in \mb{R}^{r\times p}$,} $(\m{A}_i,\m{B}_i)$ is controllable and $(\m{A}_i,\m{C}_i)$ is observable, $\forall i \in \{1,\ldots, n\}$.\footnote{{{In the agent's models (ii) and (iii), the controllability of $(\m{A}_i,\m{B}_i)$ and observability of $(\m{A}_i,\m{C}_i)$ imply that we can find $\m{K}_i,\m{H}_i$ to arbitrarily assign the eigenvalues of the matrices $\m{A}_i+\m{B}_i\m{K}_i$ and $\m{A}_i+\m{H}_i\m{C}_i$. In designing distributed algorithm that ensures the states to achieve a matrix-scaled consensus, we only need $(\m{A}_i,\m{B}_i)$ to be stabilizable and $(\m{C}_i,\m{A}_i)$ to be detectable\cite{Ma2010necessary}. Also, as all matrices $\m{A}_i, \m{C}_i$ has the same dimension, the necessary condition for the $n$-agent system to achieve state MSC, as discussed in \cite[Thm.~6]{Wieland2009internal}, is satisfied.}}}
\end{enumerate}

We associate to each agent a scaling matrix $\m{S}_i \in \mb{R}^{d\times d}$ ($d\ge 2$) and a state vector $\m{x}_i \in \mb{R}^d$. The matrix $\m{S}_i$ is either positive definite or negative definite.  For each scaling matrix, we define a matrix signum function 
\begin{align}
\text{sign}(\m{S}_i) = \left\{ {\begin{array}{*{20}{ll}}
{1},& \text{$\m{S}_i$ is positive definite},\\
{-1},& \text{$\m{S}_i$ is negative definite}.
\end{array}} \right.
\end{align}
and an absolute matrix function $|\m{S}_i| = \text{sign}(\m{S}_i)\m{S}_i$. It is not hard to see that $\text{sign}(\m{S}_i) = \text{sign}(\m{S}_i^\top) = \text{sign}(\m{S}_i^{-1})$, and $|\m{S}_i^{-1}| = |\m{S}_i|^{-1}$ \cite{NicholsonLAA}. 

Let $\m{x} = \text{vec}(\m{x}_1,\ldots,\m{x}_n)$, $\m{S}=\text{blkdiag}(\m{S}_1,\ldots,\m{S}_n)$, $\text{sign}(\m{S}) = \text{diag}(\text{sign}(\m{S}_1),\ldots,\text{sign}(\m{S}_n))$, and $|\m{S}|=\text{blkdiag}(|\m{S}_1|,\ldots,|\m{S}_n|)$. {The $n$-agent system (or the network) is said to achieve a matrix-scaled consensus (MSC) if and only if $
\m{x} \in \mc{C}_S$, where}
\begin{align} \label{eq:msc_space}
\mc{C}_S = \{\m{x} \in \mb{R}^{dn}|~ \m{S}_i\m{x}_i = \m{S}_j \m{x}_j,~\forall i, j = 1,\ldots, n\}.
\end{align}
is the MSC {space. In this paper, our objective is designing distributed algorithms so that $\m{x}(t)$ globally asymptotically approaches $\mc{C}_S$ (i.e., $\forall\m{x}(0),~\m{x}(t)\to \mc{C}_S$, as $t\to+\infty$). Equivalently, $\forall \m{x}(0) \in \mb{R}^{dn}$, we would like $\lim_{t\to+\infty}(\m{S}\m{x}_i-\m{S}_j\m{x}_j) = \m{0}_d,~\forall i,j = 1,\ldots,n$.}

{
\section{Matrix-scaled consensus space and matrix-scaled Laplacian}
\label{sec:3}
\subsection{Geometry of the matrix-scaled consensus space in $\mb{R}^2$}
This subsection is devoted to study the matrix-scaled consensus space $\mc{C}_S$. We restricted the study for $d=2$ (the $Oxy$ coordinate system) because in this case, we can fully examine elementary linear transformations associated with a $2\times 2$ positive/negative definite scaling matrix. By this way, we will characterize the geometry of the matrix-scaled consensus space and sketch a guideline on designing of the scaling matrices for a particular objective. First, we recall several results from linear algebra, whose proofs are straight forward and will be omitted.
\begin{lemma}\label{lem:geometry} \cite[Section~4.5]{NicholsonLAA}
The matrix $\m{S}_i = \begin{bmatrix} a & b\\ c & d\end{bmatrix} \in \mb{R}^{2\times 2}$ is positive definite (resp., negative definite) if and only if $a+d>0$ and $4ad > (b+c)^2$ (resp., $a + d < 0$ and $4ad > (b+c)^2$). Particularly,
\begin{enumerate}
\item[(i)] an axis-scaling matrix $\m{S}_{\text{sc}} = \begin{bmatrix} a & 0\\ 0 & d\end{bmatrix}$, where $a$ and $d$ are respectively the $x$- and $y$-scaling factors, is positive definite (resp., negative definite) if and only if $a>0$ and $d > 0$ (resp., $a < 0$ and $d<0$).
\item[(ii)] a rotation matrix $\mathbf{R}(\theta) = \begin{bmatrix}
\cos(\theta)&-\sin(\theta)\\\sin(\theta) & \cos(\theta)
\end{bmatrix}$ with $\theta \in [0,2\pi)$ is positive definite (resp., nagative definite) if and only if $\theta\in [0,\frac{\pi}{2}) \cup (\frac{3\pi}{2},2\pi)$ (resp. $\theta \in (\frac{\pi}{2},\frac{3\pi}{2})$).
\item[(iii)] $x$-, $y$-shear transformation matrices $\m{S}_{\text{shx}}= \begin{bmatrix} 1 & c\\ 0 & 1\end{bmatrix}$ and $\m{S}_{\text{shy}}= \begin{bmatrix} 1 & c\\ 0 & 1\end{bmatrix}$ are positive definite if and only if $|c|<2$.
\item[(iv)] a matrix $\m{S}_{\text{trans}}' = \begin{bmatrix}
1 & 0 & a\\0 & 1 & b\\0 & 0 & 1
\end{bmatrix}$ corresponding to a translation in O$xy$, given that the homogeneous coordinate $\m{x}_i'=[\m{x}_i^\top,1]^\top,~i=1,\ldots,n,$ is adopted. The matrix $\m{S}_{\text{trans}}'$ is positive definite if and only if $a^2 + b^2 < 4$.
\item[(v)] $\m{S}_i' = \left[ {\begin{array}{*{20}{c}}
  \m{S}_i &\vline & \begin{array}{*{20}{c}}
  a_i \text{sign}(\m{S}_i) \\ 
  b_i \text{sign}(\m{S}_i) 
\end{array} \\ 
\hline
\begin{array}{*{20}{c}}
{0}  &  {0} 
\end{array}&\vline & \text{sign}(\m{S}_i) 
\end{array}} \right]$, where $\m{S}_i \in \mb{R}^{2 \times 2}$ belongs to one of type (i)-- (iii) matrices in this lemma and $a^2_i+b^2_i<4$, has the same positive/negative definiteness property as the matrix $\m{S}_i$.
\end{enumerate}
\end{lemma}
Equipped with Lemma~\ref{lem:geometry}, we can now characterize the geometry of the MSC space. From the definition in Eqn.~\eqref{eq:msc_space}, a point $\m{x}=\text{vec}(\m{x}_1,\ldots,\m{x}_n) \in \mb{R}^{2n}$ belongs to $\mc{C}_S$ if and only if there exists $\m{x}_0 \in \mb{R}^2$ such that $\m{S}_i\m{x}_i = \m{x}_0$, $\forall i = 1, \ldots, n$. Thus, given that the system has achieved a MSC, one can firstly specify the \emph{virtual consensus point} $\m{x}_0=[x_{01},x_{02}]^\top \in \mb{R}^2$, and then subsequently determine $n$ clusters $\m{x}_i = \m{S}_i^{-1}\m{x}_0$, $\forall i = 1, \ldots, n$. The matrix $\m{S}_i^{-1}$ specifies an inverse linear transformation with regard to the linear transformation defined by $\m{S}_i$. Thus, if $\m{x}_0 \neq \m{0}_2$, we can visualize the geometry of $\mc{C}_S$ via individual equations $\m{x}_i = \m{S}_i^{-1}\m{x}_0$. Figure~\ref{fig:geometryMSCS} visualizes the geometry of the $\mc{C}_S$ for different types of the scaling matrix $\m{S}_i$, each corresponding to a basic linear transformation.
}

\begin{figure}[t]
    \centering
    \subfloat[]{\includegraphics[width=.35\linewidth]{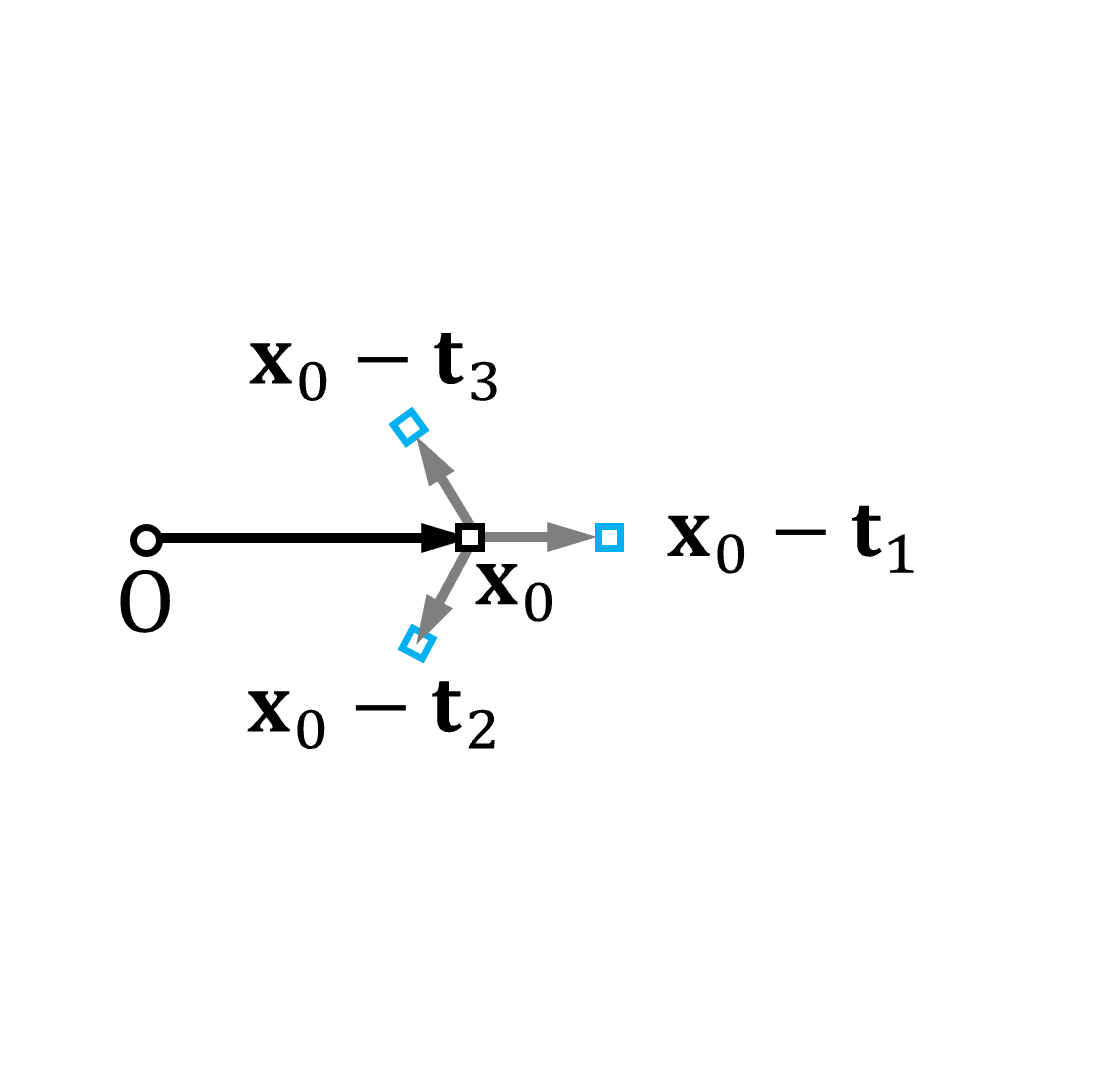}}\qquad\qquad
    \subfloat[]{\includegraphics[width=.35\linewidth]{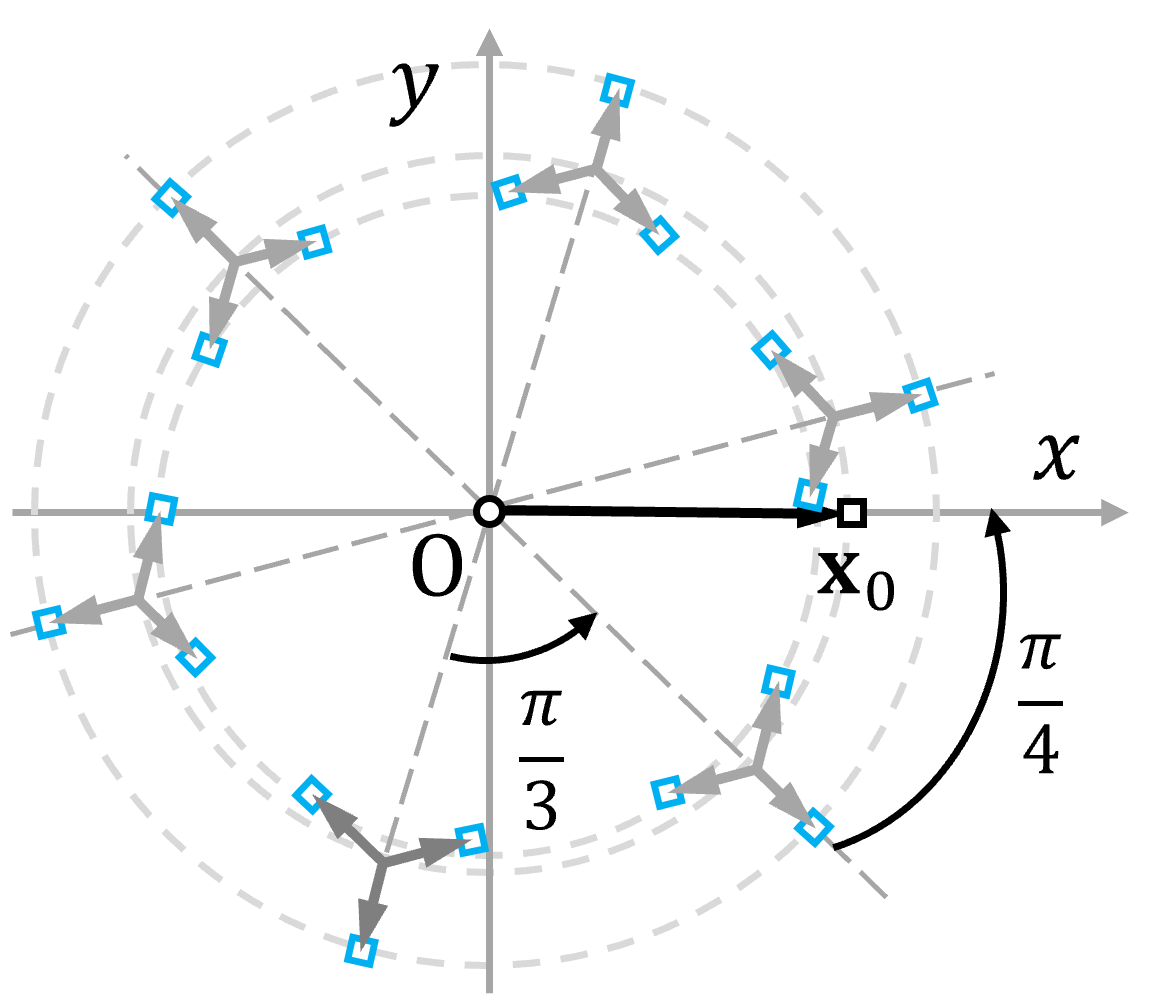}}
    \caption{{A snowflake-like formation in $\mc{C}_S$ corresponding to $\m{S}_i$ in Example~\ref{eg:snowflake}.}}
    \label{fig:snow}
\end{figure}

{While the geometric interpretation of Figs.~\ref{fig:geometryMSCS}a--d are clear, Fig.~\ref{fig:geometryMSCS}e requires a modification of our setup. Each agent has a state vector $\m{x}_i = [x_{i1},x_{i2}]^\top \in \mb{R}^2$ (the real coordinate). For realizing a translation with regard to the virtual consensus point $\m{x}_0$, each agent's state vector is appended with a \emph{signed homogeneous coordinate}, i.e., $\m{x}_i'= [x_{i1},x_{i2},\text{sign}(\m{S}_i)]^\top \in \mb{R}^3$. The MSC space is defined with regard to the equations $\m{S}_i'\m{x}_i'=\m{S}_j'\m{x}_j'=\m{x}_0'$,~$i,j=1,\ldots,n,~i\neq j$, where $\m{x}_0' = \begin{bmatrix}
    \m{x}_0 \\ 1
\end{bmatrix}$ and $\m{S}_i' = \left[ {\begin{array}{*{20}{c}}
  \m{S}_i &\vline & \begin{array}{*{20}{c}}
  a_i \text{sign}(\m{S}_i) \\ 
  b_i \text{sign}(\m{S}_i) 
\end{array} \\ 
\hline
\begin{array}{*{20}{c}}
{0}  &  {0} 
\end{array}&\vline & \text{sign}(\m{S}_i) 
\end{array}} \right]= \text{sign}(\m{S}_i) \left[ {\begin{array}{*{20}{c}}
  |\m{S}_i| &\vline & \begin{array}{*{20}{c}}
  a_i  \\ 
  b_i 
\end{array} \\ 
\hline
\begin{array}{*{20}{c}}
{0}  &  {0} 
\end{array}&\vline & 1
\end{array}} \right]$ as in Lemma~\ref{lem:geometry}(v). Note that $a_i=b_i=0$ if the agent $i$ does not wish to take a translation. In the following example, we will design the MSC space for a specific shape.}
{\begin{example}[A snowflake formation]\label{eg:snowflake} For $n=18$, $\theta = \pi/3$, the snowflake shape in Fig.~\ref{fig:snow} belongs to the MSC space for $\m{S}_i' = \left[ {\begin{array}{*{20}{c}}
  \m{S}_i &\vline & \begin{array}{*{20}{c}}
  \m{t}_i \text{sign}(\m{S}_i)
\end{array} \\ 
\hline
\begin{array}{*{20}{c}}
{0}  &  {0} 
\end{array}&\vline & \text{sign}(\m{S}_i) 
\end{array}} \right]$, $i=1,\ldots,18$, where  $\m{S}_{3k+1}=\m{S}_{3k+2}=\m{S}_{3k+3}=\m{R}(\pi/4 + k\theta)$, $\m{t}_{3k+1}=\begin{bmatrix}
    1\\ 0
\end{bmatrix}$, $\m{t}_{3k+2}=\m{R}(2\theta)\begin{bmatrix}
    1\\ 0
\end{bmatrix}$, and $\m{t}_{3k+3}=\m{R}(4\theta)\begin{bmatrix}
    1\\ 0
\end{bmatrix}$,~$\forall k=0,\ldots,5$. As equations $\m{S}_i\m{x}_i = \m{x}_0-\m{t}_i$ hold $\forall i$, once $\m{x}_0$ is available, we can find $\m{x}_0-\m{t}_i,~i=1,2,3$ as depicted in Fig.~\ref{fig:snow}(a). Then, the position of agent $i$ are determined by rotating these vectors angles $\frac{\pi}{4}+k\frac{\pi}{6}$ (rad),$~k=0,\ldots,5,$ clockwise (Fig.~\ref{fig:snow}(b)).
\end{example}}
{
\subsection{Algebratic properties of the matrix-scaled Laplacian}
Defining the matrix $\bm{\Omega}=(\text{sign}(\m{S})\m{L}\otimes \m{I}_d)\m{S}$, we will refer to $\bm{\Omega}$ as the \emph{matrix-scaled Laplacian} since its structure resembles a Laplacian with block matrix weights $\text{sign}(\m{S}_i)l_{ij}\m{S}_j$ \cite{Trinh2018matrix}. The following lemmas will be used throughout this paper.}
\begin{lemma} \label{lem:omega} The matrix $\bm{\Omega}$ has $d$ zero eigenvalues and $dn-d$ eigenvalues with positive real parts. The left and right kernels of $\bm{\Omega}$ are spanned by columns of $(\text{sign}(\m{S})\m{1}_n)\otimes \m{I}_d$ and $\m{S}^{-1}(\m{1}_n\otimes \m{I}_d)$, respectively.
\end{lemma}
{
\begin{IEEEproof} 
Since $|\m{S}|^\top = \m{S}^\top(\text{sign}(\m{S})\otimes\m{I}_d)$ is positive definite, the eigenvalues of $\bm{\Omega}$ are the same as those of $|\m{S}|^\top\bm{\Omega}=\m{S}^\top\bar{\m{L}}\m{S}$, where $\bar{\m{L}}=\m{L}\otimes\m{I}_d$. Indeed, if $\m{v} \in \mb{R}^{dn}$ is a left-eigenvector of $\bm{\Omega}$ corresponding to an eigenvalue $\Lambda$, then $|\m{S}|^{-1}\m{v}$ is also a left-eigenvector of $\m{S}^\top\bar{\m{L}}\m{S}$ corresponding to an eigenvalue $\Lambda$ \cite{NicholsonLAA}. Because $\m{S}^\top\bar{\m{L}}\m{S}$ is symmetric, $\bar{\m{L}}$ has $d$ zero eigenvalues and $dn-d$ positive eigenvalues. Thus,  $\m{S}^\top\bar{\m{L}}\m{S}$ and $\bm{\Omega}$ also have $d$ zero eigenvalues and $dn-d$ positive eigenvalues.\\
\quad By direct computation, we can check that $(\m{1}_n^\top\text{sign}(\m{S})\otimes \m{I}_d)\bm{\Omega} = \m{0}_{d\times dn}$ and $\bm{\Omega} (\m{S}^{-1}(\m{1}_n\otimes \m{I}_d) = \m{0}_{dn \times d}$. Furthermore, as we have shown that dim$(\text{ker}(\bm{\Omega}))=d$, it follows that range$(\text{sign}(\m{S})\m{1}_n)\otimes \m{I}_d) = $ker$(\bm{\Omega})$ and range$(\m{S}^{-1}(\m{1}_n\otimes \m{I}_d))= $ker$(\bm{\Omega}^\top)$. 
\end{IEEEproof}
}

{
\begin{lemma} \cite{Trinh2022matrix} Let $\m{V} = [\m{v}_1,\ldots,\m{v}_{dn}]$ and $\m{V}^{-1}=\m{Z}^\top$ be orthogonal matrices (which can be chosen with real entries) such that $\m{V}_{[1:d]}=\m{S}^{-1}(\m{1}_n\otimes \m{P})$, $\m{Z}_{[1:d]}=(\text{sign}(\m{S})\m{1}_n\otimes \m{I}_d)$,\footnote{We use $\m{V}_{[k:r]}$, with $k<r,$ to denote a sub-matrix containing columns $k,\ldots,r$ of $\m{V}$.} and $\m{P} = \left(\sum_{i=1}^n|\m{S}_i|^{-1}\right)^{-1} \in \mb{R}^{d\times d}$. Then, 
\begin{align}
\m{V}^{-1}\bm{\Omega}\m{V} = \begin{bmatrix}
\m{0}_{d\times d} & \m{0}_{d\times d(n-1)}\\
\m{0}_{d(n-1)\times d} & \bm{\Omega}'
\end{bmatrix},
\end{align}
where $-\bm{\Omega}'= -\m{Z}_{[d+1:d(n-1)]}^\top \bm{\Omega} \m{V}_{[d+1:dn]}$ is Hurwitz. 
\end{lemma}
}

\begin{remark} \label{rem:2.1}
If we further assume that $\m{S}_i=\m{S}_i^\top,~\forall i = 1,\ldots,n$, based on \cite{Ostrowski1959}[Thm.~3], all eigenvalues $\Lambda_1,\ldots,\Lambda_{dn}$ of $\bm{\Omega}$ are real and 
    \begin{align} \label{eq:eigenvalue_omega}
    \begin{split}
        \Lambda_1 &= \ldots=\Lambda_d = 0,\\
        \Lambda_k &= \theta_k \lambda_{\lceil k/d\rceil},~k=d+1,\ldots,dn,
    \end{split}
    \end{align}
where $\lceil x \rceil$ denotes the {smallest integer} that is greater than or equal to $x$, $\lambda_k$ is the $k$-th smallest eigenvalue of $\m{L}$ and $\theta_k \in [p_{\min},p_{\max}]$, $p_{\min}$ and $p_{\max}$ are respectively the smallest and the largest eigenvalue of all $|\m{S}_i|$.

For example, consider $\mc{G} = \mc{C}_6$ - the {undirected} cycle of length six. The scaling matrices are   $\m{S}_1=\m{S}_2= \begin{bmatrix}
        2 & -\frac{\sqrt{3}}{4}\\
        -\frac{\sqrt{3}}{4} & \frac{7}{4}
    \end{bmatrix}$, $\m{S}_3=\m{S}_4 = \begin{bmatrix}
        2 & 0\\
        0 & 1
    \end{bmatrix}$, $
    \m{S}_5=\m{S}_6 = \begin{bmatrix}
        -3 & 0\\
        0 & -1
    \end{bmatrix}.$
Then, $0=\lambda_1<\lambda_2=\lambda_3=1<\lambda_4=\lambda_5=3<\lambda_6=4$, $p_{\min}=0.5$, $p_{\max}=3$, and the {spectrum of} $\bm{\Omega}$ is $\{0,~0,~1.059,~1.264,~2.088,2.387,3.406,3.477,~5.051,~6.657,$
$~7.388,~10.222\}$. {Then, $\lambda_2 p_{\min} = 0.5$, $\lambda_6 p_{\max}=12$, and it is easy to check that all nonzero eigenvalues of $\bm{\Omega}$  lie entirely on the interval $[0.5,~12]$.}
\end{remark}

\section{Matrix-scaled consensus algorithms for single-integrator agents}
\label{sec:single_integrator}
\subsection{Matrix-scaled consensus of single-integrators}
We firstly consider the ideal case, where agents are modeled by single integrators. The matrix-scaled consensus algorithm is proposed for each agent $i=1,\ldots,n$ as follows
\begin{align} \label{eq:msc_single_integrator}
\dot{\m{x}}_i &=\m{u}_i = - \text{sign}(\m{S}_i) \sum_{j \in \mc{N}_i} {w_{ij}}(\m{S}_i\m{x}_i - \m{S}_j \m{x}_j).
\end{align}
The $n$-agent system can be expressed in the matrix form as follows:
\begin{align}
\dot{\m{x}} = -(\text{sign}(\m{S})\m{L}\otimes \m{I}_d)\m{S}\m{x} = - \bm{\Omega} \m{x}.
\end{align}
The asymptotic behavior of the system \eqref{eq:msc_single_integrator} is summarized in the following theorem, whose proof can be found in \cite{Trinh2022matrix}. 
{
\begin{theorem} \cite{Trinh2022matrix} \label{thm:msc_single_integrator} Under the MSC algorithm \eqref{eq:msc_single_integrator}, $\m{x}(t)$ exponentially converges to $\m{S}^{-1}(\m{1}_n\otimes\m{x}_{0}) \in \mc{C}_S$, where $\m{x}_{0}:=\m{P}(\m{1}_n^\top\text{sign}(\m{S})\otimes \m{I}_d)\m{x}(t) = \m{P} \sum_{i=1}^n \text{sign}(\m{S}_i)\m{x}_i(0)$ is a virtual consensus point.
\end{theorem}
}

\begin{remark} \label{rem:lyap_msc} Note that the proof that $\lim_{t\to+\infty}\m{x}(t)=\m{S}\m{x}_{{0}}$ can also be shown by considering the Lyapunov candidate function $V=\m{x}^\top\m{S}^\top\bar{\m{L}}\m{S}\m{x}$.
\end{remark}
{
\begin{remark}
In the MSC algorithm~\eqref{eq:msc_single_integrator}, each agent needs $d^2$ multiplications and $d(d-1)$ additions to obtain $\m{S}_i\m{x}_i$. Then, $|\mc{N}_i|d$ additions and $d$ multiplications are additionally required to determine the input $\m{u}_i$. In sparse network (each vertex does not have a lot of neighbors), the complexity of the MSC algorithm for each agent is $\mc{O}(d^2)$ (polynomial complexity), and for the whole system is $\mc{O}(nd^2)$. Note that the MSC algorithm~\eqref{eq:msc_single_integrator} is \emph{scalable} with regard to network size, i.e., the complexity of the algorithm does not increases for each agent as the size of the network grows. 
\end{remark}
}
{
The next theorem expands the class of interaction functions which can be used in a MSC algorithm. 
\begin{theorem} \label{thm:msc-nonlinear} Let $\m{f}: \mb{R}^N \to \mb{R}^N\,(N\in\mb{N}_+)$ be a {Lipschitz continuous} function that satisfies 
$\m{y}^\top\m{f}(\m{y}) > \m{0},~\forall \m{0}_N\neq \m{y}\in \mc{D} \subseteq \mb{R}^N,\, \m{0}_d \in \mc{D},~\m{f}(\m{0}_N)=\m{0}_N.$ 
Consider the following nonlinear MSC algorithm
\begin{align} \label{eq:msc_nonl}
    {\m{u}}_i = -\text{sign}(\m{S}_i) \sum_{j\in \mc{N}_i} {w_{ij}}\m{f}(\m{S}_i\m{x}_i-\m{S}_j\m{x}_j),~i=1,\ldots,n.
\end{align}
\end{theorem}}
\begin{IEEEproof}
    The matrix-scaled consensus {system with generalized interaction} \eqref{eq:msc_nonl} can be expressed in matrix form as
\begin{align}
\dot{\m{x}} = -(\text{sign}(\m{S}){\m{H}}^\top\m{W} \otimes \m{I}_d)\m{f}((\m{H}\otimes \m{I}_d)\m{S}\m{x}).
\end{align}
It follows that
\begin{align*}
(\m{1}_n^\top\text{sign}(\m{S})\otimes \m{I}_d)\dot{\m{x}} =
-(\m{1}_n^\top\m{H}^\top\m{W}\otimes \m{I}_d)\m{f}((\m{H}\otimes \m{I}_d)\m{S}\m{x})
=\m{0}_d
\end{align*}
which implies that $\m{x}_{{0}}$ as defined in Theorem~\ref{thm:msc_single_integrator} is {a constant vector under \eqref{eq:msc_nonl}.}

For stability analysis, {let $\bar{\m{W}}=\m{W}\otimes \m{I}_d$, $\bar{\m{H}}=\m{H}\otimes\m{I}_d$, and} consider the Lyapunov function \cite{Khalil2015nonlinear}
\[V(\m{x}) = \int_{\m{0}_{dn}}^{\m{x}}\m{f}(\bar{\m{H}}\m{S}\m{y})^\top \bar{\m{W}}\bar{\m{H}}\m{S} d\m{y},\] 
which is positive semidefinite and continuously differentiable in $\mc{D}$. Along any trajectory of \eqref{eq:msc_nonl}, we have
\begin{align*}
    \dot{V} &=-\m{f}(\bar{\m{H}}\m{S}\m{x})^\top \bar{\m{W}}\bar{\m{H}}|\m{S}|\bar{\m{H}}^\top\bar{\m{W}}\m{f}(\bar{\m{H}}\m{S}\m{x}) \\
&= -\m{f}(\bar{\m{H}}\m{S}\m{x})^\top \bar{\m{W}}\bar{\m{H}}(|\m{S}|+|\m{S}|^\top)\bar{\m{H}}^\top\bar{\m{W}}\m{f}(\bar{\m{H}}\m{S}\m{x}) \leq 0.
\end{align*}
Based on LaSalle's invariance principle, each trajectory of the system approaches the largest invariant set in $\Omega = \{\m{x}|~\dot{V} = 0\}$. Solving $\dot{V}=0$ gives $\bar{\m{H}}^\top\bar{\m{W}}\m{f}(\bar{\m{H}}\m{S}\m{x})=\m{0}$, which further implies that $\m{x}^\top\m{S}^\top\bar{\m{H}}^\top\bar{\m{W}}\m{f}(\bar{\m{H}}\m{S}\m{x})={0}$, or $\m{S}\m{x}\in \text{im}(\m{1}_n\otimes \m{I}_d)$. Combining with $\m{x}_{{0}}(t)=\m{x}_{{0}}(0)$ yields $\m{S}\m{x} \equiv \m{1}_n \otimes \m{x}_{{0}}$. Thus, $\m{x}(t) \to \m{S}^{-1}(\m{1}_n\otimes\m{x}_{{0}}) \in \mc{C}_S$, as $t\to +\infty$.
\end{IEEEproof}

In Theorem~\ref{thm:msc-nonlinear}, if the input of the agents  $\|\dot{\m{x}}\|_{\infty}=\|\m{u}\|_{\infty}$ must be upper bounded by a constant $\beta>0$ {for all $t\geq 0$}, we may choose $\m{f}(\cdot)=\frac{\beta}{\max_{i\in \{1,\ldots,n\}}|\mc{N}_i|}\tanh(\cdot)$. Moreover, a larger class of nonlinearly output-coupled systems can be considered, for examples, the Kuramoto oscillators model has $\m{f}(\cdot)=\sin(\cdot)$ and $\mc{D}=\left(-\frac{\pi}{2},\frac{\pi}{2}\right)$.

{
\begin{remark}
Suppose that $\m{S}_i,~i=1,\ldots,n,$ are symmetric, the following finite-time MSC algorithm is proposed
\begin{align} \label{eq:msc_FT}
\m{u}_i = - \text{sign}(\m{S}_i)\text{sig}^{\alpha}\Big(\sum_{j \in \mc{N}_i} {w_{ij}}(\m{S}_i\m{x}_i - \m{S}_j \m{x}_j)\Big),
\end{align}
where $\alpha\in(0,1)$, the function $\text{sig}^{\alpha}(\cdot)$ is defined component-wise for a vector, and for $x\in \mb{R}$,  $\text{sig}^\alpha(x):=\text{sgn}(x)|x|^\alpha$. The finite time convergence can be shown by considering the Lyapunov function $V=\m{x}^\top\m{S}^\top\bar{\m{L}}|\m{S}|^{-1}\bar{\m{L}}\m{S}\m{x}$. Without the assumption on the symmetry of the scaling matrices, the problem becomes more complicated. The following modified MSC algorithm 
\begin{align} \label{eq:msc_FT1}
\m{u}_i = - \text{sign}(\m{S}_i)|\m{S}_i|^{-1}\text{sig}^{\alpha}\Big(\sum_{j \in \mc{N}_i} {w_{ij}}(\m{S}_i\m{x}_i - \m{S}_j \m{x}_j)\Big)
\end{align}
guarantees the network to achieve finite time convergence to a point in $\mc{C}_S$. The detailed proofs are omitted to save space.
\end{remark}
}
\subsection{Adaptive matrix-scaled consensus of single-integrators with parametric uncertainties}
Let each agent be modeled by the single integrator with {uncertain parameters} \eqref{eq:model}. The following adaptive matrix-scaled consensus law is designed based on certainty equivalence {principle}
\begin{subequations}
\begin{align}
\m{u}_i &= - \text{sign}(\m{S}_{i}) \sum_{j\in \mc{N}_i} {w_{ij}}(\m{S}_{i}\m{x}_i - \m{S}_{j}\m{x}_j) - \bm{\phi}_i(t,\m{x}_i) \hat{\bm{\theta}}_i, \label{eq:AMWC-1}\\
\dot{\hat{\bm{\theta}}}_i &= \gamma_i \bm{\phi}_i(t,\m{x}_i)^\top \m{S}_{i}^\top \sum_{j\in \mc{N}_i} {w_{ij}}(\m{S}_{i}\m{x}_i - \m{S}_{j}\m{x}_j), \label{eq:AMWC-2}
\end{align}
\end{subequations}
where $\gamma_i>0$ are adaptive rates, $i=1,\ldots,n$. We can rewrite the system~\eqref{eq:model} under the matrix-scaled consensus algorithm \eqref{eq:AMWC-1}--\eqref{eq:AMWC-2} in matrix form as follows:
{ \begin{subequations}
\begin{align}
\dot{\m{x}} &= -\bm{\Omega}\m{x} +  \text{blkdiag}(\bm{\phi}_1,\ldots,\bm{\phi}_n) (\bm{\theta}  - \hat{\bm{\theta}} ), \label{eq:AMWC-3}\\
\dot{\hat{\bm{\theta}}} & = \text{blkdiag}(\gamma_1\bm{\phi}_1,\ldots,\gamma_n \bm{\phi}_n)^\top\m{S}^\top{\bar{\m{L}}}\m{S}\m{x}, \label{eq:AMWC-4}
\end{align}
\end{subequations}}
where $\m{x} = \text{vec}(\m{x}_1,\ldots,\m{x}_n) \in \mb{R}^{dn}$, $\bm{\theta} = \text{vec}(\bm{\theta}_1,\ldots,\bm{\theta}_n) \in \mb{R}^{nr}$, $\hat{\bm{\theta}} = \text{vec}(\hat{\bm{\theta}}_1,\ldots,\hat{\bm{\theta}}_n) \in \mb{R}^{nr}$, and $\bm{\phi}(t)=\text{blkdiag}(\bm{\phi}_1,\ldots,\bm{\phi}_n) \in \mb{R}^{dn\times dr}$.

\begin{theorem} \label{thm:msc_single_integrator_adaptive} Suppose that $\bm{\phi}(t)$ is uniformly bounded. Under the adaptive matrix-scaled algorithm \eqref{eq:AMWC-1}--\eqref{eq:AMWC-2}, 
\begin{enumerate}
\item[(i)] $\m{x}(t)\to \mc{C}_S$ and $\hat{\bm{\theta}}(t) \to \bm{\theta}^*$, as $t\to +\infty$, and 
\item[(ii)] if additionally, $\bm{\phi}_i(t)$ is  persistently exciting (PE), i.e., there exist $\mu_1\ge\mu_2>0$, $T>0$ such that for any $t\ge 0$, $\mu_1\m{I}_d \ge \int_t^{t+T} \bm{\phi}_i(\tau) \bm{\phi}_i(\tau)^\top d\tau \ge \mu_2\m{I}_d,$ then $\hat{\bm{\theta}}_i(t) \to \bm{\theta}_i,~\forall i=1, \ldots, n,$ as $t\to+\infty$.
\end{enumerate} 
\end{theorem}

\begin{IEEEproof}
(i) Consider the function $V = \frac{1}{2} \m{x}^\top \m{S}^\top\bar{\m{L}}\m{S}\m{x} + \sum_{i=1}^n \frac{1}{2\gamma_i} \|\bm{\theta}_i- \hat{\bm{\theta}}_i\|^2.$ 
Since $\m{S}$ is invertible and $\bar{\m{L}}$ is symmetric, $\m{S}^\top\bar{\m{L}}\m{S}$ has the same number of positive, negative and zero eigenvalues as $\bar{\m{L}}$. It follows that $\m{S}^\top\bar{\m{L}}\m{S}$ is symmetric positive semidefinite. As a result, $V$ is continuously differentiable, positive definite and radially unbounded with regard to $\text{vec}(\sqrt{\overline{\m{W}}}\bar{\m{H}}\m{S}\m{x}, \bm{\theta}- \hat{\bm{\theta}})$. Moreover, $\frac{1}{4}\|\sqrt{\overline{\m{W}}}\bar{\m{H}}\m{S}\m{x}\|^2 + \frac{1}{2\max_{i \in \mc{V}}\gamma_{i}}\|\bm{\theta}- \hat{\bm{\theta}}\|^2 \leq V \leq \|\sqrt{\overline{\m{W}}}\bar{\m{H}}\m{S}\m{x}\|^2 + \frac{1}{2\min_{i \in \mc{V}}\gamma_{i}}\|\bm{\theta}- \hat{\bm{\theta}}\|^2$, and along any trajectory of the system,
\begin{align*}
\dot{V}(t) = -\m{x}^\top \m{S}^\top \bar{\m{L}} \left(|\m{S}|+|\m{S}^{\top}| \right) \bar{\m{L}} \m{S}\m{x} \leq 0,
\end{align*}
which implies that $\|\bar{\m{L}}\m{S}\m{x}\|$, $\|\bm{\theta} - \hat{\bm{\theta}}\|$ are uniformly bounded, and $\lim_{t\to+\infty}V(t)$ exists and is finite. As $\bm{\theta}$ is a constant vector, it follows that $\hat{\bm{\theta}}$ is uniformly bounded. Further, $\|\dot{\m{x}}(t)\| \le \|\bar{\m{L}}\m{S}\m{x}(t)\| + \|\text{blkdiag}(\bm{\phi}_1,\ldots,\bm{\phi}_n)\| \|\bm{\theta} - \hat{\bm{\theta}}\|$ is uniformly bounded. {Since }
\begin{align*}
\ddot{V}(t) &= -2 \m{x}^\top \m{S}^\top \bar{\m{L}} \left(|\m{S}|+|\m{S}|^{\top} \right) \bar{\m{L}} \m{S} \dot{\m{x}} \\
|\ddot{V}(t)| &\le 2 {\lambda_{\max}\left(|\m{S}|+|\m{S}|^{\top} \right)} \|\bar{\m{L}} \m{S} \m{x}\| \|\bar{\m{L}}\| \|\m{S}\| \|\dot{\m{x}}\|,
\end{align*}
{$\ddot{V}(t)$} is also uniformly bounded. It follows from Barbalat's lemma \cite{Slotine1991applied} that $\lim_{t\to+\infty} \dot{V}(t) = 0$, or $\lim_{t\to+\infty} \bar{\m{L}} \m{S} \m{x}(t) = \m{0}_{dn}$. This implies that $\m{S}_i\m{x}_i \to \m{S}_j\m{x}_j$ as $t\to +\infty$, or $\m{x}(t)$ approaches the set $\mc{C}_S$ as $t\to+\infty$. As $\lim_{t\to+\infty} V(t)$ exists and $\lim_{t\to +\infty}\m{x}^\top \m{S}^\top \bar{\m{L}} \m{S}\m{x}(t) = {0}$, it follows that $\lim_{t\to +\infty}\hat{\bm{\theta}}(t) = \bm{\theta}^*$ exists and is finite.

(ii) Next, let $\tilde{\bm{\theta}} = \hat{\bm{\theta}} - \bm{\theta}$ and define 
\begin{align*}
    \m{y} &= \begin{bmatrix}
        \m{y}_{\mc{T}}\\
        \m{y}_{\mc{C}}
    \end{bmatrix}  \\
    &= \bar{\m{H}} \m{S} \m{x} = \left(\begin{bmatrix} \m{I}_{n-1} \\ \m{T} \end{bmatrix} \otimes \m{I}_d \right) (\m{H}_{\mc{T}}\otimes \m{I}_d)\m{S}\m{x} \\
    &= \left(\m{R} \otimes \m{I}_d \right) \m{y}_{\mc{T}} = \bar{\m{R}} \m{y}_{\mc{T}},
\end{align*} 
it follows that
\begin{subequations}
\begin{align}
\dot{\m{y}}_{\mc{T}} & =  \bm{\varphi}(\m{y}_{\mc{T}}) + \bar{\m{H}}_{\mc{T}} \m{S} \text{blkdiag}(\bm{\phi}_1,\ldots,\bm{\phi}_n) \tilde{\bm{\theta}}, \label{eq:AMWC-5}\\
\dot{\tilde{\bm{\theta}}} & = \bm{\psi}(t), \label{eq:AMWC-6}
\end{align}
\end{subequations}
{where $\bm{\varphi}(\m{y}_{\mc{T}})=-\bar{\m{H}}_{\mc{T}} |\m{S}|\bar{\m{L}}\m{S}\m{x}(t)$ and $\bm{\psi}(t)=\text{blkdiag}(\gamma_1\bm{\phi}_1,\ldots,\gamma_n\bm{\phi}_n)\m{S}^\top\bar{\m{L}} \m{S} \m{x}$. } Observe that 
\begin{align*}
    &\lim_{t\to +\infty} \m{y}_{\mc{T}}(t) = \m{0}_{d(n-1)},\\
    &\lim_{t\to +\infty} \bm{\varphi}(\m{y}_{\mc{T}}(t)) = - \bar{\m{H}}_{\mc{T}} |\m{S}| \lim_{t\to +\infty} \bar{\m{L}}\m{S}\m{x}(t) = \m{0}_{d(n-1)},\\
    &\lim_{t\to +\infty} \bm{\psi}(t) = -\lim_{t \to +\infty} \text{blkdiag}(\gamma_1\bm{\phi}_1,\ldots,\gamma_n\bm{\phi}_n)\m{S}^\top\bar{\m{L}} \m{S} \m{x}  = \m{0}_{nr},
\end{align*}
and
\begin{align*}
W(t)&=\int_t^{t+T} \bar{\m{H}}_{\mc{T}} \m{S} \bm{\phi}_i(\tau) \bm{\phi}_i(\tau)^\top \m{S}^\top \bar{\m{H}}_{\mc{T}}^\top d\tau \\
&= \bar{\m{H}}_{\mc{T}} \m{S} \left( \int_t^{t+T} \bm{\phi}_i(\tau) \bm{\phi}_i(\tau)^\top d\tau \right) \m{S}^\top \bar{\m{H}}_{\mc{T}}^\top 
\end{align*}
satisfies 
\begin{subequations}
\begin{align}
\mu_1 \bar{\m{H}}_{\mc{T}}\m{S}\m{S}^\top\bar{\m{H}}_{\mc{T}}^\top \le W(t) \le  \mu_2 \bar{\m{H}}_{\mc{T}}\m{S}\m{S}^\top\bar{\m{H}}_{\mc{T}}^\top,\\ 
\mu_1 \lambda_{\max}(\m{L}_e) \m{I}_{dn} \le W(t) \le  \mu_2 \lambda_{\max}(\m{L}_e) \m{I}_{dn}.
\end{align}
\end{subequations}
As $\m{L}_e = \bar{\m{H}}_{\mc{T}}\m{S}\m{S}^\top \bar{\m{H}}_{\mc{T}}^\top=\m{L}_e^\top$ is positive semidefinite, rank$(\m{L}_e)$=rank$(\bar{\m{H}}_{\mc{T}})= d(n-1)$, $\m{L}_e$ is positive definite. Since all conditions of Lemma \ref{lemma:PE} (in the Appendix \ref{append:A}) are satisfied, $\lim_{t\to+\infty}\tilde{\bm{\theta}}(t) = \m{0}_{nr}$ or $\lim_{t\to+\infty}\hat{\bm{\theta}}(t) = {\bm{\theta}}$.
\end{IEEEproof}

\section{Matrix-scaled consensus algorithms for time-invariant linear dynamical agents}
\label{sec:linear_agents}
In this section, we firstly propose two matrix scaled consensus algorithms for the simplified general linear models \eqref{eq:homogeneous_linear} and \eqref{eq:heterogeneous_linear}. The proposed algorithms and stability analyses in Subsection~\ref{subsec:4.1} prepare for Subsections \ref{subsec:4.2} and \ref{subsec:4.3}, where the general linear models \eqref{eq:homogeneous_linear} and \eqref{eq:heterogeneous_linear} will be considered. Under the proposed algorithms, we show that the multi-agent systems asymptotically {converges to a trajectory in $\mc{C}_S$}.

\subsection{Simplified linear agents}
\label{subsec:4.1}
\subsubsection{The agents' dynamics are identical}
\label{sss:1}
Let the agents {be} {governed by Eqn.~\eqref{eq:homogeneous_linear}} with $\m{B}=\m{I}_d$. We assume that each agent has information on the state variable $\m{x}_i$ and communicates the matrix-scaled state $\m{S}_i\m{x}_i$ with its neighboring agents. The following MSC algorithm is proposed
\begin{align} \label{eq:msc_synch}
\m{u}_i = -c \text{sign}(\m{S}_i) \sum_{j \in\mc{N}_i} {w_{ij}}(\m{S}_i\m{x}_i - \m{S}_j \m{x}_j),
\end{align}
where $i=1,\ldots,n,$ and $c>0$ is a {positive} coupling gain. The $n$-agent system can be written in matrix form as
\begin{align} \label{eq:simple_homo_allagents}
\dot{\m{x}} = (\m{I}_n\otimes \m{A} - c \bm{\Omega})\m{x}.
\end{align}

For stability analysis of the system \eqref{eq:simple_homo_allagents}, the following lemma whose proof is given in Appendix \ref{append:A} will be used.
{\begin{lemma}\label{lem:matrix_stability} Let $-\bm{\Omega}'$ be a Hurwitz matrix, $\bm{\Delta}_{\Omega}$ be a perturbation matrix of the same dimension, $\|\bm{\Delta}_{\Omega}\|=\delta_{\Omega}$, and $\m{Q}=\m{Q}^\top>0$ is the unique solution to the Lyapunov equation $\m{Q}\bm{\Omega}'+(\bm{\Omega}')^\top\m{Q} = \m{I}$. For any $c > 2\delta_{\Omega} \lambda_{\max}(\m{Q})$, the matrix $-c\bm{\Omega}' + \bm{\Delta}_{\Omega}$ is Hurwitz.
\end{lemma}}

Let $\m{y} = \m{V}^{-1}\m{x} \in \mb{R}^{dn}$, and define $\m{y}_{[k:r]} = [y_k,y_{k+1},\ldots,y_r]^\top$, for $k<r$, it follows that
\begin{align*}
\dot{\m{y}}_{[1:d]}&=(\m{1}_n^\top\text{sign}(\m{S}))\otimes \m{I}_d) (\m{I}_n\otimes \m{A} - c \bm{\Omega})\m{V}\m{y}  =\m{A} \m{y}_{[1:d]},\label{eq:simple_homo_agents_matrix1}\\
\dot{\m{y}}_{[d+1:dn]}&=\underbrace{\left(\m{Z}_{[d+1:dn]}^\top(\m{I}_n\otimes \m{A})\m{V}_{[d+1:dn]}-c\bm{\Omega}'\right)}_{:=\bm{\Omega}_c}\m{y}_{[d+1:dn]}. \nonumber
\end{align*}
We have the following theorem on the system  \eqref{eq:simple_homo_allagents}:
\begin{theorem} \label{thm:simple_homo_agents}
Let $\m{Q}=\m{Q}^\top>0$ be the unique solution of the Lyapunov equation $\m{Q}{\bm{\Omega}'}+(\bm{\Omega}')^\top\m{Q} = \m{I}_{d(n-1)}$ and $c> 2\|\m{A}\| \lambda_{\max}(\m{Q})$. Then, $\m{x}(t) \to \mc{C}_S$, as $t\to+\infty$. 
\end{theorem}

\begin{IEEEproof}
We have $\|\m{Z}_{[d+1:dn]}^\top(\m{I}_n\otimes \m{A})\m{V}_{[d+1:dn]}\| \leq \|\m{V}^{-1}(\m{I}_n\otimes \m{A})\m{V}\|=\|\m{I}_n\otimes \m{A}\| = \|\m{A}\|$. Based on Lemma~\ref{lem:matrix_stability}, the matrix {$\bm{\Omega}_c$} is Hurwitz. It follows that $\m{y}_{[d+1:dn]}$ globally exponentially  converges to $\m{0}_{d(n-1)}$. The solution of Eqn.~\eqref{eq:simple_homo_allagents} is 
\begin{align*}
   \m{x}(t) = \m{V}\m{y}(t) &=\m{V}
\text{blkdiag}(\text{exp}(\m{A}t), 
\text{exp}(\bm{\Omega}_c t)) \m{y}(0) \nonumber\\
&= \m{S}^{-1}(\m{1}_n\otimes \m{P})\text{exp}(\m{A}t) (\m{1}_n^\top \text{sign}(\m{S})\otimes \m{I}_d)\m{x}(0) +\bm{\xi}(t), 
\end{align*}
where $\bm{\xi}(t)=\m{V}_{[d+1:dn]} \text{exp}(\bm{\Omega}_c t)\m{Z}_{[d+1:dn]}^\top\m{x}(0) \to \m{0}_{dn}$ exponentially fast. Thus, \[\lim_{t\to+\infty} (\m{x}_i(t)-\m{S}_i^{-1}\m{P}\text{exp}(\m{A}t)\m{P}^{-1}\m{x}_{{0}}) = \m{0}_d,~\forall i=1,\ldots,n,\] or $\m{x}(t)\to \mc{C}_S$, as $t\to+\infty$. 
\end{IEEEproof}

{Unlike the consensus and adaptive consensus algorithms in \cite{Li2009consensus,Li2013distributed,Mei2021unified}, the matrices $\m{I}_{n}\otimes \m{A} - c\bm{\Theta}$ cannot be decomposed along each eigenvector of the matrix $\bm{\Theta}$ due to the scaling matrices $\m{S}_i$. Thus, the calculation of a coupling gain $c$ to stabilize $\bm{\Theta}_c$ requires information of the whole network (matrix $\bm{\Omega}$). As a result, the proposed algorithm \eqref{eq:msc_synch} not fully distributed. It is thus desired to design adaptive laws to automatically tuning $c$ based on the locally available information of each agent. Let agent $i$ maintains a coupling gain $c_i(t)$ with $c_i(0)>0$. The following MSC algorithm with adaptive gain tuning is proposed for the $n$-agent system
\begin{subequations} \label{eq:msc_synch_adaptive}
    \begin{align} 
    \dot{\m{x}} &= (\m{I}_n \otimes \m{A} - \bar{\m{C}}\bm{\Omega})\m{x}, \label{eq:msc_synch_adaptive1}\\
    \dot{c}_i &= \kappa_i \bm{\chi}_i^\top |\m{S}_i| \bm{\chi}_i, \, c_i(0)>0,~\forall i=1,\ldots,n, \label{eq:msc_synch_adaptive2} \\
    \bm{\chi}_i &= \text{exp}(-\m{A}(t-t_0)) \sum_{j\in \mc{N}_i}{w_{ij}}(\m{S}_i\m{x}_i - \m{S}_j\m{x}_j),
\end{align}
\end{subequations}
where $\m{C}=\text{diag}(c_1(t),\ldots,c_n(t))$, and $\bar{\m{C}} = \m{C}\otimes \m{I}_d$. However, due to the matrix $\bar{\m{C}}$, the state transformation $\m{y}=\m{V}^{-1}\m{x}$ cannot decompose the system \eqref{eq:msc_synch_adaptive} and the  may be even unstable for large $c_i$. In the following theorem, we prove that \eqref{eq:msc_synch_adaptive} guarantees the $n$-agent system to achieve MSC given that $\m{S}_i$ commutes with $\m{A}$.
\begin{theorem}\label{thm:msc_adaptive_tunning} Suppose that the matrices $\m{S}_i,~i=1,\ldots,n,$ commutes with $\m{A}$, then under the adaptive MSC algorithm \eqref{eq:msc_synch_adaptive}, we have $\m{x}(t)\to\mc{C}_S$, as $t \to +\infty$.
\end{theorem}
\begin{IEEEproof}
    The proof of this theorem can be found in Appendix~\ref{app:C}.
\end{IEEEproof}
}
\subsubsection{Simplified heterogeneous linear agents with unknown system matrix and full control input}
\label{sss:2}
In this subsection, we design an adaptive matrix-scaled consensus law for a network of heterogeneous linear dynamics with full control input
\begin{equation} \label{eq:linear_model}
\dot{\m{x}}_i = \m{A}_i \m{x}_i + \m{u}_i,\,i=1,\ldots,n,
\end{equation}
where $\m{A}_i \in \mb{R}^{d\times d}$ is unknown to each agent $i$. Let $\m{A}_i = \m{A} + (\m{A}_i-\m{A}) = \m{A} + \bar{\m{A}}_i$, where $\m{A}$ is a pre-selected matrix such that $\m{W}_{d+1:dn}^\top\otimes \m{A}\m{V}_{[d+1:dn]} - c\bm{\Omega}'$ is Hurwitz. 

{Let the $k$-th row of the matrix $\bar{\m{A}}_i$ be denoted by $\bar{\m{A}}_{ik}$}, we have the representation \cite{Burbano2019distributed}:
\begin{equation*}
\bar{\m{A}}_i \m{x}_i = \begin{bmatrix}
\bar{\m{A}}_{i1}^\top \m{x}_i\\
\vdots \\
\bar{\m{A}}_{in}^\top \m{x}_i
\end{bmatrix} 
= \begin{bmatrix}
\m{x}_i^\top \bar{\m{A}}_{i1} \\
\vdots \\
\m{x}_i^\top \bar{\m{A}}_{id} 
\end{bmatrix} = (\m{I}_d \otimes \m{x}_i^\top) \text{vec}(\bar{\m{A}}_i).
\end{equation*}

By setting $\bm{\phi}_i := \m{I}_d \otimes \m{x}_i^\top \in \mb{R}^{d\times d^2}$ and $\bm{\theta}_i:=\text{vec}(\bar{\m{A}}_i)$, the linear time-invariant model \eqref{eq:linear_model} is transformed into the single-integrator with uncertain parameters \eqref{eq:model}. 

The system achieves MSC if the uncertainty is being compensated faster than the evolution of $\|\m{x}\|$ (exponential rate). The following MSC algorithm is proposed
\begin{subequations} \label{eq:msc_dtb_obsv}
\begin{align}
    \dot{\m{z}}_i &= \m{A}\m{x}_i - c \text{sign}(\m{S}_i)\sum_{j\in\mc{N}_i}{w_{ij}}(\m{S}_i\m{x}_i-\m{S}_j\m{x}_j),\\
    \m{u}_i &= - c \text{sign}(\m{S}_i)\sum_{j\in\mc{N}_i}{w_{ij}}(\m{S}_i\m{x}_i-\m{S}_j\m{x}_j) +\beta_1 \text{sgn}(\m{e}_i) + \beta_2 \text{diag}(\text{sgn}(\m{e}_i))(\m{I}_d\otimes |\m{x}_i|^\top)\m{1}_{d^2},
\end{align}
\end{subequations}
where $|\m{x}_i| := [|x_{i1}|,\ldots,|x_{id}|]^\top$, $\m{e}_i = {\m{z}}_i - \m{x}_i,$ $\beta_1, \beta_2 >0$ are control gains, and sgn$(\cdot)$ denotes the signum function. The following theorem will be proved.

\begin{theorem} \label{thm:msc_dtb_obsv}
    Let $\mc{G}$ be connected and $\beta_2\geq \max_{i=1,\ldots,n}\|\text{vec}(\m{\bar{A}}_i)\|_{\infty}$. Under the algorithm \eqref{eq:msc_dtb_obsv}, $\m{x}(t) \to \mc{C}_S$ as $t\to +\infty$.
\end{theorem}

\begin{IEEEproof}
We have 
\begin{align}
    \dot{\m{e}}_i &= \bar{\m{A}}_i\m{x}_i - \beta_1 \text{sgn}(\m{e}_i) - \beta_2 \text{sgn}(\m{e}_i)(|\m{x}_i|^\top \otimes \m{I}_d) \m{1}_{d^2} \nonumber\\
    &= - \beta_1 \text{sgn}(\m{e}_i) + \bm{\phi}_i(t) \bm{\theta}_i - \beta_2 \text{diag}(\text{sgn}(\m{e}_i))(\m{I}_d \otimes |\m{x}_i|^\top) \m{1}_{d^2}\nonumber
\end{align}
As the right-hand side of the above equation is discontinuous, the solution of  $\m{e}_i$ is understood in Filippov sense. Using the Lyapunov function $V=\frac{1}{2}\|\m{e}_i\|^2$, we have
\begin{align}
\dot{V} &\in^{\text{a.e.}} \tilde{V} = \bigcap_{\bm{\nu}\in \partial V} \bm{\nu}^\top K[\dot{\m{e}}_i] \nonumber\\
    &= \m{e}_i^\top \Big(-\beta_1 \text{sgn}(\m{e}_i)  + \bm{\phi}_i(t) \bm{\theta}_i - \beta_2 \text{diag}(\text{sgn}(\m{e}_i))(|\m{x}_i|^\top \otimes \m{I}_d) \m{1}_{d^2}\Big) \nonumber\\
    &\leq -\beta \|\m{e}_i\|_1 - \beta_2 \sum_{k=1}^d|\m{e}_{ik}| \|\m{x}_i\|_{1} + \sum_{k=1}^d |\m{e}_{ik}| \|\m{x}_i\|_{1} \|\bm{\theta}_i\|_{\infty} \nonumber\\
    & \leq -\beta \|\m{e}_i\|_1 \nonumber\\ &\leq - \epsilon V^{1/2},
\end{align}
for $\epsilon= \beta\sqrt{2}$, and this implies that $V(t)\to 0$ in finite time \cite{Haddad2008finite}. Therefore, there exists $T>0$ such that $\m{e}_i = \m{0}_d$ or $\m{x}_i = \m{z}_i$ for $t\geq T$. For $t\geq T$, $\m{x}_i$ {can be considered to be governed by the simplified homogeneous linear agent \eqref{eq:simple_homo_allagents}. Thus, $\m{x}\to \mc{C}_S$ as $t\to +\infty$.}
\end{IEEEproof}


\subsection{Homogeneous general linear agents}
\label{subsec:4.2}
In this subsection, we consider the $n$-agent system with identical general linear model \eqref{eq:homogeneous_linear}. The following observer-based  matrix-scaled consensus algorithm is proposed
\begin{subequations}
\label{eq:homo_obsv}
\begin{align}
&\dot{\hat{\m{x}}}_i = \m{A}\hat{\m{x}}_i +  \m{B}\m{u}_i + \m{H}(\m{C}{\hat{\m{x}}}_i-\m{y}_i), \label{eq:homo_obsv_i1}\\
&\dot{\bm{\eta}}_i = \m{A}{\bm{\eta}}_i + \m{B}\m{u}_i + \m{H}(\m{C}{\hat{\m{x}}}_i-\m{y}_i) + c\text{sign}(\m{S}_i) \sum_{j\in \mc{N}_i}{w_{ij}}(\m{S}_i(\hat{\m{x}}_i -\bm{\eta}_i) {-} \m{S}_j(\hat{\m{x}}_j - \bm{\eta}_j)), \label{eq:homo_obsv_i2}\\
&\m{u}_i = \m{K} \bm{\eta}_i,\, i=1,\ldots, n, \label{eq:homo_obsv_i4}
\end{align}
\end{subequations}
where the matrices $\m{H}$ and $\m{K}$ are designed so that $\m{A}+\m{H}\m{C}$ and $\m{A}+\m{B}\m{K}$ are Hurwitz, and $c>0$ is a coupling gain. 

In the proposed algorithm, \eqref{eq:homo_obsv_i1} is a Luenberger observer which estimates the state $\m{x}$. The MSC algorithm is conducted via the auxiliary state variables ${\bm{\zeta}}_i=\hat{\m{x}}_i-\bm{\eta}_i$. It is worth noting that $c\text{sign}(\m{S}_i)(\m{S}_i(\hat{\m{x}}_i -\bm{\eta}_i) {-} \m{S}_j(\hat{\m{x}}_j - \bm{\eta}_j))=c\text{sign}(\m{S}_i)(\m{S}_i\bm{\zeta}_i - \m{S}_j\bm{\zeta}_j)$ acts as a proportional control input to make {$\bm{\zeta}_i$} converge to $\mc{C}_S$. When the coupling gain $c$ is sufficiently large, it becomes a high-gain controller and {the matrix-scaled variables $\m{S}_i\bm{\zeta}_i(t)$ are forced to a solution of the system ${\dot{\bar{\bm{\zeta}}} = \m{P}\m{A}\m{P}^{-1}{\bar{\bm{\zeta}}}}$.} Finally, by forcing $\m{\bm{\eta}}$ converge to $\m{0}_{dn}$ under the control law \eqref{eq:homo_obsv_i4}, $\m{x}\to \mc{C}_S$ as $t\to +\infty$.

For stability analysis, from Eqn.~\eqref{eq:homo_obsv}, we can write
\begin{subequations}
\begin{align}
\dot{\tilde{\m{x}}}_i &= (\m{A}+\m{H}\m{C}){\tilde{\m{x}}}_i,\\
\dot{\bm{\zeta}}_i &= \m{A}{\bm{\zeta}}_i-c\text{sign}(\m{S}_i)\sum_{j\in \mc{N}_i} {w_{ij}}(\m{S}_i\bm{\zeta}_i - \m{S}_j\bm{\zeta}_j),
\end{align}
\end{subequations}
where $i=1,\ldots,n,$ and $\tilde{\m{x}}_i = \hat{\m{x}}_i - \m{x}_i$. Denoting $\bm{{\tilde{\m{x}}}}=\text{vec}({\tilde{\m{x}}}_1,\ldots,{\tilde{\m{x}}}_n)$, ${\bm{\zeta}}=\text{vec}({\bm{\zeta}}_1,\ldots,{\bm{\zeta}}_n)$, and ${\bm{\eta}}=\text{vec}({\bm{\eta}}_1,\ldots,{\bm{\eta}}_n)$, we will prove a theorem regarding the following system
\begin{subequations}
\begin{align}
\dot{\tilde{\m{x}}} &=\m{I}_n\otimes(\m{A}+\m{H}\m{C}) \tilde{\m{x}}, \label{eq:tildex_matrix}\\
\dot{\bm{\zeta}} &= (\m{I}_n\otimes \m{A}-c\bm{\Omega}){\bm{\zeta}}, \label{eq:zeta_matrix}\\
\dot{\bm{\eta}} &= (\m{I}_n\otimes (\m{A}+\m{B}\m{K})){\bm{\eta}} + c(\m{I}_n\otimes(\m{H}\m{C})) \tilde{\m{x}}+\bm{\Omega}{\bm{\zeta}}. \label{eq:eta_matrix}
\end{align}
\end{subequations}

\begin{theorem} \label{thm:homo_general_linear}
{Let $\m{H},\m{K}$ be designed so that $\m{A}+\m{H}\m{C}$ and $\m{A}+\m{B}\m{K}$} are Hurwitz, and the control gain $c$ is chosen so that $\bm{\Theta}_c = \m{Z}_{[d+1:dn]}^\top(\m{I}_n\otimes \m{A})\m{V}_{[d+1:dn]}-c\bm{\Omega}'$ is Hurwitz. Under the algorithm \eqref{eq:homo_obsv}, $\m{x}(t) \to \mc{C}_S$ as $t \to +\infty$. More specifically,  $\m{S}_i\m{x}_i(t),~i=1,\ldots,n,$ asymptotically converge to a common solution of ${\dot{\bar{\bm{\zeta}}} = \m{P}\m{A}\m{P}^{-1}{\bar{\bm{\zeta}}}}$.
\end{theorem}

\begin{IEEEproof}
(i) Because $\m{A}+\m{H}\m{C}$ is Hurwitz, ${\tilde{\m{x}}}$ converges to $\m{0}_{dn}$ exponentially fast.

(ii) It follows from Thm.~\ref{thm:simple_homo_agents} that $\bm{\zeta}(t)$ exponentially converges to $\mc{C}_S$. More specifically, 
\[\lim_{t\to+\infty} (\bm{\zeta}_i(t)-\m{S}_i^{-1}\m{P}\text{exp}(\m{A}t)\m{P}^{-1}\bm{\zeta}_{{0}}) = \m{0}_d,~\forall i=1,\ldots,n,\]
where
\[\bm{\zeta}_{{0}} = (\m{1}_n^\top \text{sign}(\m{S})\otimes \m{P})\bm{\zeta}(0)= \m{P} \sum_{i=1}^n \text{sign}(\m{S}_i)\bm{\zeta}_i(0).\]

(iii) Consider equation \eqref{eq:eta_matrix} with $\tilde{\m{x}}=\m{0}_{dn}$ and $\bm{\zeta} \in \mc{C}_S$, we have the unforced system 
\[\dot{\bm{\eta}} = (\m{I}_n\otimes (\m{A}+\m{B}\m{K})) \bm{\eta}.\] 
{Because} $\m{A}+\m{B}\m{K}$ is Hurwitz, the unforced system is globally exponentially stable, and \eqref{eq:eta_matrix} is input-to-state stable \cite{Khalil2015nonlinear}. As the external input $(\m{I}_n\otimes(\m{H}\m{C})) \tilde{\m{x}}+\bm{\Omega}{\bm{\zeta}}$ converges to $\m{0}_{dn}$ exponentially fast according to (i) and (ii),  ${\bm{\eta}} \to \m{0}_{dn}$ exponentially fast.

(iv) Finally, as $\bm{\zeta} = \hat{\m{x}} - \bm{\eta}= \m{x}-\tilde{\m{x}}-\bm{\eta}$, \[\lim_{t\to+\infty}\bm{\eta}(t) = \lim_{t\to+\infty}\tilde{\m{x}}(t) = \m{0}_{dn},\] 
it follows that 
\[\lim_{t\to+\infty} (\m{x}_i(t)-\m{S}_i^{-1}\m{P}\text{exp}(\m{A}t)\m{P}^{-1}\bm{\zeta}_{{0}}) = \m{0}_d,\]
for all $i=1,\ldots,n.$ 

Thus, we conclude that $\m{x}(t)\to \mc{C}_S$, as $t\to +\infty$, and the convergence rate is exponential.
\end{IEEEproof}

\subsection{Heterogeneous general linear agents}
\label{subsec:4.3}
Finally, we design a MSC algorithm for the system with $n$ heterogenenous linear agents \eqref{eq:heterogeneous_linear}. The main idea is combining the control strategy in subsection~\ref{subsec:4.2} and the disturbance observer in~\ref{sss:2}. Consider the algorithm 
\begin{subequations} \label{eq:msc_heterogeneous}
\begin{align}
&\dot{\hat{\m{x}}}_i = \m{A}_i\hat{\m{x}}_i +  \m{B}_i\m{u}_i + \m{H}_i(\m{C}_i{\hat{\m{x}}}_i-\m{y}_i), \label{eq:het_obsv_i1}\\
&\dot{\bm{\eta}}_i = \m{A}_i{\bm{\eta}}_i + \m{B}_i\m{u}_i + \m{H}_i(\m{C}_i{\hat{\m{x}}}_i-\m{y}_i) - \hat{\m{u}}_i, \label{eq:het_obsv_i2}\\
&\dot{\m{z}}_i = \m{A}\bm{\zeta}_i - c \text{sign}(\m{S}_i)\sum_{j\in\mc{N}_i}{w_{ij}}(\m{S}_i\bm{\zeta}_i-\m{S}_j\bm{\zeta}_j),\\
&\hat{\m{u}}_i = - c \text{sign}(\m{S}_i)\sum_{j\in\mc{N}_i}{w_{ij}}(\m{S}_i\bm{\zeta}_i-\m{S}_j\bm{\zeta}_j) +\beta_1 \text{sgn}(\m{e}_i) + \beta_2 \text{diag}(\text{sgn}(\m{e}_i))(\m{I}_d\otimes |\bm{\zeta}_i|^\top)\m{1}_{d^2}, \label{eq:het_obsv_i3}\\
&\m{u}_i = \m{K}_i \bm{\eta}_i,~ i=1,\ldots, n, \label{eq:het_obsv_i4}
\end{align}
\end{subequations}
where $\bm{\zeta}_i = \hat{\m{x}}_i - \bm{\eta}_i$, $\m{e}_i = {\m{z}}_i - \bm{\zeta}_i$, $\m{A}_i+\m{B}_i\m{K}_i$ and $\m{A}_i + \m{H}_i\m{C}_i$ are Hurwitz. We have the following theorem. 

\begin{theorem} \label{thm:heterogeneous_general_linear}
{Let $\m{H}_i,\m{K}_i$ be designed so that $\m{A}_i+\m{H}_i\m{C}_i$ and $\m{A}_i+\m{B}_i\m{K}_i$} are Hurwitz, $\forall i = 1,\ldots, n$, the control gain $c$ is chosen so that $\bm{\Theta}_c = \m{Z}_{[d+1:dn]}^\top(\m{I}_n\otimes \m{A})\m{V}_{[d+1:dn]}-c\bm{\Omega}'$ is Hurwitz, and $\beta_2\geq \max_{i}\|\bar{\m{A}}_i\|_{\infty}$. Under the matrix scaled consensus algorithm \eqref{eq:msc_heterogeneous}, $\m{x}(t) \to \mc{C}_S$ as $t \to +\infty$. 
\end{theorem}

\begin{IEEEproof}
We have
\begin{subequations} \label{eq:het_msc_dtb_obs}
    \begin{align} 
        \dot{\bm{\zeta}}_i &= \m{A}_i {\bm{\zeta}}_i + \hat{\m{u}}_i, \label{eq:het_msc_dtb_obs1} \\
        \dot{\m{z}}_i &= \m{A}\bm{\zeta}_i - c \text{sign}(\m{S}_i)\sum_{j\in\mc{N}_i}{w_{ij}}(\m{S}_i\bm{\zeta}_i-\m{S}_j\bm{\zeta}_j), \label{eq:het_msc_dtb_obs2}\\
        \dot{\m{e}}_i &= \bar{\m{A}}_i\bm{\zeta}_i - \beta_1\text{sgn}(\m{e}_i) - \beta_2 \text{diag}(\text{sgn}(\m{e}_i))(\m{I}_d\otimes |\bm{\zeta}_i|^\top)\m{1}_{d^2}.
    \end{align}
\end{subequations}
By a similar analysis as in the proof of Thm.~\ref{thm:msc_dtb_obsv}, $\m{e}_i = \m{z}_i - \bm{\zeta}_i= \m{0}_d$ for $t\ge T$. Thus, for $t\geq T$, we may consider $\bm{\zeta}_i$ to evolve according to the equation $\dot{\bm{\zeta}}_i = \m{A} \bm{\zeta}_i - c \text{sign}(\m{S}_i)\sum_{j\in\mc{N}_i}(\m{S}_i\bm{\zeta}_i-\m{S}_j\bm{\zeta}_j),~i=1,\ldots,n.$ 

Similar to the proof of Thm.~\ref{thm:homo_general_linear}, it follows that ${\bm{\zeta}}_i(t)$ asymptotically achieves matrix-scaled consensus to a solution of the system $\dot{\bm{\zeta}}_0 = \m{P}\m{A}\m{P}^{-1}\bm{\zeta}_0$. As $\m{A}_i + \m{B}_i \m{K}_i$ and $\m{A}_i+\m{H}_i\m{C}_i$ are Hurwitz, $\tilde{\m{x}}_i = \m{x}_i - \hat{\m{x}}_i \to \m{0}_d$ exponentially fast, and the unforced system 
$\dot{\bm{\eta}}_i = (\m{A}_i + \m{B}_i\m{K}_i)\bm{\eta}_i$
is globally exponentially stable. Moreover, the system \eqref{eq:het_msc_dtb_obs2} has the external input $\m{H}_i\m{C}_i\tilde{\m{x}}_i - \hat{\m{u}}_i$, which vanishes to $\m{0}_d$ exponentially fast. Thus, $\bm{\eta}_i \to \m{0}_d$, as $t\to +\infty$. 
\begin{figure}[t]
    \centering
    \includegraphics[width=.4\linewidth]{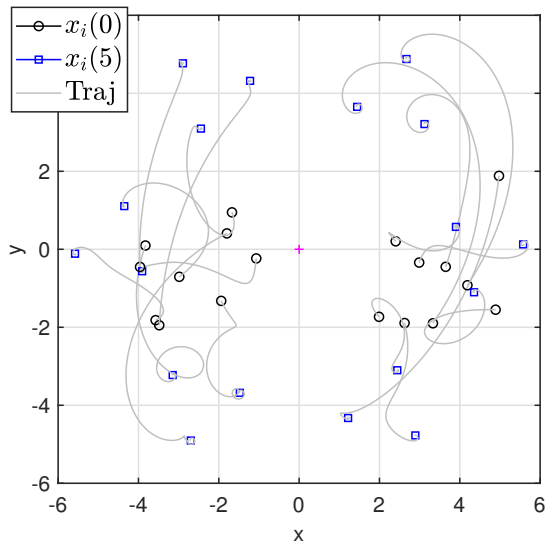}
    \caption{{Simulation of the 18-agent system in Example~\ref{eg:snowflake} under the MSC algorithm \eqref{eq:msc_single_integrator}.}}
    \label{fig:snowflake}
\end{figure}

Finally, from $\m{x}_i - \bm{\zeta}_i = \tilde{\m{x}}_i + \bm{\eta}_i$, it follows that $\m{x}_i$ asymptotically achieves matrix-scaled consensus with regard to a solution of $\dot{\bm{\zeta}}_0 = \m{P}\m{A}\m{P}^{-1}\bm{\zeta}_0$.
\end{IEEEproof}

{
\begin{remark}
We have introduced a novel approach to dealing with heterogeneous dynamics by fast compensating the heterogeneity using the help of a sliding-mode control input. By this way, the MSC algorithm can be designed as for identical linear systems after some finite time and an exact MSC can be exponentially achieved. It is hard to perfectly realize the proposed algorithm \eqref{eq:msc_heterogeneous} because of the signum function. Also, using a nonsmooth algorithm may reduce the durability of each agent, especially when the gap between pre-selected and the agent's matrices, $\|\m{A}_i-\m{A}\|$, are significant. As far as we know, the approach has not been proposed in the literature.
\end{remark}
}
\section{Simulation results}
\label{sec:5}
\begin{figure*}[t]
\centering
\subfloat[$\m{x}_i(t)$]{\includegraphics[width=0.22\linewidth]{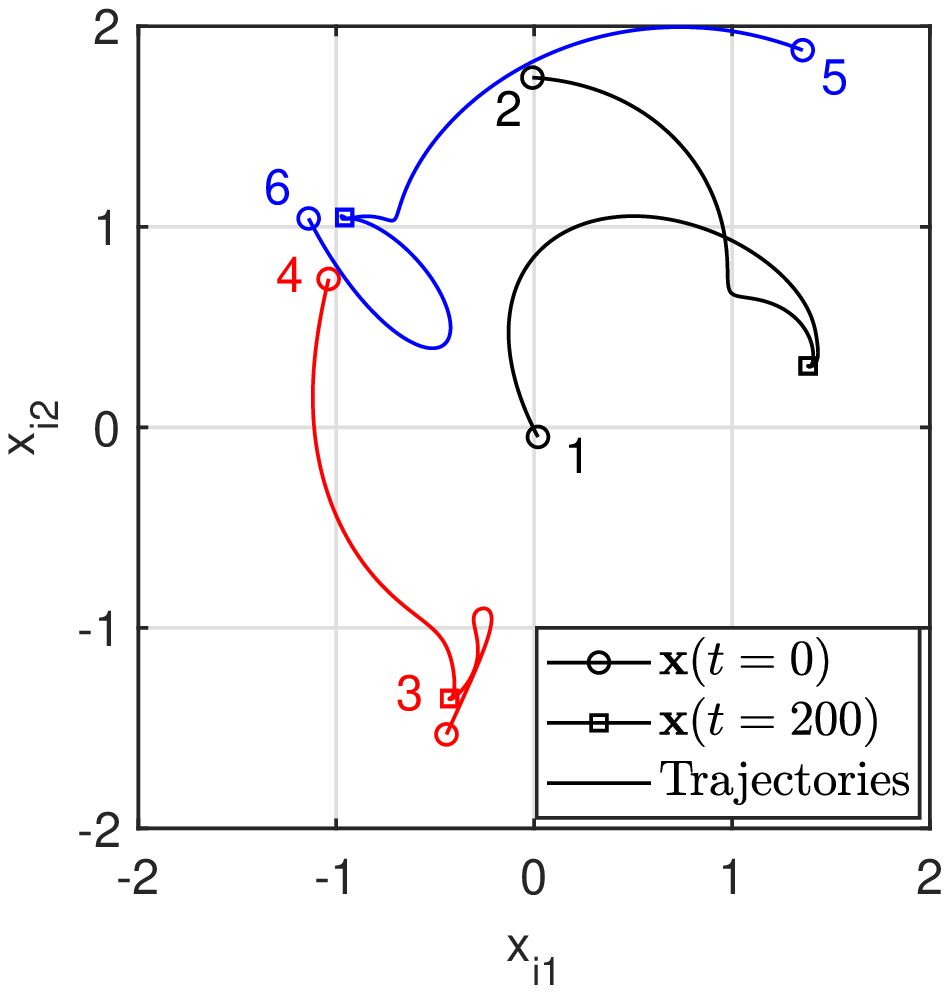}}
\hfill
\subfloat[$x_{i1}(t)$]{\includegraphics[width=0.22\linewidth]{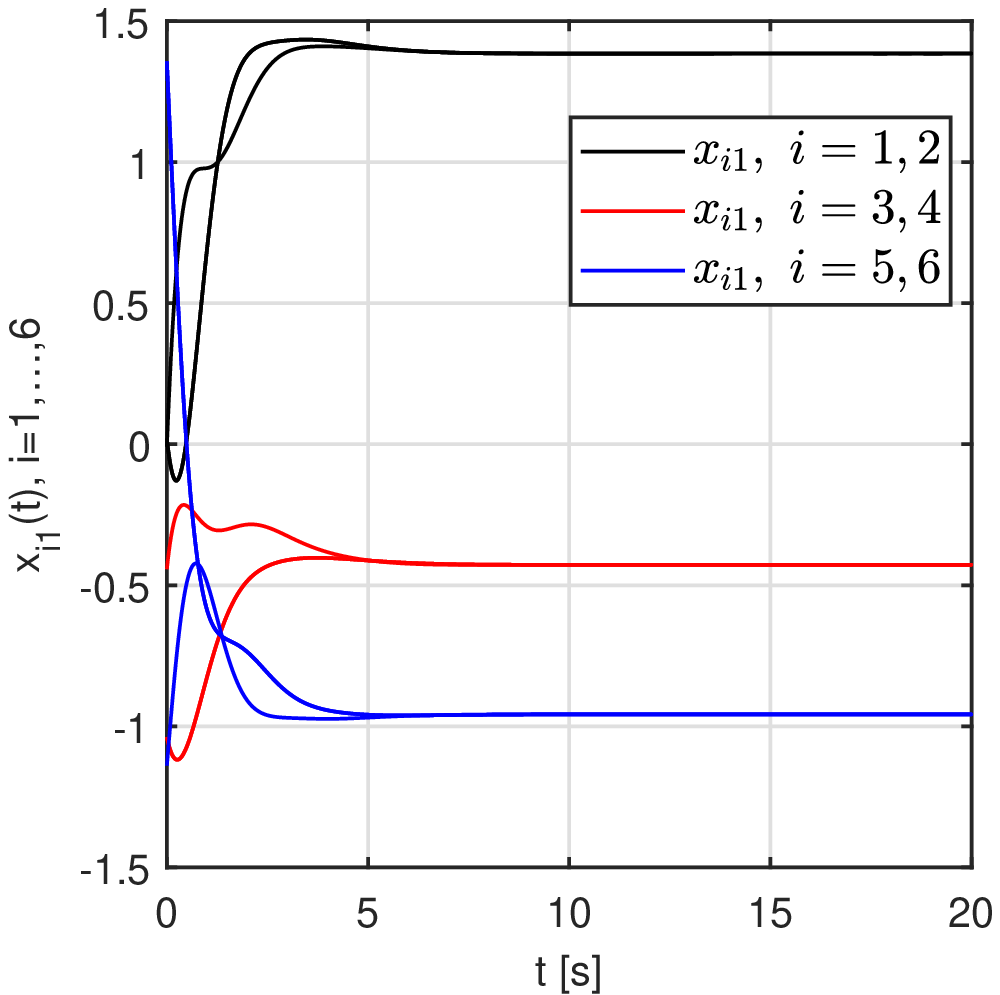}}
\hfill
\subfloat[$x_{i2}(t)$]{\includegraphics[width=0.22\linewidth]{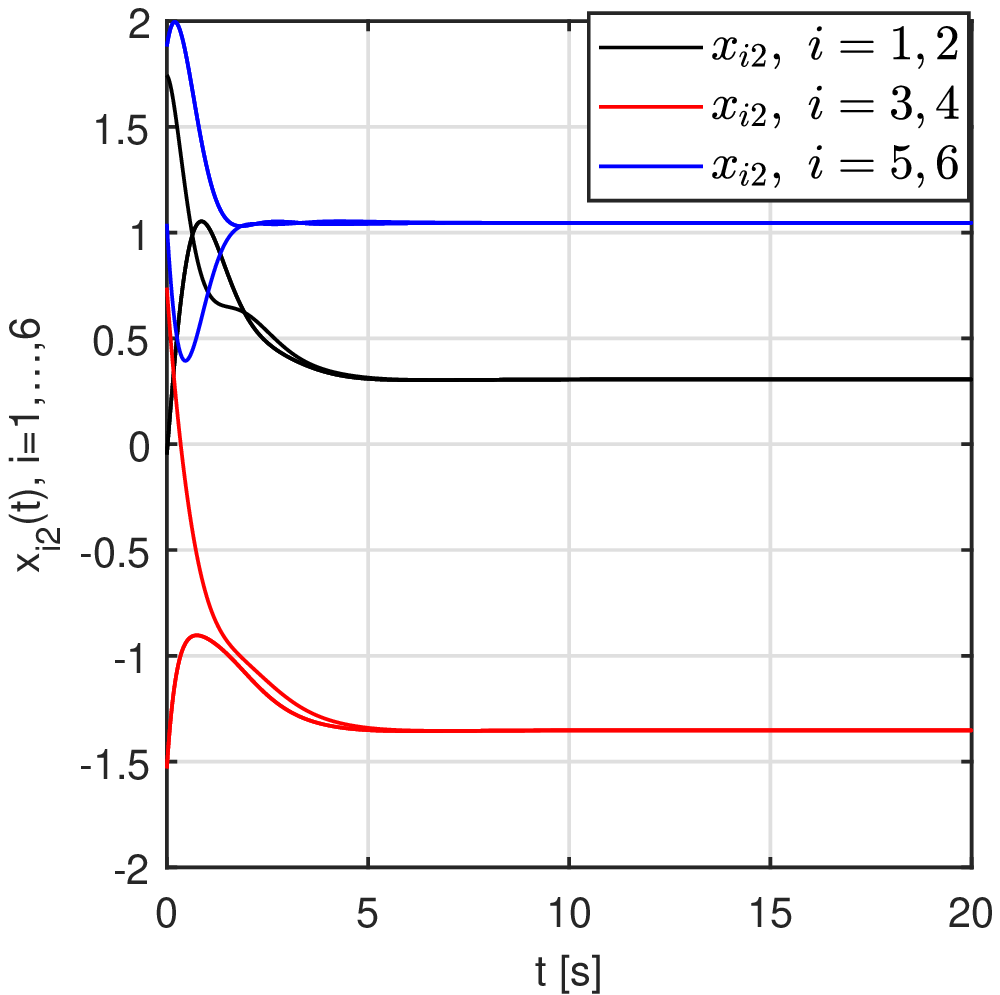}}\hfill
\subfloat[$\|\m{u}\|_{\infty}$]{\includegraphics[width=0.22\linewidth]{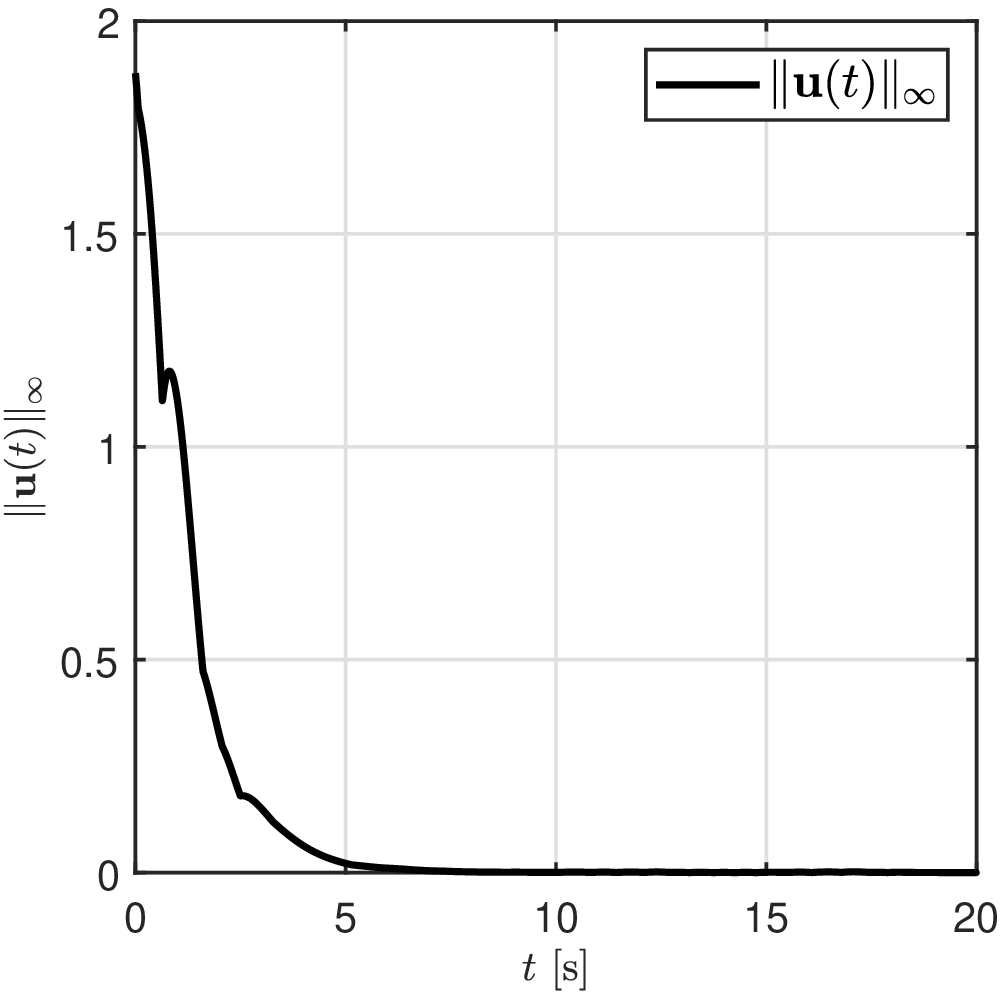}}\\
\subfloat[$\m{x}_i(t)$]{\includegraphics[width=0.22\linewidth]{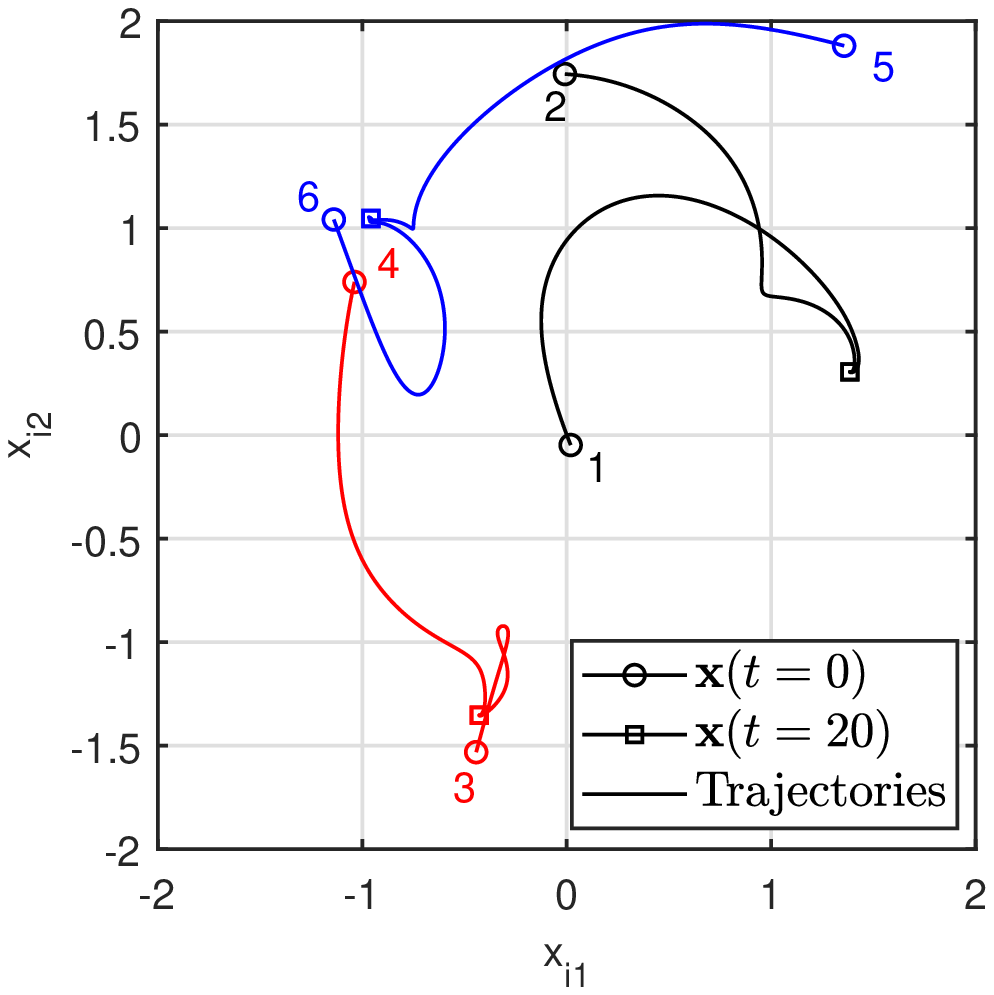}}
\hfill
\subfloat[$x_{i1}(t)$]{\includegraphics[width=0.22\linewidth]{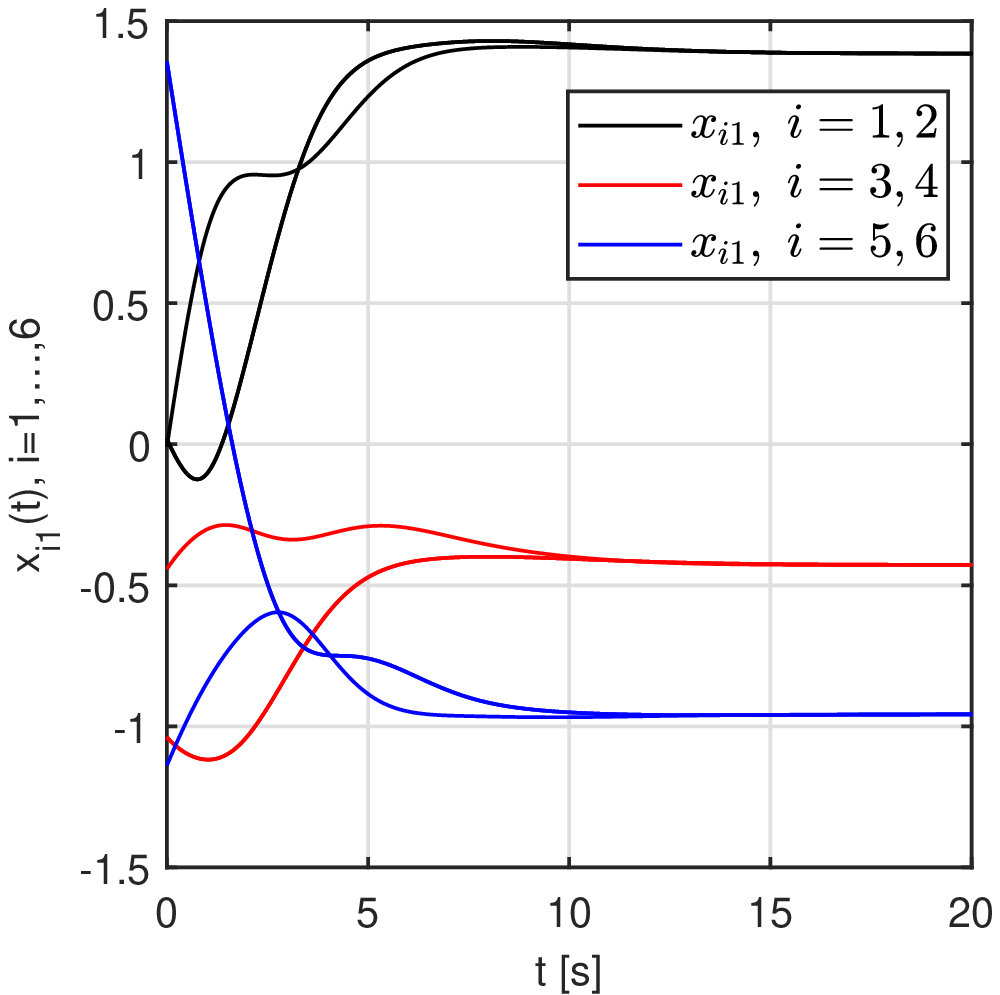}}
\hfill
\subfloat[$x_{i2}(t)$]{\includegraphics[width=0.22\linewidth]{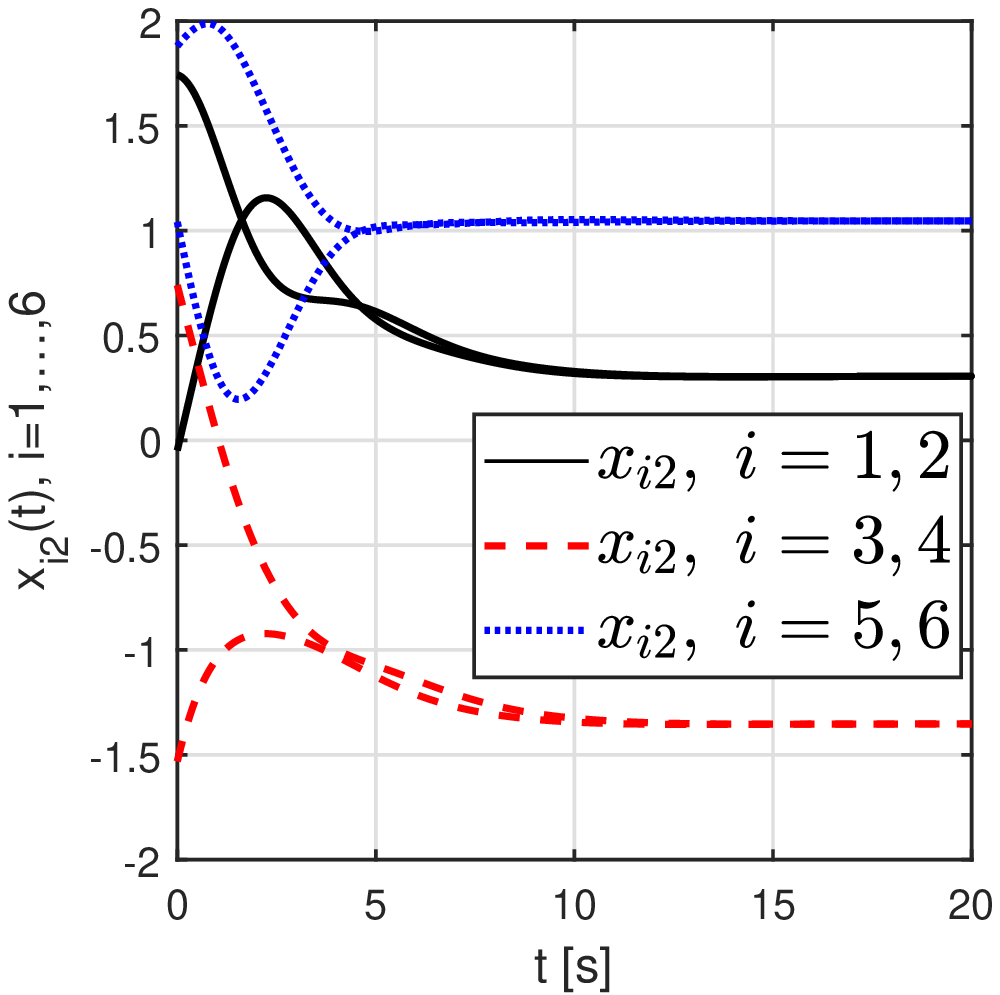}}\hfill
\subfloat[$\|\m{u}\|_{\infty}$]{\includegraphics[width=0.22\linewidth]{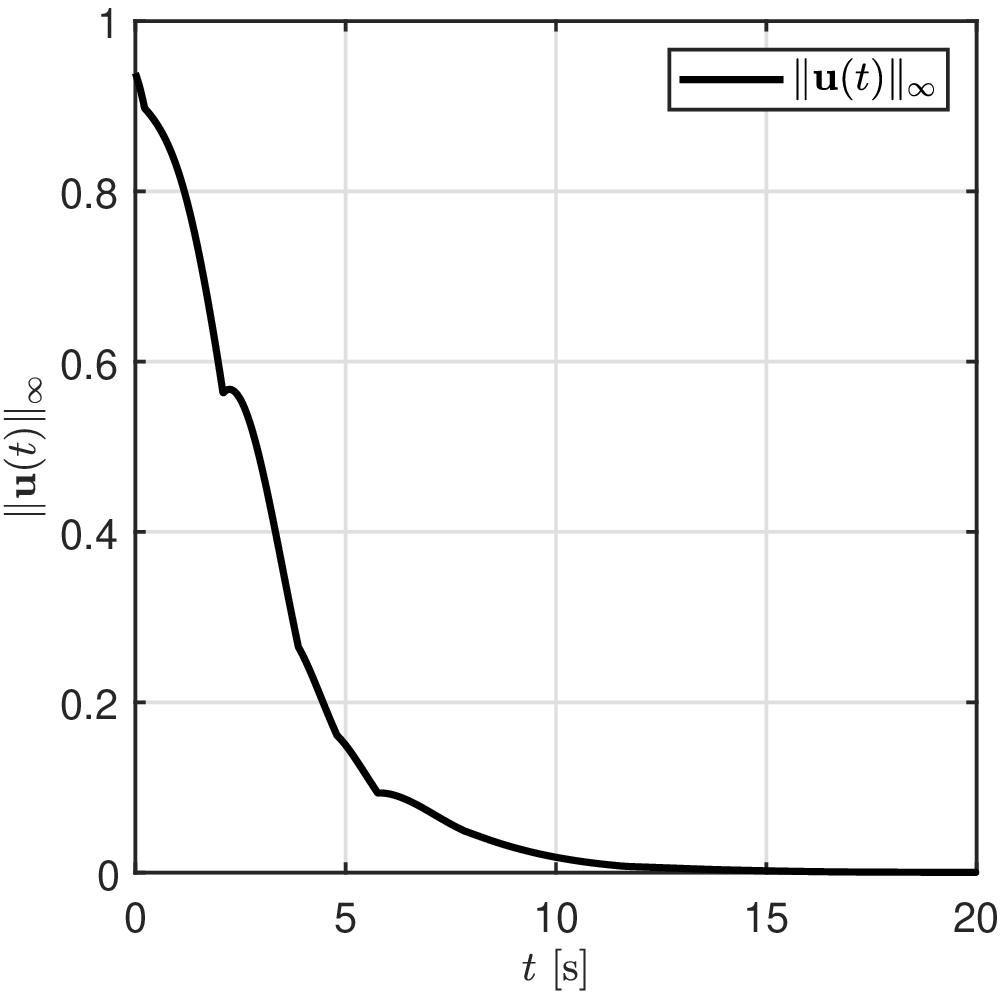}}
\caption{The agents converge to three clusters in 2D under the matrix-scaled consensus algorithms \eqref{eq:msc_single_integrator} (Figs. (a)--(d)) and \eqref{eq:msc_nonl} (Figs. (e)--(h)).}
\label{fig:msc_sim1}
\end{figure*}
\begin{figure*}
\centering
\subfloat[$\m{x}_i(t)$]{\includegraphics[width=0.22\linewidth]{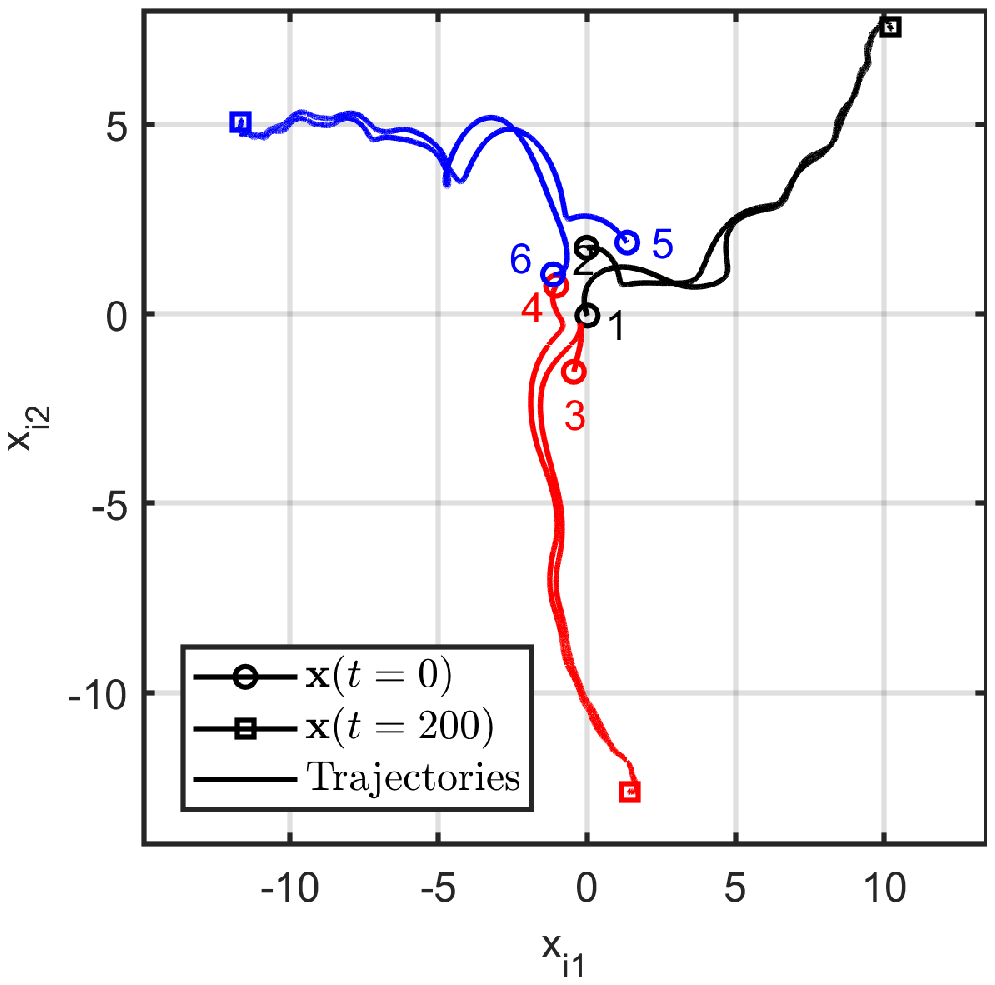}}
\hfill
\subfloat[$\bm{\theta}_i(t),\hat{\bm{\theta}}_i(t)$]{\includegraphics[width=0.22\linewidth]{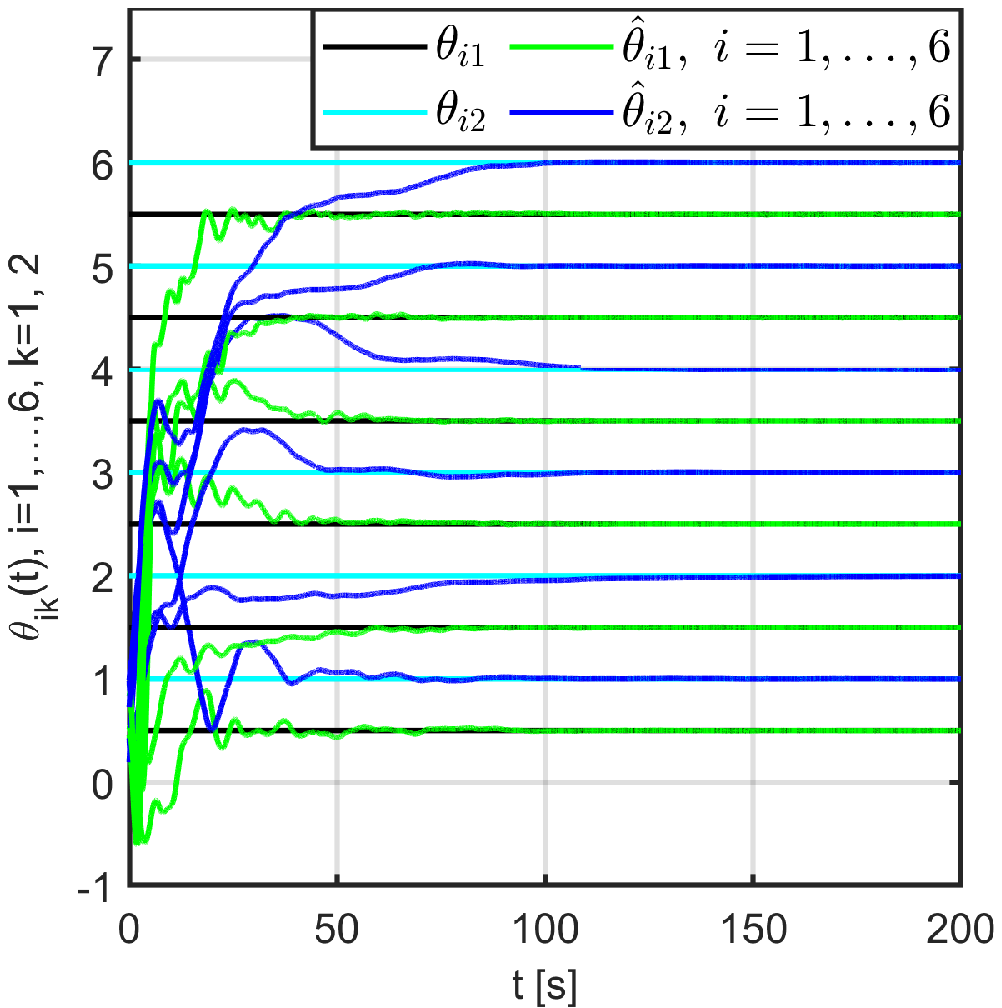}}
\hfill
\subfloat[$x_{i1}(t)$]{\includegraphics[width=0.22\linewidth]{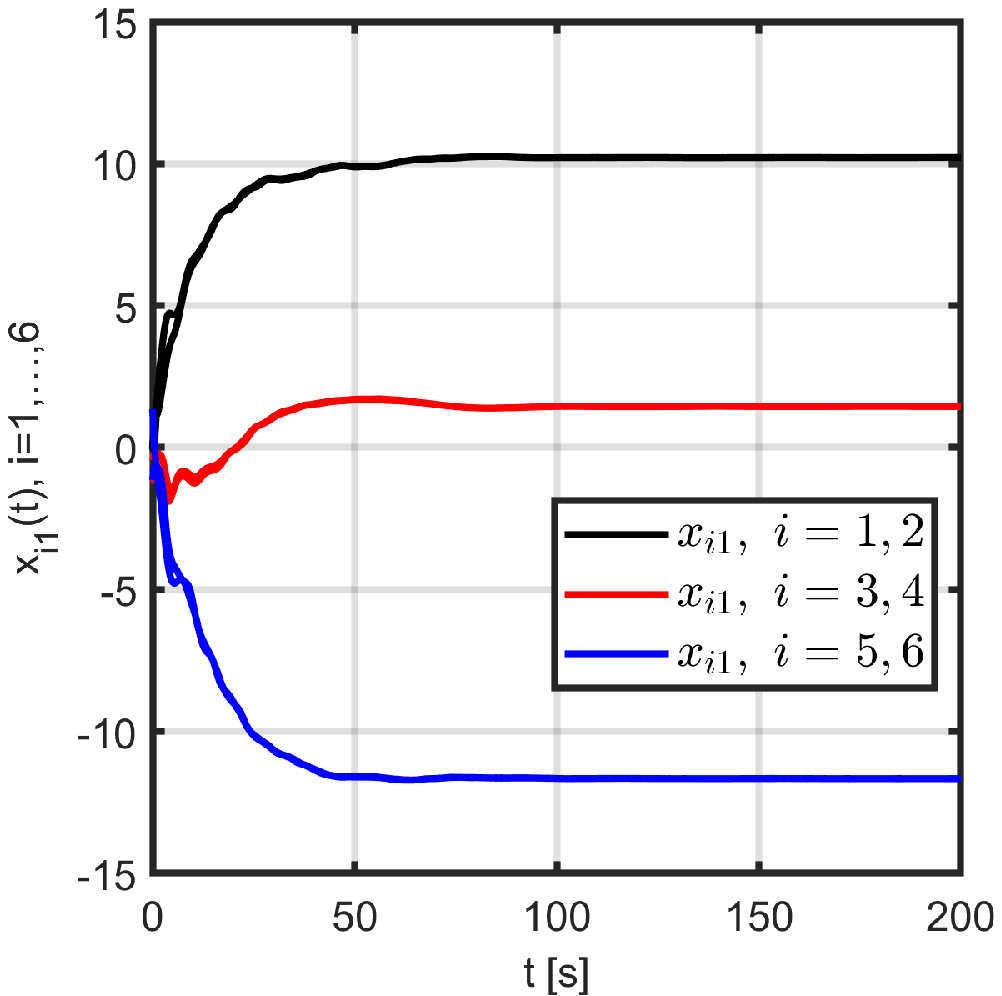}}
\hfill
\subfloat[$x_{i2}(t)$]{\includegraphics[width=0.22\linewidth]{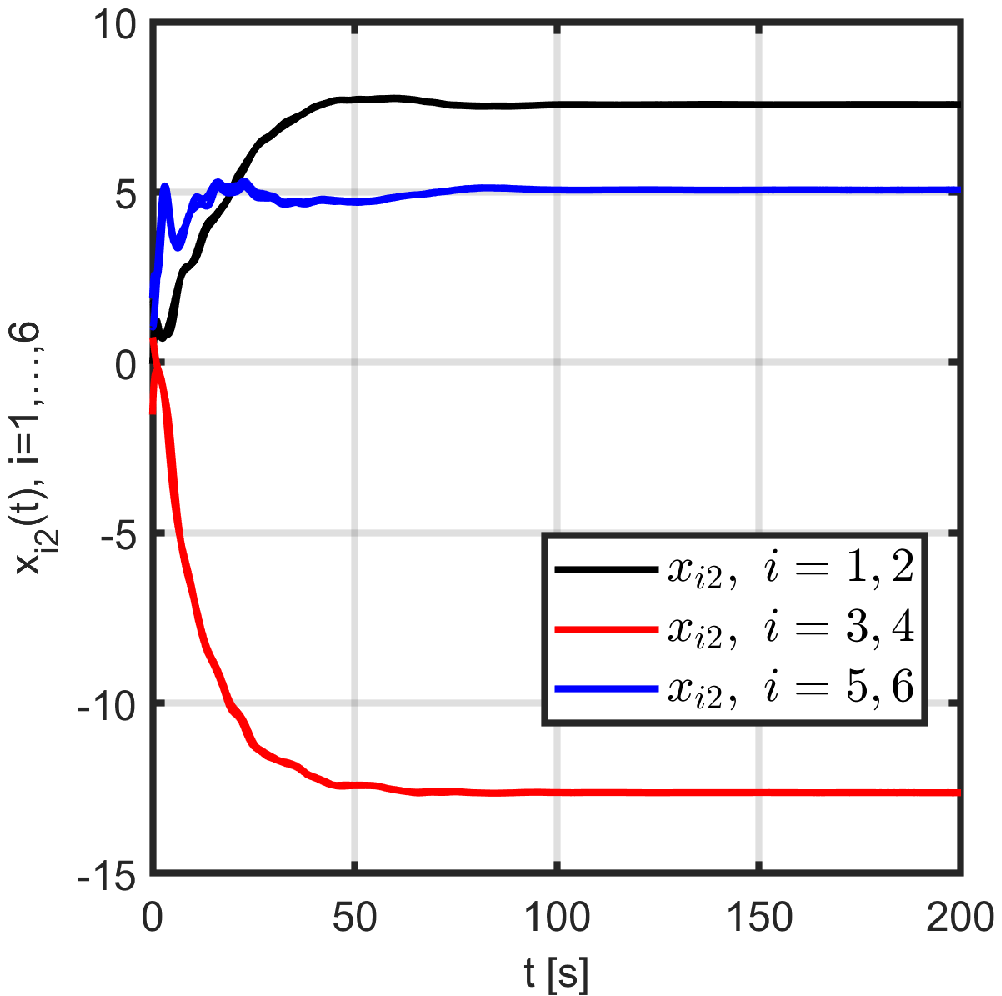}}
\caption{The six-agent system under the adaptive MSC algorithm \eqref{eq:AMWC-1}--\eqref{eq:AMWC-2}.}
    \label{fig:msc_sim2_adaptive}
\end{figure*}
In this section, we provide simulations of the theoretical results in Sections \ref{sec:single_integrator}, \ref{sec:linear_agents}. {Except for the first simulation in subsection \ref{subsec:single-integrator}, in which we simulate the Example~\ref{eg:snowflake},} in all other simulations, a six-agent system in the two dimensional space ($d=2$) is considered. The interaction between agents are captured by an undirected cycle of six vertices {with unity edge weights ($w_{ij}=1,~\forall (i,j)\in \mc{E}$).}  Denote the SO(2) rotation matrix of angle $\theta$ (rad) by {$\mathbf{R}(\theta) = \begin{bmatrix}
\cos(\theta)&-\sin(\theta)\\\sin(\theta) & \cos(\theta)
\end{bmatrix} $}. The scaling matrices are chosen as $\mathbf{S}_1 =  \mathbf{S}_2=\mathbf{R}(\frac{\pi}{3})$ (positive definite), $\mathbf{S}_3 =  \mathbf{S}_{4}=-\mathbf{I}_2$ (negative definite), and $\mathbf{S}_{5} = \mathbf{S}_{6}=\mathbf{R}(\frac{5\pi}{3})$ (positive definite).

\subsection{Matrix-scaled consensus of single integrators}
\label{subsec:single-integrator}
{
First, we simulate a 18-agent system in Example~\ref{eg:snowflake} under the MSC algorithm \eqref{eq:msc_single_integrator} over a connected network. The initial position of each agent $i\in \{1,\ldots,18\}$ is randomly generated by $\m{x}_i'(0) = \begin{bmatrix}
    3\text{sign}(\m{S}_i) + \varepsilon_{i1}\\
    \varepsilon_{i2} \\
    1
\end{bmatrix}$, where $\varepsilon_{i1},\varepsilon_{i2}$ are random variables which follows a uniformly distribution on interval $[-2,2]$. Fig.~\ref{fig:snowflake} depicts the evolution of the system projected onto the Oxy plane. A snowflake shape is asymptotically taken up by the agents after 10 seconds, as designed in Example~\ref{eg:snowflake}.}

Second, Figs.~\ref{fig:msc_sim1} (a)--(d) show a simulation of the six-agent system discussed in the beginning of this section under the MSC algorithm \eqref{eq:msc_single_integrator} for 20 seconds. The initial condition $\m{x}(0)$ was randomly selected. The agents' trajectories asymptotically converge to three clusters in 2D, which are vertices of an equilateral triangle. Next, we simulate the six-agent system under the nonlinear MSC algorithm \eqref{eq:msc_nonl} with the same initial state $\m{x}(0)$. We would like the input of each agent to satisfy $\|\m{u}_i(t)\|_{\infty}<\beta=1$. This objective is achieved by selecting $\m{f}(\cdot)=(\max_i|N_i|)^{-1}\text{tanh}(\cdot)=0.5\text{tanh}(\cdot)$. Simulation results are depicted in Figs.~\ref{fig:msc_sim1}(e)--(h). Observe that $\m{x}(t)$ converges to the same point as the previous simulation, $\|\m{u}_i\|_{\infty}\le 1,\forall t\ge 0$, the settling time (which is defined as the first time $\m{x}(t)$ enters without escaping $B_{5\%}=\{\|\m{x}(t)-\lim_{t\to+\infty}\m{x}(t)\|<0.05\}$) becomes larger (approx. 10 sec in comparison with 5 sec). 

\subsection{Matrix-scaled consensus of single-integrators with uncertain parameters}
\label{subsec:single-integrator-disturbance}
\begin{figure*}[t]
    \centering
\subfloat[$\m{A}$ is Hurwitz.]{\includegraphics[width=0.3\linewidth]{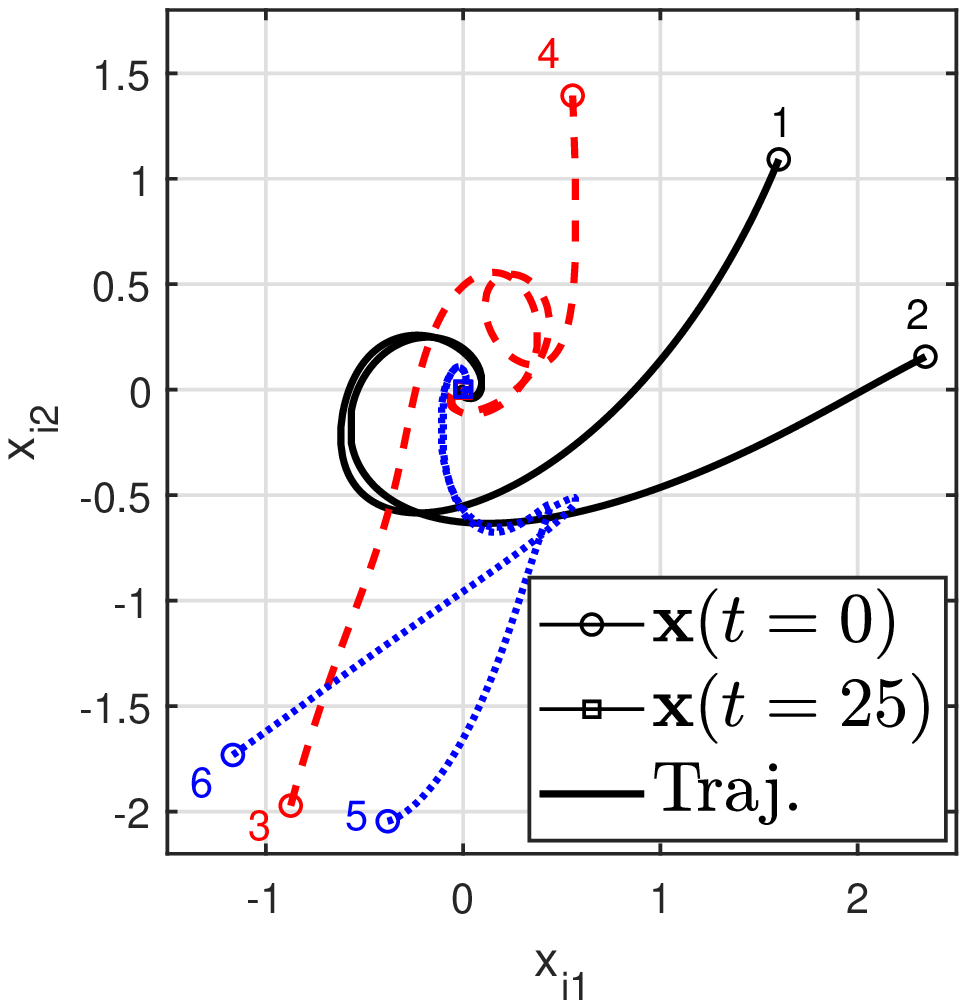}}
\hfill
\subfloat[$\m{A}$ has a pair of imaginary eigenvalues, $\m{A}+\m{A}^\top \neq \m{0}_{d\times d}$.]{\includegraphics[width=0.3\linewidth]{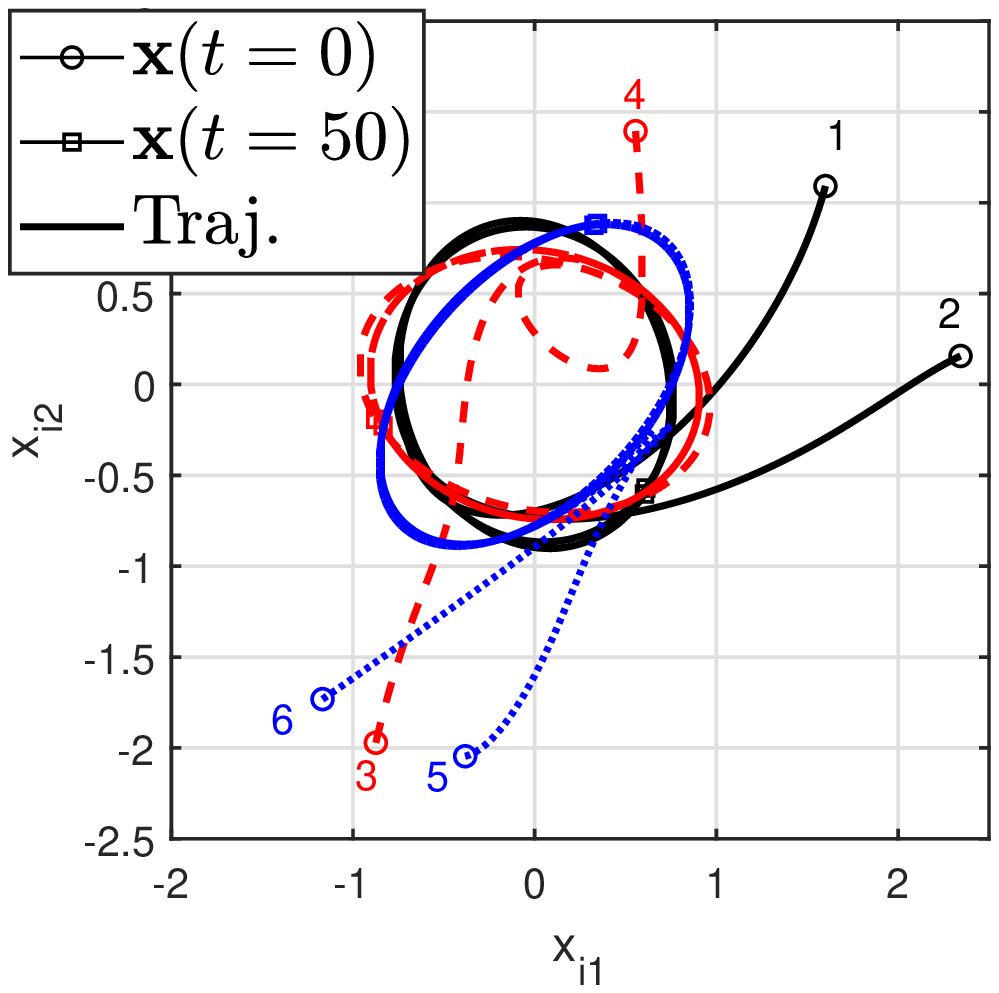}}
\hfill
\subfloat[$\m{A}$ is unstable.]{\includegraphics[width=0.3\linewidth]{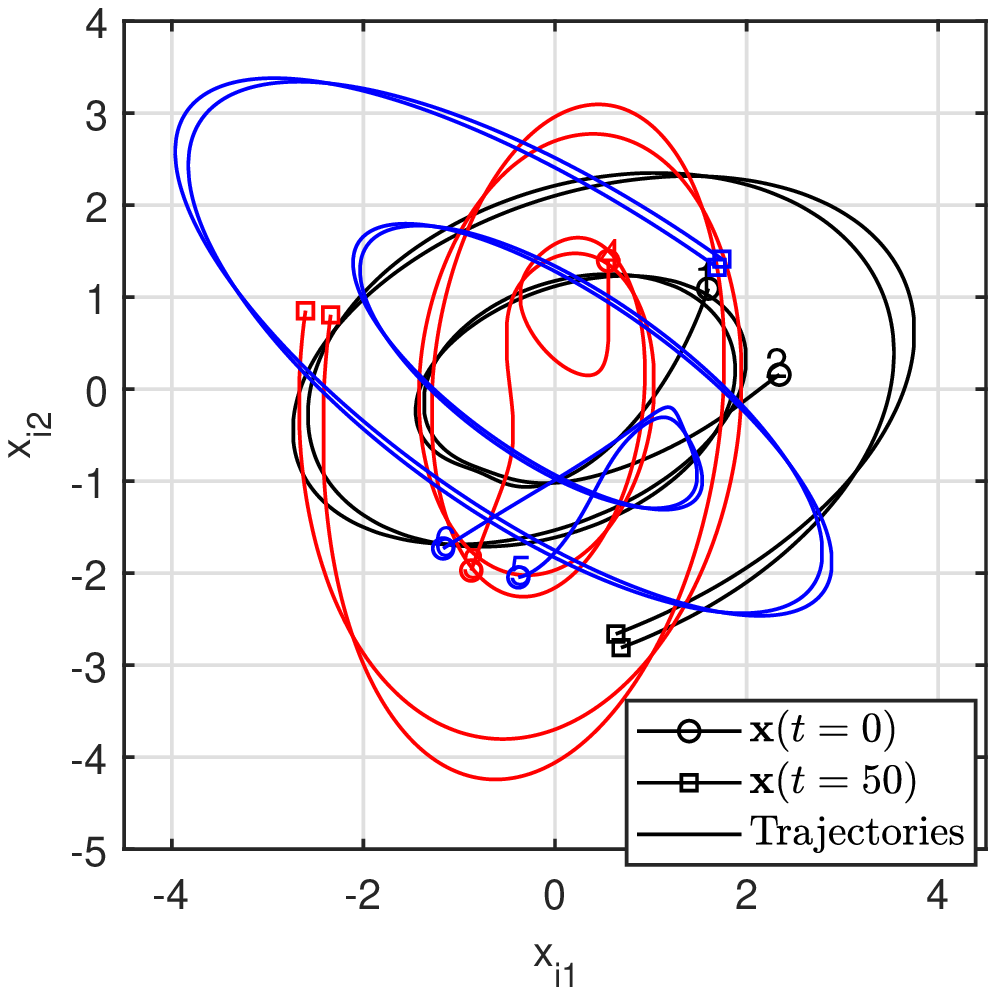}}\\
\subfloat[$\m{A}=-\m{A}^\top$ and has a pair of imaginary eigenvalues.]{\includegraphics[width=0.3\linewidth]{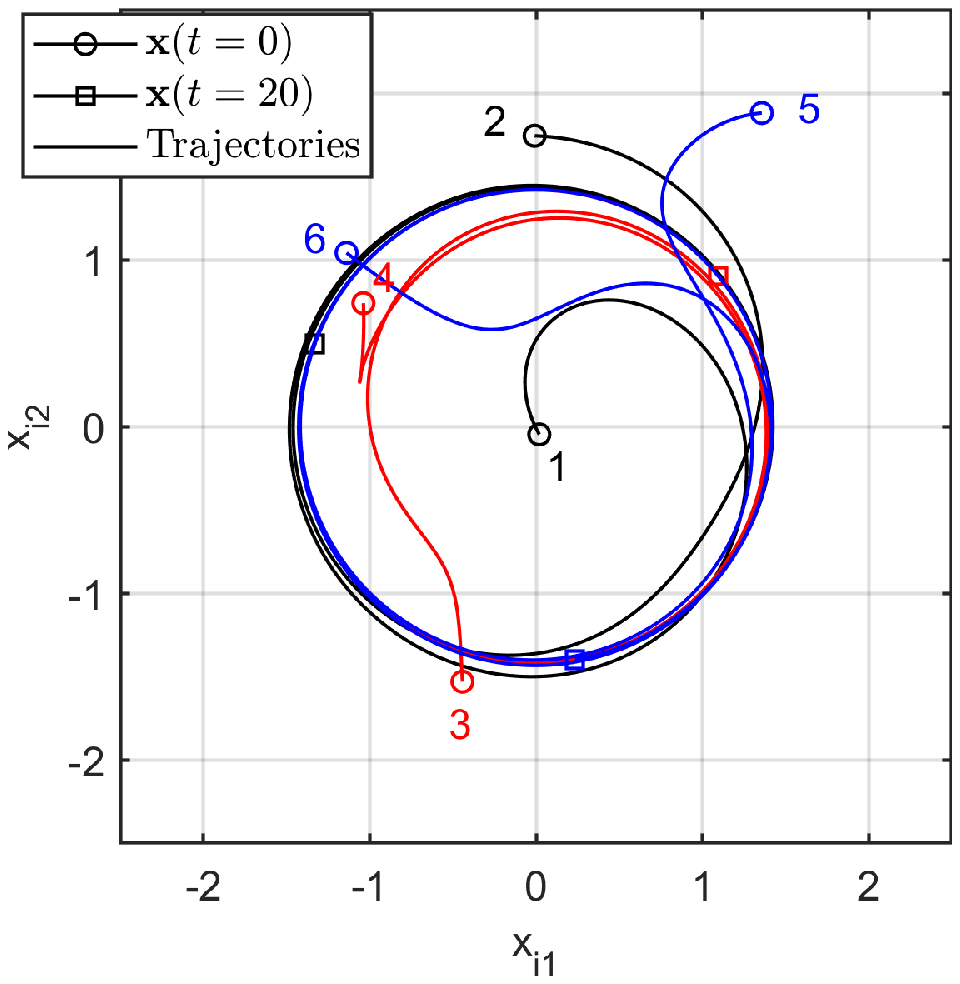}}
\hfill
\subfloat[State variables $x_{i1}(t)$ corresponding to trajectories in (d).]{\includegraphics[width=0.3\linewidth]{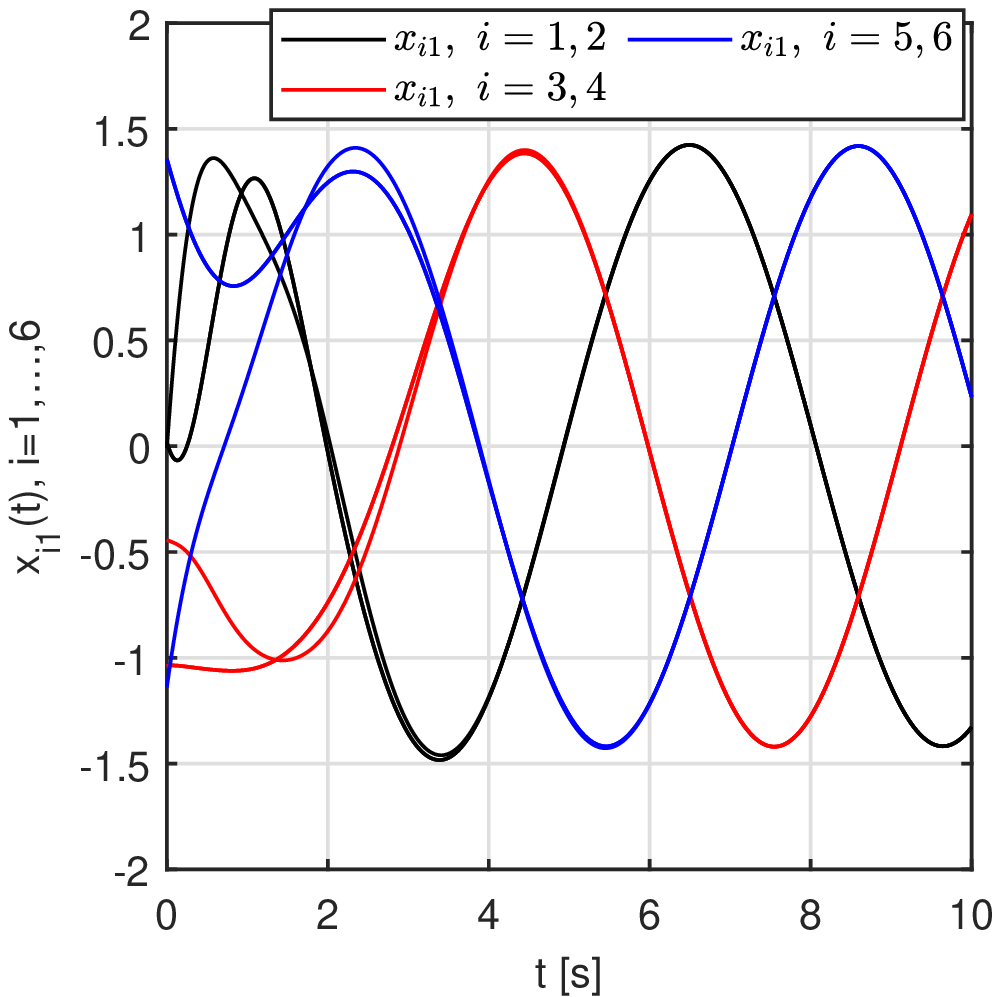}}
\hfill
\subfloat[State variables $x_{i2}(t)$ corresponding to trajectories in (d).]{\includegraphics[width=0.3\linewidth]{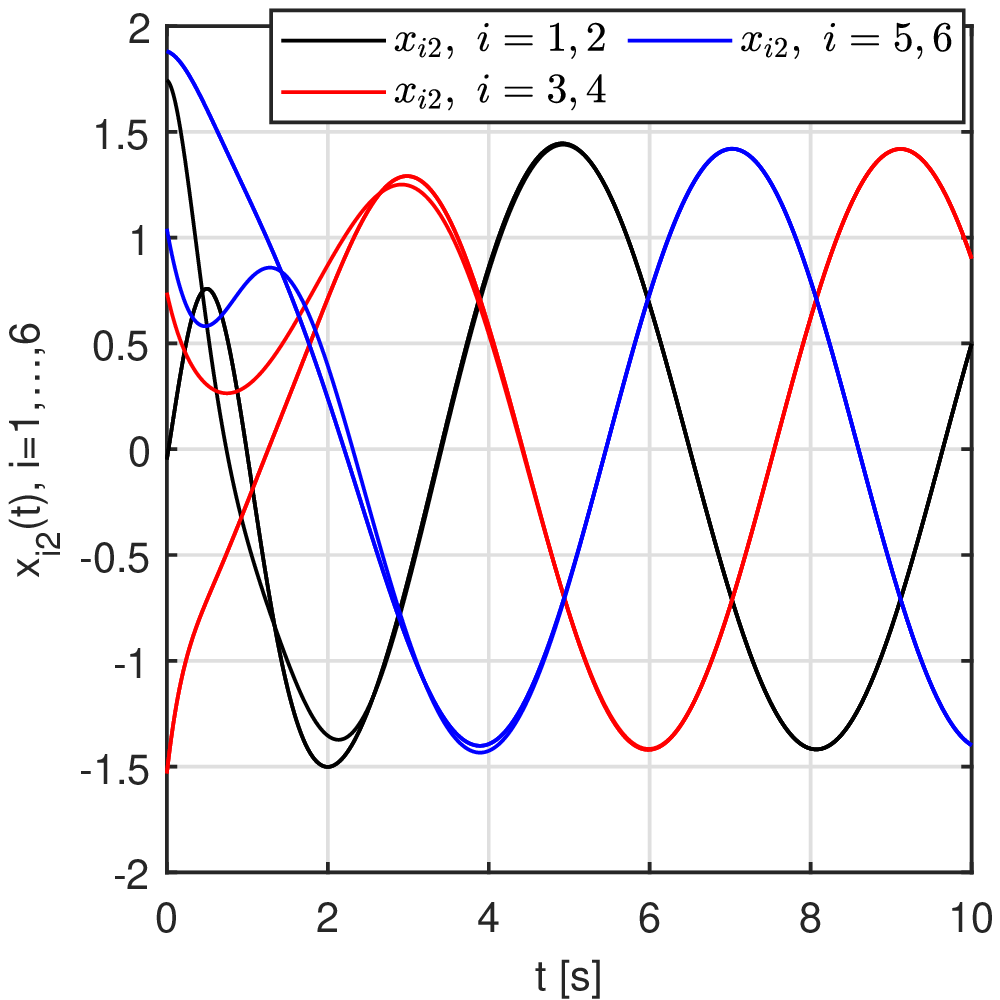}}
\caption{The six-agent system modeled by the simplified linear model \eqref{eq:simple_homo_allagents} with different choices of the matrix $\m{A}$.\label{fig:msc_sim4}}
\end{figure*}
Next, we consider a system consisting of 6-single-integrator agents with parametric uncertainties 
\begin{align*}
    \bm{\phi}_i(t)= \begin{bmatrix}
        0.2\sin(t) & 0.5-0.2\sin(\frac{it}{\pi})\\
        -0.2\sin(\frac{t}{i\pi}) & 0.1\cos(\frac{t}{\pi})
    \end{bmatrix},~i=1,\ldots, 6,
\end{align*}
which satisfy the persistently exciting condition, and $\gamma_i>0$ are randomly chosen between $[1,10]$. The constant unknown parameters are $\bm{\theta}_i = [i-0.5,~i]^\top$, and the initial estimates $\hat{\bm{\theta}}_i(0)$ are randomly generated. Simulation results depicted in Fig.~\ref{fig:msc_sim2_adaptive} show that $\m{x}_i \to \mc{C}_{S}$, $\hat{\bm{\theta}}_i(t) \approx \bm{\theta}_i$ at $t=200$ seconds (Fig.~\ref{fig:msc_sim2_adaptive} (b)). Due to the unknown parameters, the convergence rate, however, is much slower in comparison with the ideal case.
\subsection{Matrix-scaled consensus of simplified linear dynamical agents}
We simulate the matrix-scaled consensus algorithm for the six-agent system with the simplified linear model \eqref{eq:simple_homo_allagents} for  different matrices $\m{A}$ as follows
\begin{itemize}
    \item { $\m{A}=0.5\begin{bmatrix}
      -1 & 1 \\ -1 & 0  
    \end{bmatrix}$} has eigenvalues $-0.25\pm \frac{\sqrt{3}}{4}\imath$,
    \item { $\m{A}=\begin{bmatrix}
      0 & 1 \\ -1 & 0  
    \end{bmatrix}$} is skew-symmetric, has eigenvalues $\pm \imath$,
    \item { $\m{A}=0.5\begin{bmatrix}
      0 & 1 \\ -0.5 & 0  
    \end{bmatrix}$} has a pair of imaginary eigenvalues, $\m{A}$ is not skew-symmetric, 
    \item { $\m{A}=0.5\begin{bmatrix}
      1 & 1 \\ -1 & 0  
    \end{bmatrix}$} has eigenvalues $0.25\pm \frac{\sqrt{3}}{4}\imath$,
\end{itemize}
where $\imath^2 = -1$. In this subsection, the coupling gain is chosen as $c=2$. 

Simulation results are depicted in Fig.~\ref{fig:msc_sim4}. The agents' trajectories are affected by both the unforced dynamics (determined by $\m{A}\m{x}_i$) and the matrix-scaled consensus algorithm. Clearly, when $\m{A}$ is stable (unstable), all agents' states converge to 0 (resp., grow unbounded), as shown in Fig.~\ref{fig:msc_sim4}(a) (resp. Fig.~\ref{fig:msc_sim4}(c)). In case $\m{A}$ is marginally stable, the agents asymptotically reach a MSC with regard to a trajectory of the system $\dot{\m{r}}=\m{P}\m{A}\m{P}^{-1}\m{r}$. In case $\m{A}$ is skew-symmetric {(Fig.~\ref{fig:msc_sim4}(d))}, agents move on a circle centered at the origin, and if $\m{A}$ is not skew-symmetric, agents with the same scaling matrix move on a same elliptical trajectory {(Fig.~\ref{fig:msc_sim4}(b))}. Each trajectory of $\m{x}_i(t)$ differs from a common solution of $\dot{\m{r}}=\m{P}\m{A}\m{P}^{-1}\m{r}$ by the scaling matrix $\m{S}_i$.
\subsection{Matrix-scaled consensus of general homogeneous linear dynamical agents}
\begin{figure*}[t]
    \subfloat[$\|\hat{\m{x}}_i(t)\|$]{\includegraphics[width=0.3\linewidth]{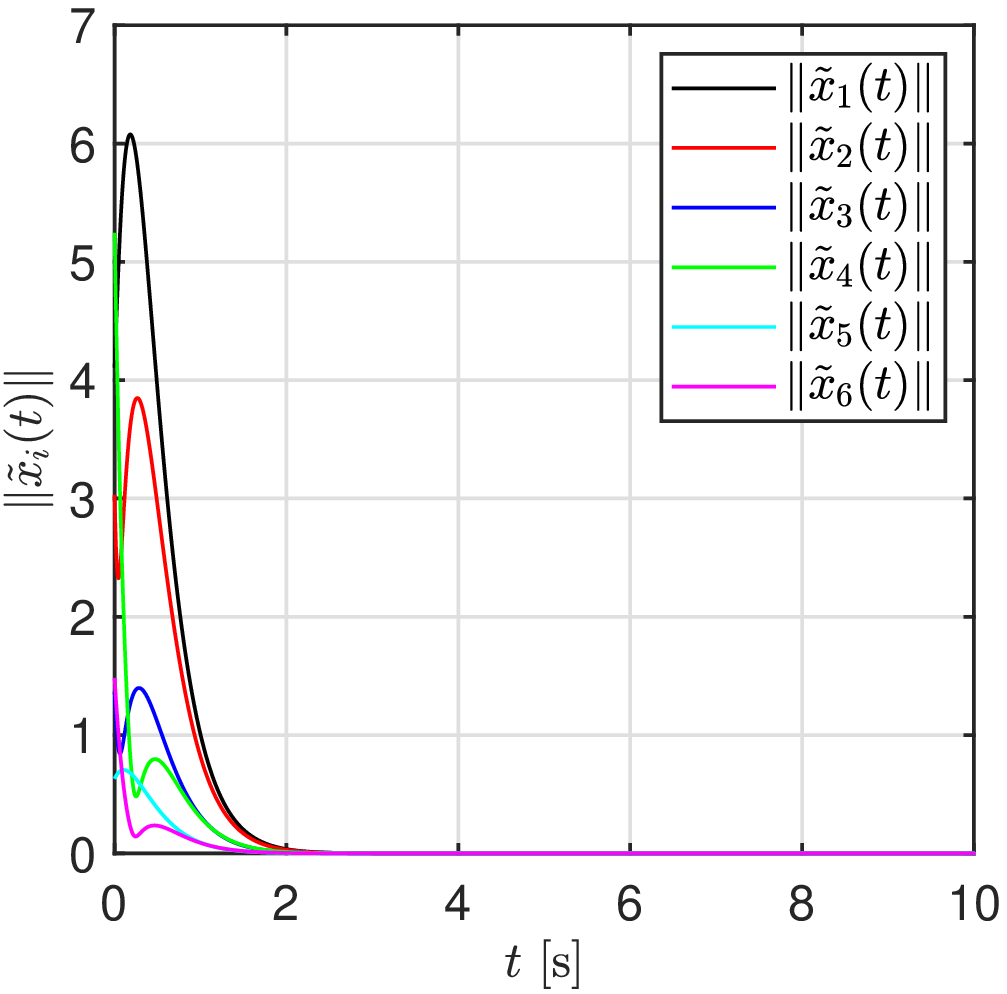}}\hfill
    \subfloat[$\eta_{i1}(t)$]{\includegraphics[width=0.3\linewidth]{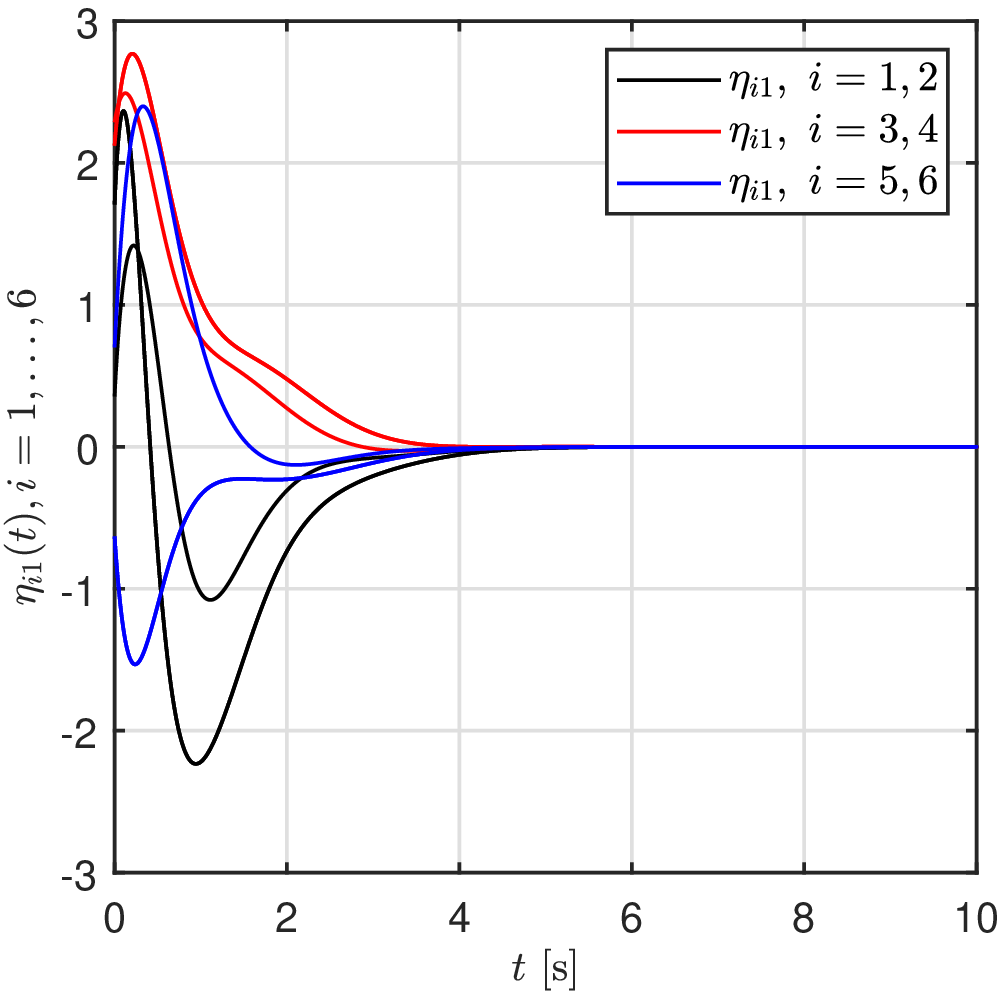}}\hfill
    \subfloat[$\eta_{i2}(t)$]{\includegraphics[width=0.3\linewidth]{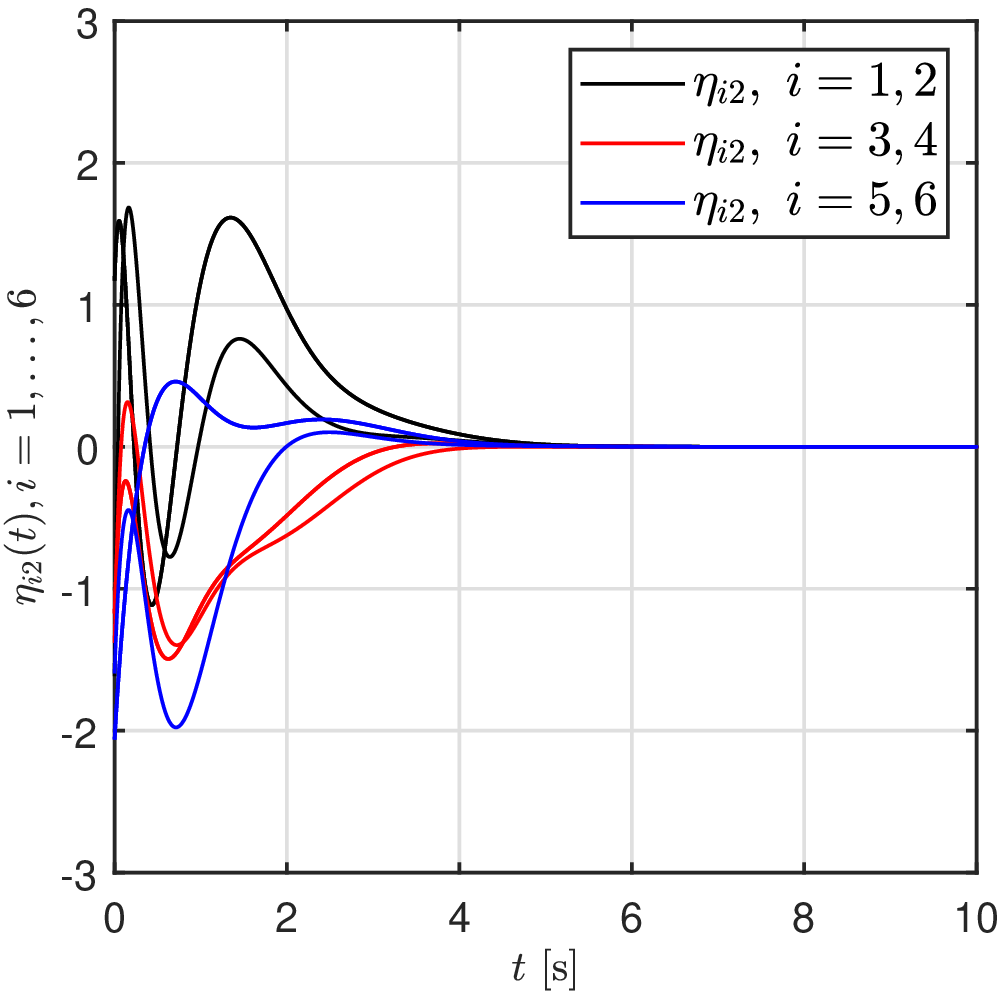}}\hfill\\
    \subfloat[$\m{x}_i$]{\includegraphics[width=0.3\linewidth]{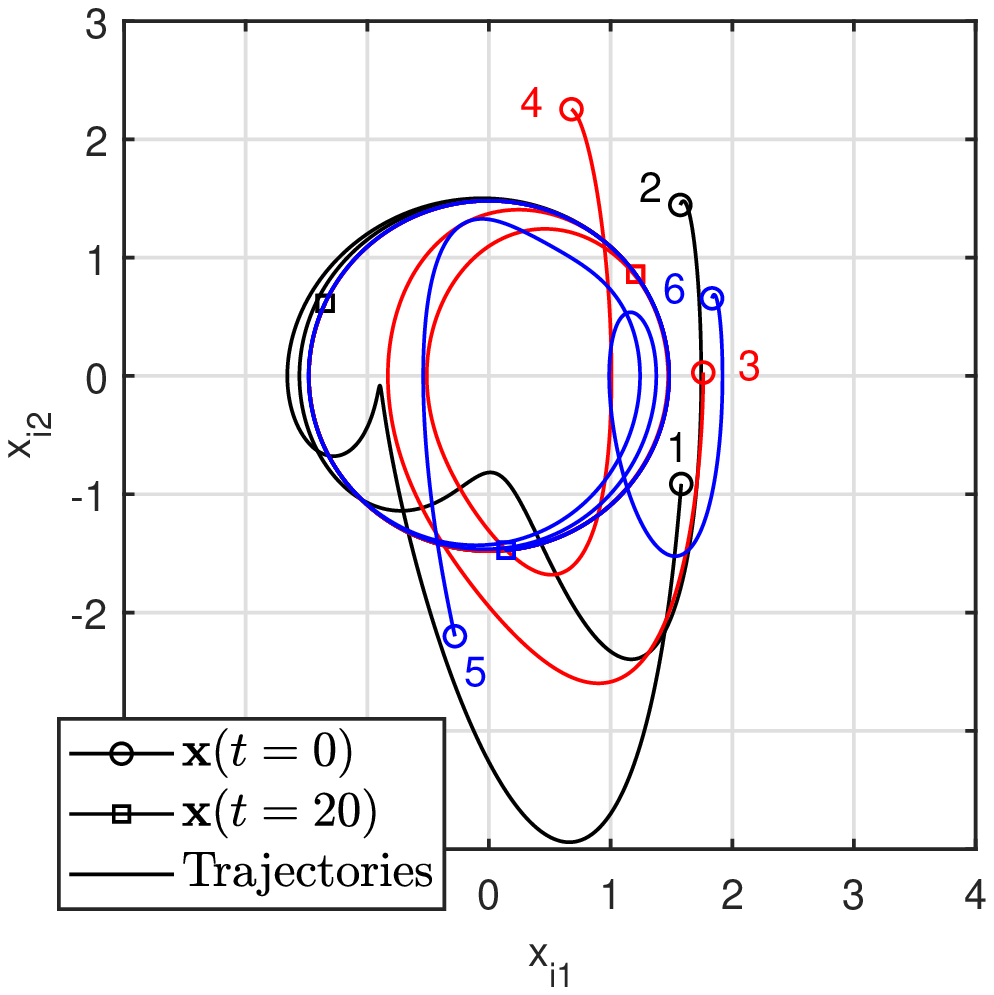}}\hfill
    \subfloat[${x}_{i1}$]{\includegraphics[width=0.3\linewidth]{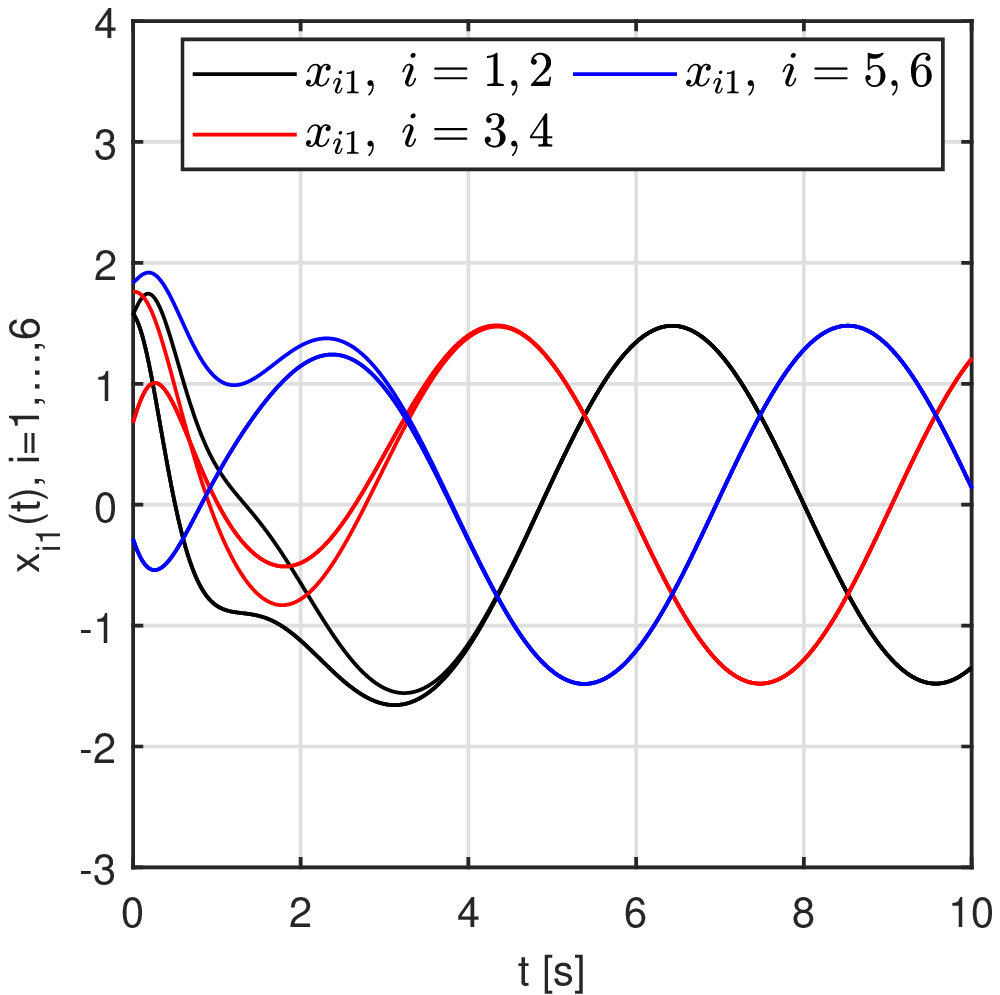}}\hfill
    \subfloat[${x}_{i2}$]{\includegraphics[width=0.3\linewidth]{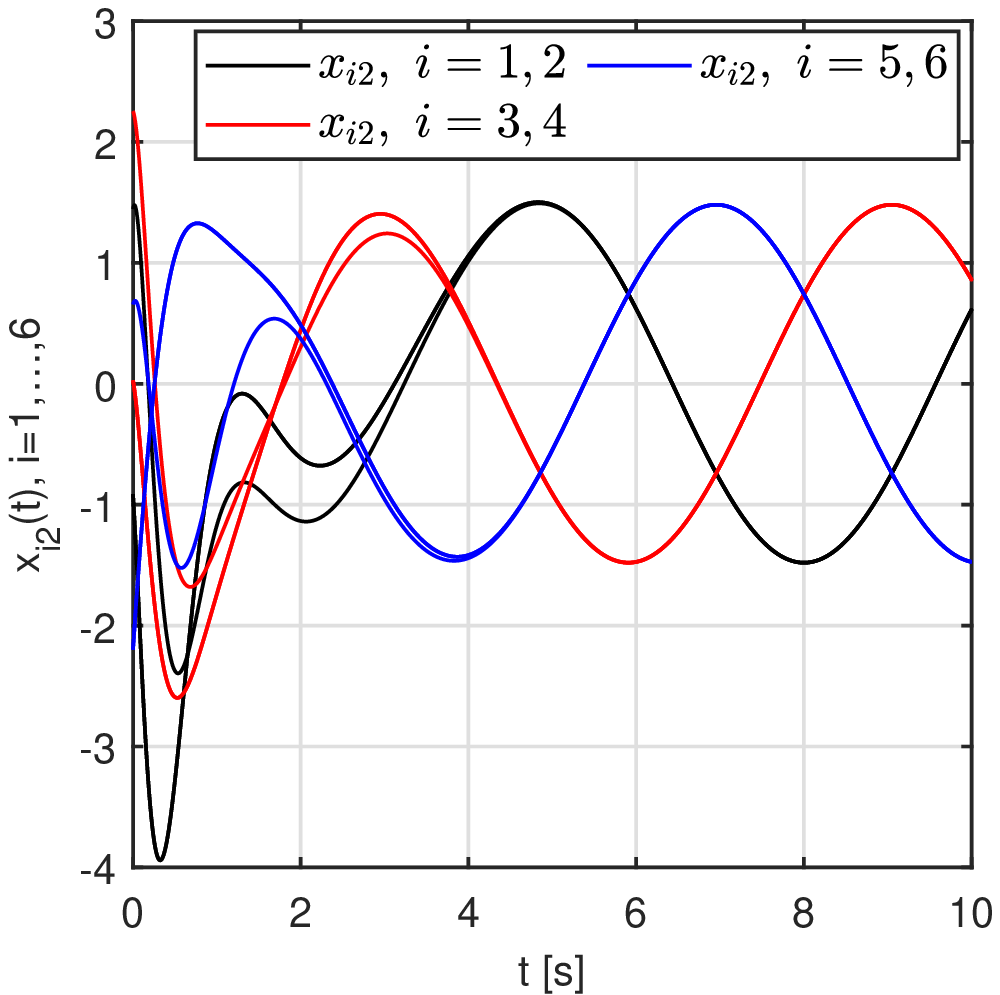}}
    \caption{Simulations of the homogeneous linear agent system~\eqref{eq:sim_homo} under the matrix-scaled consensus algorithm \eqref{eq:homo_obsv}.\label{fig:sim_homo}}
\end{figure*}

In this subsection, let the agent be modeled by a linear oscillator with a control input
\begin{subequations}\label{eq:sim_homo}
\begin{align}
    \dot{x}_{i1} &= x_{i2}, \\
    \dot{x}_{i2} &= -\omega^2 x_{i1} + u_i,\\
    y_{i} &=x_{i1},~\, i=1,\ldots, 6,
\end{align}
\end{subequations}
and $\omega = 1$. The corresponding system matrices are $\m{A} = \begin{bmatrix}
      0 & 1 \\ -1 & 0  
\end{bmatrix}$, $\m{B} = \begin{bmatrix}
      0 \\ 1
\end{bmatrix}$, $\m{C} = [1,~0]$, $\m{K}=[-3,-4]$, $\m{H}=[-8, -15]^\top$, and $c=2$. Simulation corresponding to a randomly generated initial condition is displayed in Fig.~\ref{fig:sim_homo}. The state estimate error $\tilde{\m{x}}_i=\hat{\m{x}}_i -\m{x}_i$ and the auxiliary variable $\bm{\eta}$ vanish exponentially fast. Figures \ref{fig:sim_homo}(d), (e), (f) depict the trajectories of 6 agents. It can be observed that after the transient time, each agent's trajectory differs from a common solution of $\dot{\m{r}}=\m{P}\m{A}\m{P}^{-1}\m{r}$ by its scaling matrix.

\subsection{Matrix-scaled consensus of general heterogeneous linear dynamical agents}
Consider a system of six heterogeneous linear agents, where matrix $\m{A}$ is the same as in the previous subsection, $\bar{\m{A}}_i$ are randomly generated with entries belong to the interval $[-0.2,~0.2]$. The matrices $\m{B}_i=\m{B} + \bm{\Delta}_B$, where $\bm{\Delta}_B=\begin{bmatrix}
    0 & \delta_{b_i}
\end{bmatrix}^\top$, with $\delta_{b_i}$ be randomly generated to take values on the interval $[-0.5,0.5]$, and $\m{C}_i = \m{C},~~\forall i=1,\ldots,n$. The matrices $\m{K}_i$ and $\m{H}_i$ are designed so that each matrix $\m{A}_i+\m{B}_i\m{K}_i$ have eigenvalues $-2,-2$ and each matrix $\m{A}_i+\m{H}_i\m{C}_i$ have eigenvalues $-4, -4$, for $i=1,\ldots,6$. The control gains are given as $c=2$, $\beta_1=2$ and $\beta_2=5$. 

\begin{figure*}
\centering
    \subfloat[$\m{x}_i$]{\includegraphics[width=0.3\linewidth]{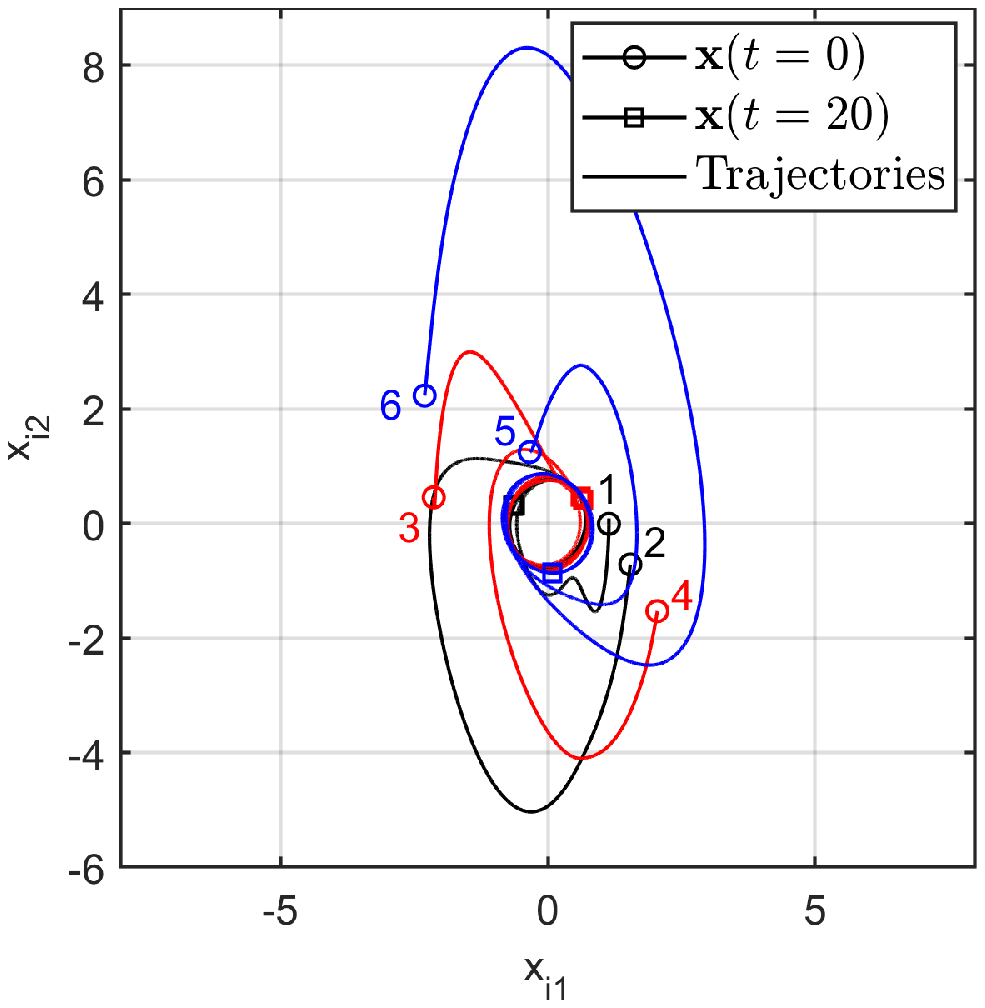}} \hfill
    \subfloat[$x_{i1}$]{\includegraphics[width=0.3\linewidth]{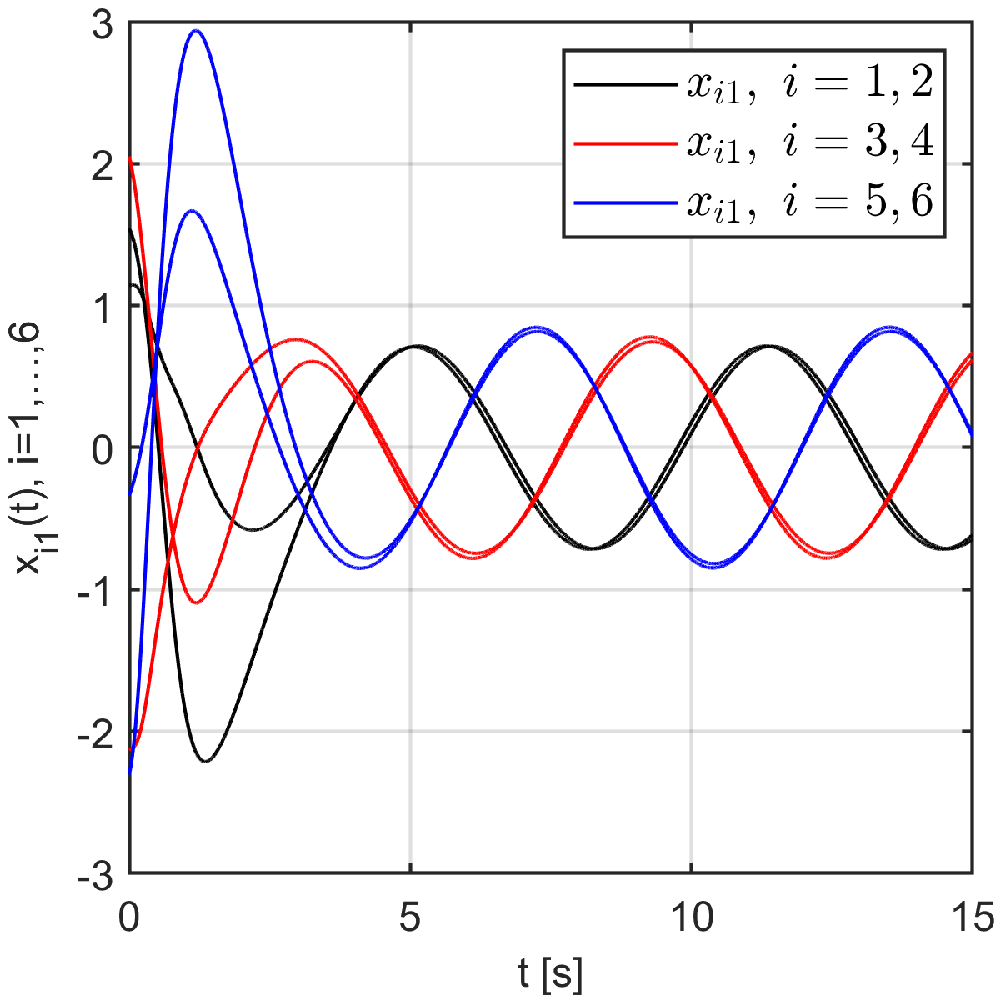}} \hfill
    \subfloat[$x_{i2}$]{\includegraphics[width=0.3\linewidth]{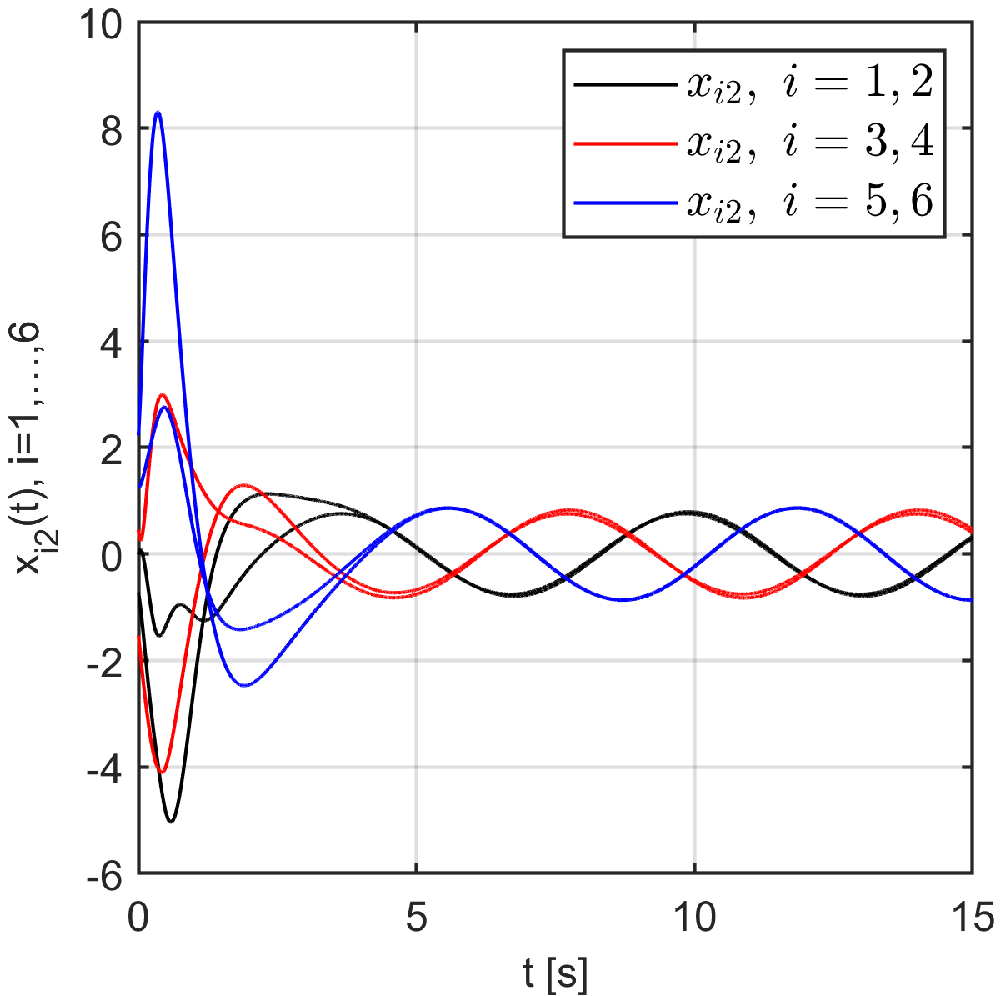}} \\
    \subfloat[$\m{x}_i$]{\includegraphics[width=0.3\linewidth]{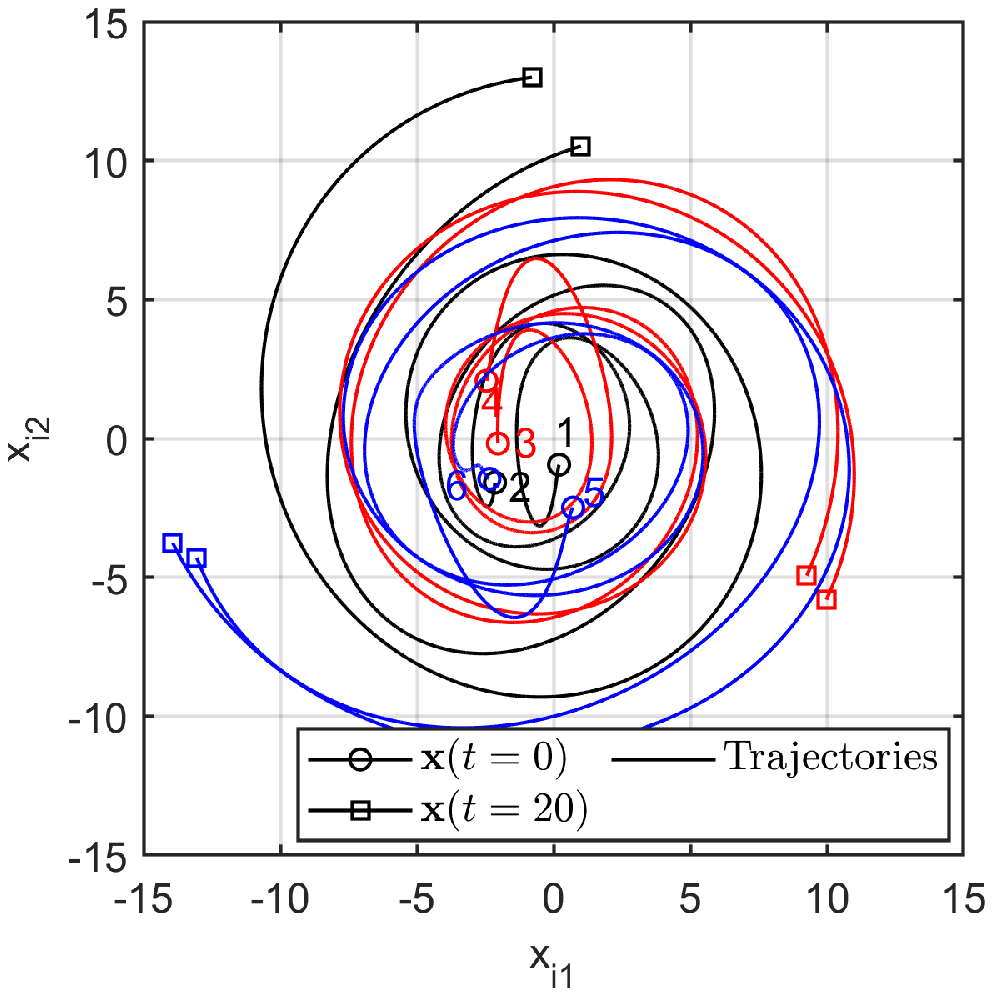}} \hfill
    \subfloat[$x_{i1}$]{\includegraphics[width=0.3\linewidth]{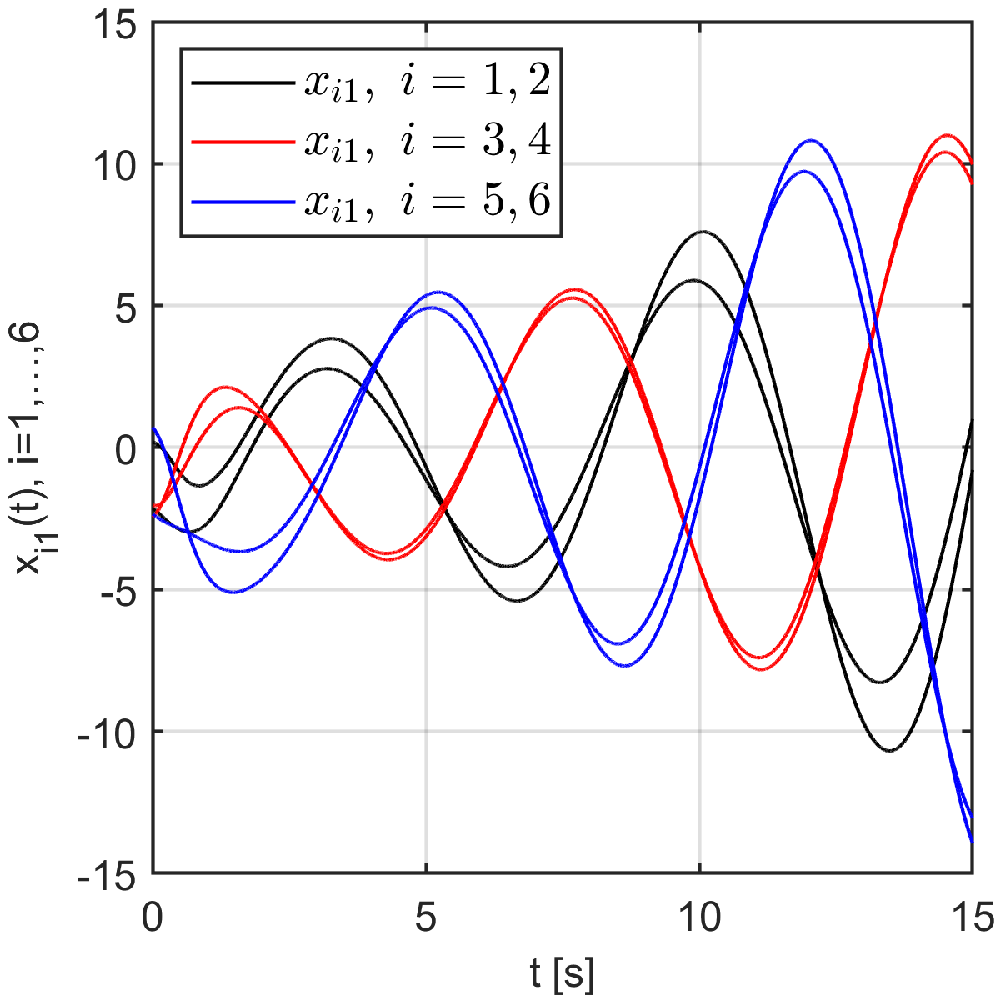}} \hfill
    \subfloat[$x_{i2}$]{\includegraphics[width=0.3\linewidth]{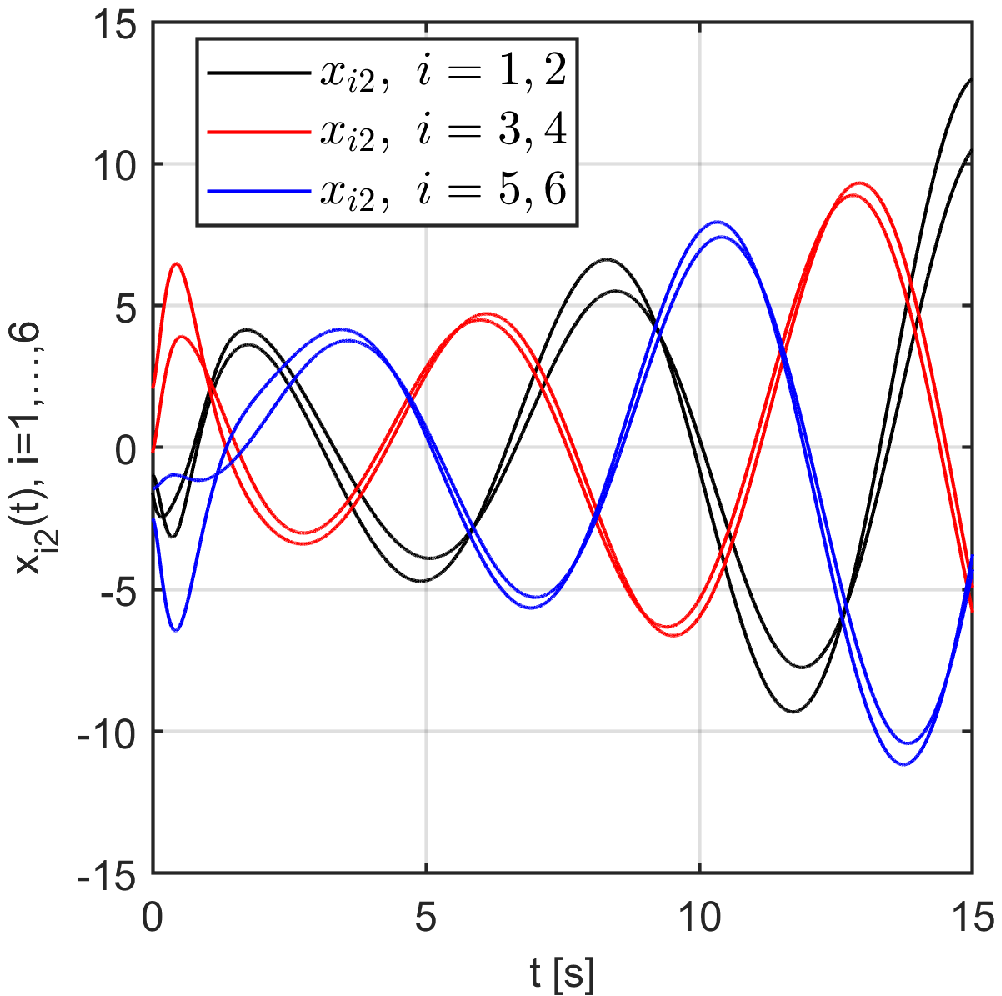}}
    \caption{Simulation of the heterogeneous linear systems (i) under the algorithm \eqref{eq:het_msc_dtb_obs}: (a), (b), (c); and (ii) under the algorithm \eqref{eq:het_msc_dtb_obs} without the signum term in $\hat{\m{u}}$: (d), (e), (f).}
    \label{fig:sim_heterogeneous_systems}
\end{figure*}

Figures~\ref{fig:sim_heterogeneous_systems}(a), (b), (c) show that $\m{x}(t) \to \mc{C}_S$ asymptotically under the algorithm \eqref{eq:het_msc_dtb_obs}. The state vector of each agent $i$ asymptotically converges to a trajectory, which differ from a trajectory of $\dot{\m{r}}=\m{P}\m{A}\m{P}^{-1}\m{r}$ by the matrix scaling weight $\m{S}_i^{-1}$. 

For comparison, the same heterogeneous system is simulated without using the control law \eqref{eq:het_msc_dtb_obs}. Without the signum terms, Fig.~\ref{fig:sim_heterogeneous_systems}(d), (e), (f) show that the system diverges. Thus, simulation results are consistent with the analysis.

\section{Conclusions}
\label{sec:6}
{In this paper, several matrix-scaled consensus algorithms were designed for single-integrators and linear time-invariant agents interacting over an undirected network. The asymptotic behaviors of the proposed algorithms associate with the algebraic properties of the matrix-scaled Laplacian. The main challenges in the stability analysis comes from the asymmetry and multi-dimensionality of the matrix-scaled Laplacian. The matrix-scaled consensus space has been characterized in detail in case $d=2$, which allows these algorithms to be applicable in formation control. Simulation results have also been extensively provided to demonstrate the algorithm. For a simulation of the MSC algorithm over a large-scale network, we refer readers to \cite{Trinh2024simulation}.}

{As the study of the matrix-scaled consensus algorithm in this paper only focused on undirected graphs, our further work will focus on studying the matrix-scaled consensus algorithm with directed or signed graphs.}

\appendices
\section{Asymptotic behavior of an adaptive system}
\label{append:A}
\begin{lemma}\cite{Besanccon2000remarks} \label{lemma:PE} Given the system in the following form
\begin{subequations}
\begin{align}
\dot{\m{x}}_1 & = \m{h}(t) \m{x}_2 + \bm{\varphi}(t),~\m{x}_1\in \mb{R}^n,~\m{x}_2\in \mb{R}^m,\\
\dot{\m{x}}_2 & = \bm{\psi}(t),
\end{align}
\end{subequations}
where the functions $\m{h}: \mb{R}^+ \to \mb{R}^{n\times m}$, $\bm{\varphi}: \mb{R}^+ \to \mb{R}^n$, and $\bm{\psi}: \mb{R}^+ \to \mb{R}^m$ are such that (i) $\lim_{t\to +\infty} \|\m{x}_1(t)\| = \lim_{t\to +\infty} \|\bm{\varphi}(t)\| = \lim_{t\to +\infty} \|\bm{\psi}(t)\| = {0}$, (ii) $\m{h}(t)$, $\dot{\m{h}}(t)$ are bounded and (iii) there exist positive constants $T, \mu_1, \mu_2$ and $l$ such that for all $t\ge 0$, $\mu_1 \m{I}_n \ge \int_t^{t+T}\m{h}(\tau)^\top\m{h}(\tau) d\tau \ge \mu_2 \m{I}_n$, then $\lim_{t\to +\infty} \|\m{x}_2(t)\| = 0$.
\end{lemma}

{
\section{Proof of Lemma~\ref{lem:matrix_stability}}
\label{append:B}
Let $\bm{\Omega}_c=-c\bm{\Omega}' + \bm{\Delta}_{\Omega}$, and consider the dynamical system 
\begin{align*}
\dot{\m{x}} = \bm{\Omega}_c\m{x} = (-c\bm{\Omega}' + \bm{\Delta}_{\Omega})\m{x}.
\end{align*}
Consider the Lyapunov function $V=\m{x}^\top \m{Q}\m{x}$, we have, $\dot{V} = \m{x}^\top(\m{Q}\bm{\Omega}_c + \bm{\Omega}_c^\top\m{Q})\m{x} 
= - c\|\m{x}\|^2 + 2\m{x}^\top\m{Q}\bm{\Delta}_{\Omega}\m{x} 
\leq -(c - 2 \|\m{Q}\|\|\bm{\Delta}_{\Omega}\|) \|\m{x}\|^2 
\leq -(c - 2 {\delta}_{\Omega} \lambda_{\max}(\m{Q}))\lambda_{\min}^{-1}(\m{Q})V.$ 
Thus, $\m{x}(t)$ is globally exponentially stable. Equivalently, the matrix $\bm{\Omega}_c$ is Hurwitz. \hfill\ensuremath{\blacksquare}
}

{
\section{Proof of Theorem~\ref{thm:msc_adaptive_tunning}}
\label{app:C}
Consider the state transformation \[\m{y}_i(t) = \text{exp}(-\m{A}(t-t_0))\m{S}_i\m{x}_i(t), \,\forall i = 1,\ldots,n.\]
Without loss of generality, we can set $w_{ij}=1,~\forall (i,j)\in \mc{E}$ in the proof. Let $\bar{\m{A}} = \m{I}_n\otimes\m{A}$, we have $\m{y} = (\m{I}_n \otimes \text{exp}(-{\m{A}}(t-t_0)))\m{S}\m{x} = \text{exp}(-\bar{\m{A}}(t-t_0))\m{S}\m{x}$, and thus
\begin{align}
    \dot{\m{y}} &= -\bar{\m{A}}\text{exp}(-\bar{\m{A}}(t-t_0))\m{S}\m{x} + \text{exp}(-\bar{\m{A}}(t-t_0))\m{S}(\bar{\m{A}} - \text{sign}(\m{S})\bar{\bm{\m{C}}}(\m{L}\otimes\m{I}_d)\m{S})\m{x} \nonumber\\
    &=-|\m{S}|\bar{\m{C}}\bar{\m{L}}\text{exp}(-\bar{\m{A}}(t-t_0))\m{S}\m{x} =-|\m{S}|\bar{\m{C}}\bar{\m{L}}\m{y}, \label{eq:appC1}
\end{align}
where we have used the fact that $\bar{\m{A}}$, exp$(\bar{\m{A}}(t-t_0)$, $\m{S}$, and $\bar{\m{C}}$ are pair-wise commute. The adaptive tuning rule for $c_i$ can also be rewritten in the new state variable as follows
\begin{align}
    \dot{c}_i &= \kappa_i \sum_{j\in \mc{N}_i}(\m{y}_i - \m{y}_j)^\top |\m{S}_i| \sum_{j\in \mc{N}_i}(\m{y}_i - \m{y}_j), \label{eq:appC2}
\end{align}
where $c_i(0)>0$, $\forall i = 1,\ldots, n$. }

{Consider the Lyapunov function $V = \frac{1}{2}\m{y}^\top \bar{\m{L}}\m{y} + \sum_{i=1}^n \frac{(c_i - c^*)^2}{2\kappa_i}$, where $c^*>0$ is a constant gain. Clearly $V\geq 0$ and the derivative of $V$ along a trajectory of \eqref{eq:appC1}--\eqref{eq:appC2} is
\begin{align*}
    \dot{V} &= -\m{y}^\top \bar{\m{L}}|\m{S}|\bar{\m{C}}\bar{\m{L}}\m{y} + \sum_{i=1}^n \sum_{j\in \mc{N}_i}(\m{y}_i - \m{y}_j)^\top  (c_i - c^*)|\m{S}_i| \sum_{j\in \mc{N}_i}(\m{y}_i - \m{y}_j)\\
    &= -\m{y}^\top \bar{\m{L}}|\m{S}|\bar{\m{C}}\bar{\m{L}}\m{y} + \m{y}^\top \bar{\m{L}} (\bar{\m{C}}-c^*\m{I}_{dn})|\m{S}|\bar{\m{L}}\m{y}\\
    &=-\frac{c^*}{2} \m{y}^\top \bar{\m{L}}(|\m{S}|+|\m{S}|^\top)\bar{\m{L}}\m{y} \leq 0
\end{align*}
It follows that $V(t)\leq V(0)$ and thus $\bar{\m{L}}\m{y}$ and $c_i$ are bounded. By LaSalle's invariance principle, $\bar{\m{L}}\m{y}\to \m{0}$, or equivalently $(\m{y}_i - \m{y}_j) \to \m{0}_d$, $\forall i,j =1,\ldots,n$.
It follows that exp$(\m{A}(t-t_0))(\m{y}_i - \m{y}_j) = (\m{S}_i\m{x}_i - \m{S}_j\m{x}_j) \to \m{0}_d$, as $t\to+\infty$, or the $n$-agent asymptotically achieves a MSC. Furthermore, from Eqn.~\eqref{eq:appC2}, $c_i(t)$ are non-decreasing $\forall t \geq 0$. Combining with the fact that $c_i$ are upper bounded, it follows that $\lim_{t\to+\infty} c_i(t)$ exists and is finite.\hfill\ensuremath{\blacksquare}}

\bibliographystyle{IEEEtran}
\bibliography{minh2024}

\balance
\end{document}